%% file: paper.tex
\documentclass[format=sigconf, review=false]{acmart}
\usepackage{booktabs} 
\usepackage{titlesec}

\usepackage[ruled]{algorithm2e} 

\SetAlFnt{\small}
\SetAlCapFnt{\small}
\SetAlCapNameFnt{\small}
\SetAlCapHSkip{0pt}
\IncMargin{-\parindent}
\setcitestyle{acmnumeric}

\AtBeginDocument{%
  \providecommand\BibTeX{{%
    \normalfont B\kern-0.5em{\scshape i\kern-0.25em b}\kern-0.8em\TeX}}}

\newif\ifEditMode

\EditModefalse

\usepackage{booktabs}
\usepackage{enumitem}
\usepackage{float}
\usepackage{graphicx}
\usepackage[utf8]{inputenc}
\usepackage{grffile}
\usepackage{multirow}
\usepackage[group-separator={,},per-mode=symbol]{siunitx}
\usepackage{subcaption}
\usepackage{xcolor}
\usepackage{xspace}
\usepackage{wrapfig}

\usepackage{hyperref}
\usepackage{dsfont}

\hypersetup{pdfstartview=FitH,
  pdfpagelayout=SinglePage,
  bookmarksnumbered=true,
  bookmarksopen=true,
  colorlinks=true,
  citecolor=blue,
  linkcolor= blue,
  urlcolor=blue}

\theoremstyle{plain}
\newtheorem{theorem}{Theorem}		
\newtheorem{corollary}{Corollary}		
\newtheorem*{corollary*}{Corollary}		

\theoremstyle{remark}
\newtheorem*{remark*}{Remark}

\newcommand{\R}{\mathbb{R}}		
\newcommand{\E}{\mathbb{E}}		
\newcommand{\Mset}{\mathcal{M}}		
\newcommand{\lmean}{\bar{l}}		
\newcommand{\lb}{\boldsymbol{l}}		
\newcommand{\lti}{\tilde{l}}
\newcommand{\ltib}{\tilde{\lb}}
\newcommand{\hb}{\boldsymbol{h}}		
\newcommand{\hti}{\tilde{h}}
\newcommand{\htib}{\tilde{\hb}}
\newcommand{\btau}{\bar{\tau}}
\newcommand{\ud}{\mathop{}\!\mathrm{d}}
\newcommand{\ind}{\mathds{1}}
\newcommand{\eff}{\mathds{\eta}}

\newcommand{\term}[1]{\emph{#1}}
\newcommand{\newterm}[1]{\emph{#1}}
\newcommand{\stress}[1]{\textit{#1}}

\newcommand{\parai}[1]{\noindent{\textit{#1}}\quad}
\newcommand{\paraib}[1]{\noindent{\textit{\textbf{#1}}}\quad}

\newcommand{\figcap}[1]{\caption{\textit{#1}}}
\newcommand{\sfigcap}[1]{\vspace*{-19pt}\caption{\textit{\small #1}}}

\newcommand{\tabcap}[1]{\caption{\textit{\small #1}}}

\newcommand{\ums}[1]{\SI{#1}{\milli\second}}
\newcommand{\us}[1]{\SI{#1}{\second}}

\newcommand{\pow}{{PoW}\xspace}

\newlength{\onecolgrid}
\newlength{\twocolgrid}
\newlength{\threecolgrid}
\newlength{\fourcolgrid}
\ifEditMode
\newcommand{\editlabel}[1]{\raisebox{.35ex}{\tiny \scshape[#1]}}

\newcommand{\REVIEW}[1]{
  \textcolor[HTML]{377eb8}{\editlabel{Review} #1}\xspace{}}

\xspace{}

\newcommand{\PENDING}[1]{
  \textcolor[HTML]{e67e22}{$\cdots$ \editlabel{Pending} \textit{#1} $\cdots$}\xspace{}}

\newcommand{\TODO}[1]{\textcolor[HTML]{e41a1c}{{(#1)}}}

\newcommand{\FIXME}[2]{
  \textcolor[HTML]{c0392b}{\editlabel{FIXME(#1)} \textbf{#2}}}
\newcommand{\MISSING}[1]{\textcolor[HTML]{de2d26}{\textbf{#1}}}

\newcommand{\REMOVE}[1]{
  {\small \textcolor[HTML]{a65628}{#1\\--- Consider deleting.}}}

\newcommand{\krishna}[1]{\textcolor{blue}{#1}}
\newcommand{\patrick}[1]{\textcolor{brown}{#1}}
\newcommand{\john}[1]{\textcolor{lime}{#1}}
\newcommand{\bala}[1]{\textcolor{green}{#1}}
\newcommand{\mohamed}[1]{\textcolor{magenta}{#1}}

\else
\makeatletter

\newcommand{\REVIEW}[1]{#1}

\newcommand{\PENDING}[1]{\@bsphack\@esphack}
\newcommand{\TODO}[1]{\@bsphack\@esphack}
\newcommand{\FIXME}[2]{\@bsphack\@esphack}
\newcommand{\MISSING}[1]{\@bsphack\@esphack}
\newcommand{\REMOVE}[1]{\@bsphack\@esphack}

\newcommand{\krishna}[1]{\@bsphack\@esphack}
\newcommand{\patrick}[1]{\@bsphack\@esphack}
\newcommand{\john}[1]{\@bsphack\@esphack}
\newcommand{\bala}[1]{\@bsphack\@esphack}
\newcommand{\mohamed}[1]{\@bsphack\@esphack}

\makeatother
\fi

\setcopyright{none}
\renewcommand\footnotetextcopyrightpermission[1]{} 

\graphicspath{{./figures/}}

\title{Modeling Coordinated vs. P2P Mining: An Analysis of Inefficiency and Inequality in Proof-of-Work Blockchains}
\ifEditMode
\fi
\author{Mohamed Alzayat}
\affiliation{
  \institution{MPI-SWS}
  \country{Germany}
}
\author{Johnnatan Messias}
\affiliation{
  \institution{MPI-SWS} 
  \country{Germany}
  }
\author{Balakrishnan Chandrasekaran}
\affiliation{
  \institution{MPI-INF and VU Amsterdam}
  \country{Netherlands}
  }
\author{Krishna P. Gummadi}
\affiliation{
  \institution{MPI-SWS}
  \country{Germany}
  }
\author{Patrick Loiseau}
\affiliation{
  \institution{Univ. Grenoble Alpes, Inria, CNRS, Grenoble INP, LIG}
  \country{France}
}

\begin{abstract}
	\input{sections/abstract}

\end{abstract}

\begin{document}\sloppy


\maketitle


\pagestyle{plain}

\input{sections/intro}

\input{sections/related}

\input{sections/model}

\input{sections/simulation}
\input{sections/eff}
\input{sections/ineq}
\input{sections/conclusion}

\input{sections/acknowledgments}

\bibliographystyle{ACM-Reference-Format}
\bibliography{references}

\newpage
\onecolumn
\appendix
\input{sections/app-table}
\input{sections/app-proofs}

\end{document}

%% file: sections/abstract.tex
We study efficiency in a proof-of-work blockchain with non-zero latencies, focusing in particular on the (inequality in) individual miners' efficiencies. Prior work attributed differences in miners' efficiencies mostly to attacks, but we pursue a different question: \emph{Can inequality in miners' efficiencies be explained by delays, even when all miners are honest?} Traditionally, such efficiency-related questions were tackled only at the level of the overall system, and in a \emph{peer-to-peer (P2P) setting} where miners directly connect to one another. Despite it being common today for miners to pool compute capacities in a mining pool managed by a centralized coordinator, efficiency in such a \emph{coordinated setting} has barely been studied.

In this paper, we propose a simple model of a proof-of-work blockchain with latencies for both the P2P and the coordinated settings. We derive a closed-form expression for the efficiency in the coordinated setting with an arbitrary number of miners and arbitrary latencies, both for the overall system and for each individual miner. We leverage this result to show that inequalities arise from variability in the delays, but that if all miners are equidistant from the coordinator, they have equal efficiency irrespective of their compute capacities. We then prove that, under a natural consistency condition, the overall system efficiency in the P2P setting is higher than that in the coordinated setting. Finally, we perform a simulation-based study to demonstrate that even in the P2P setting delays between miners introduce inequalities, and that there is a more complex interplay between delays and compute capacities.

%% file: sections/intro.tex

\section{Introduction} \label{s:intro}

At its core, a blockchain is a timestamped append-only chain of blocks, each of
which record a set of transactions.
Many widely used blockchains, e.g., Bitcoin and Ethereum, are based on a
proof-of-work (\pow{}) scheme, where a set of globally distributed miners
simultaneously try to solve a cryptographic puzzle based on the current chain.
When a miner succeeds---which acts as proof of the miner's work---they bundle a set
of pending transactions to create a new block that is added on top of the chain.
Miners then restart mining a new cryptographic puzzle based on the new chain.
While these terms and definitions may vary slightly across different
blockchains, the central role of miners remains essentially the same:
They grow the chain by using their compute capacity to solve cryptographic
puzzles, and they maintain it by exchanging updates about the chain.

Due to communication delays between miners, a blockchain may occasionally \emph{fork} into two or more parallel branches. 
In such scenarios, the \pow{} protocol specifies that miners should accept and build upon the \emph{longest} branch.
The blocks in the other shorter, ignored branches represent lost or unrewarded work for the miners that generated these blocks, and lowers their efficiency.
%

In this work, we study the efficiency with which miners' computational work is recorded in a blockchain, and we focus more specifically on the unfairness or inequality in the efficiency of individual miners.
We illustrate the questions motivating our study using data from the Ethereum blockchain (see ~\S\ref{subsec:p2p-setting} for more details).
The efficiency of the top five Ethereum mining pools (computed as the fraction of blocks generated by the individual mining pools that are included in the main chain) are observed to be 93.58\%, 93.45\%, 93\%, 90.52\%, and 88.68\%, respectively. 
The 5\% disparity in efficiency between the largest and fifth largest mining pool raises the question: {\it What explains such a difference in their observed efficiencies?}
%
Prior work attribute the differences to dishonest miners, who can
intentionally deviate from the protocol to get more than their ``fair share'' of
blocks---that is, proportional to their compute capacity---included in the longest
chain \cite{eyal2014majority}.
Whether such unfairness can naturally arise even when all miners truthfully follow the \pow{} protocol remains, however, an unexplored question.

To derive high-level insights, we propose a simple model that abstracts away the technical details of \pow{} blockchains and focuses on delay modeled as a latency parameter between any two miners. Then, using simulations of our model, we show that when miners form a peer-to-peer (P2P) network to exchange information, significant inequality in their efficiencies can arise, even when all miners truthfully follow the protocol. While some prior work modeled the effects of latency on mining efficiency in this traditional \emph{P2P setting} (e.g., \cite{sompolinsky2015secure}), they focused primarily on the overall system efficiency and ignored whether the overall efficiency is equitably shared across miners---the only exception is a partial result for a simple case with two miners~\cite{lewenberg2015bitcoin}.

In practice, miners increasingly prefer to join mining pools\footnote{to reduce temporal variance in their mining rewards} and the P2P nodes would correspond to mining pools rather than individual miners. Then, our observation suggests the possibility for a miner to be strategic when choosing which mining pool to join. The Ethereum data above appears, for instance, to suggest that joining a larger pool results in greater efficiency than a smaller one. However, within mining pools, a centralized coordinator typically harnesses the aggregate compute capacities of the participating miners. This \emph{coordinated setting} is quite different from the P2P setting and opens up the following question: {\it Do all miners in a mining pool achieve equitable efficiency?}

To answer this question, we modify our model to include a coordinator in addition to the miners. In that case, we analytically derive the individual miners' efficiency, even in the most general setting. We infer that a miner's inefficiency in coordinated settings varies strictly monotonically with the miner's latency to the coordinator. Consequently, any (dis)parity in a miner's latency to the coordinator will result in (dis)parity in the miner's efficiency. Mining efficiency in coordinated settings has not been studied before, except for the overall system efficiency in the simple case where each miner is equidistant from the coordinator \cite{lewenberg2015bitcoin}.


Finally, using our models of P2P and coordinated settings, we investigate the following question: {\it Would mining pools have a higher overall efficiency if the participating miners interacted in a P2P rather than a coordinated setting (everything else being kept constant)?} We answer this question in the affirmative. Our finding suggests that the coordinated setting impacts both the overall efficiency and its distribution across miners, and sheds light on the underlying mechanisms behind this effect. 

In summary, we propose a simple model of a \pow blockchain with non-zero latencies for both the P2P and the coordinated settings. Our model has three components: the miners' compute capacities, the puzzle hardness, and the inter-miner latency matrix. We focus on two metrics of efficiency: the overall system efficiency and the individual efficiency; they capture the number of blocks included in the longest chain relative to the compute capacity of the whole system and that of an individual miner, respectively. The individual efficiency allows us, in particular, to analyze fairness in the efficiency distribution. Then we have the following results:
\begin{trivlist}
\item[1.] 
We provide an analytical solution for both the overall system efficiency and the individual efficiency of any miner in the coordinated setting, in the most general case with an arbitrary number of miners and arbitrary delays. Our computations are based on renewal theory and provide closed-form expressions that may be used as an input to other research questions. 

\item[2.]
Our analytical solution enables us to infer a number of interesting implications. \emph{First}, as expected, we observe that increasing delays decreases the overall system efficiency. \emph{Second}, we show that for arbitrary delays miners have unequal efficiencies---miners that are more distant from the coordinator have lower efficiency. 
\emph{Last}, we show that if all miners are equidistant from the coordinator, they have equal efficiency \emph{irrespective of the compute capacity distribution}. Our observations suggest that a way to remove inequality in efficiency would be by placing the coordinator at an equidistant position from all miners. We also discuss the issue of optimizing the coordinator position to maximize the overall system efficiency. 

\item[3.]
We prove that the overall system efficiency in the P2P setting is always higher than that in the coordinated setting (under a natural assumption that a link through a coordinator cannot have lower delay than a direct connection between miners). To our knowledge, this result is the first comparison between the P2P and coordinated settings in terms of efficiency. It shows that the cost of reducing inequality in efficiency may be to lose on the overall system efficiency. Interestingly, it also gives a lower bound for the overall system efficiency in the P2P setting---which is notably intractable. 

\item[4.]
We perform extensive simulations to complement our analytical results. In particular, we use simulations to investigate the inequality in individual miner efficiency for the P2P setting and, in general cases, we connect the observed inequality in efficiency empirically to the relative centrality of the miners in the latency graph. Our simulations are based on a new discrete-event simulator that we developed to faithfully emulate our theoretical models. We release the simulator~\cite{our-simulator} as an open source software, which may be of independent interest to the community. 

\end{trivlist}

Overall, our results shed light on the impact of latency on individual efficiency in blockchains. They may be of interest for further studies, in particular on various strategic aspects related to miners' choices. For instance, many recent work analyze strategic considerations in a miner's choice of the compute capacity they exert~\cite{Fiat@EC19,Goren@EC19,Kiayias@EC16,Noda@EC20}. All these work, however, ignore latency and assume that each miner, conditioned on their choice of compute capacity, gets their fair share of reward. Our work shows that this assumption does not hold with non-zero latency. As such, it offers valuable input to refine studies of strategic considerations.

%% file: sections/related.tex
\section{Related Work} \label{s:related}

There is a rich body of work on security and scalability aspects relevant to the decentralized nature of \pow{} blockchains. It is well known, for instance, that a majority of honest nodes is insufficient to maintain a consistent ledger~\cite{decker2013information,eyal2014majority,Bahack13,Garay_Kiayias_Leonardos_2015,Pass_Seeman_Shelat_2017,Zhang_Preneel_2019,sompolinsky2015secure}. These studies show that \pow{} blockchains are subject to attacks by dishonest miners. In contrast, our work explores the impact of (inevitable) network delays on the overall system efficiency and the inequality in the (individual) efficiency of completely honest miners. 
%

\parai{Delays in blockchains.}
Delay has been the focus in many prior work, but from a different perspective. For instance, Bitcoin Relay Network~\cite{BRN-URL2017}, FIBRE~\cite{FIBRE-URL2019}, Falcon~\cite{Basu-Talk2016}, and more recently BDN~\cite{Kuzmanovic-QUEUE2019} focus on improving the underlying (P2P) network by reducing latencies. From the security standpoint, a propagation advantage (i.e., reduced latency) can help adversaries in launching selfish-mining, double-spending, feather-forking, and other attacks~\cite{decker2013information,eyal2014majority,gervais2016security,Zhang_Preneel_2019, Garay_Kiayias_Leonardos_2015}. Interestingly, the Bitcoin Backbone Protocol~\cite{Garay_Kiayias_Leonardos_2015} formalized the \newterm{chain quality} property, as a measure of how mined blocks are distributed amongst miners with respect to their hashing power, and explicitly mentions that the chain quality is trivially ideal if all miners are honest---which, as we shall see, our results contradict.
The interaction between puzzle hardness and network delays has a huge impact on the security of blockchains, as discussed in \cite{sompolinsky2013accelerating,sompolinsky2015secure,gervais2016security,Garay_Kiayias_Leonardos_2015}---we use the same ratio metric. 

\parai{Efficiency of blockchains.}
Many prior work studied efficiency as a global notion to capture wasted resources by miners~\cite{decker2013information,eyal2014majority,gervais2016security,Zhang_Preneel_2019,Garay_Kiayias_Leonardos_2015}. The identification of forks as a measure of wasted efforts, in particular, was clearly explained in~\cite{decker2013information}. \REVIEW{Similarly,  \cite{Kiffer-FC2021} looked at wasted work of different miners in ethereum, and, similar to what we show in \S~\ref{subsec:p2p-setting}, they highlight that the top miners are more likely to have their blocks as part of the main chain rather than ending up as an ``uncle'' or prune due to having a network advantage}.
The fork-measure was used as an indicator to study the resilience of a blockchain against different classes of attacks (e.g., double spending and selfish mining). We prefer the term efficiency as it precisely conveys our intent: capture wasted work from an altruistic point of view to understand its effects on honest miners. 

\parai{Selfish mining strategies and inequality.}
Selfish-mining strategies~\cite{eyal2014majority, sapirshtein2016optimal} and other attacks \cite{gervais2016security,Zhang_Preneel_2019, Garay_Kiayias_Leonardos_2015} show that inequality in efficiency in blockchains may happen when some miners are dishonest; they do not consider the \stress{intrinsic} inequality in a system where all miners are honest but experience different network delays. 
Network delays were taken into account by \cite{sapirshtein2016optimal} and \cite{bitcoinDynamics}, albeit still with a focus on selfish mining: \cite{sapirshtein2016optimal} provides an algorithm that computes an optimal selfish mining strategy and shows that it is profitable even for attackers with less than 25\% of the collective hashing capacity. \cite{bitcoinDynamics} shows an example where selfish mining strategies allow pooled selfish miners with a collective hashing capacity higher than 25\% to gain more than their fair share of the total reward, but the strategy decreases the total reward such that it is not profitable (under the assumption that all miners are honest, equally powerful, and randomly spatially-distributed).

\parai{Inequality in efficiency.}
A few prior work discuss efficiency and inequality in efficiency outside selfish mining. The work of \cite{Gencer-FC2018} uses well-established networking techniques to measure the latency across Bitcoin nodes and discuss its impact on the efficiency of the blockchain. They also discuss the inequality in miners' efficiencies (or fairness) in Bitcoin and Ethereum networks. They do not, however, analyze the root causes of such inequality and instead state this fairness issue as a measured observation; they also do not consider the coordinated setting. \cite{lewenberg2015inclusive} and \cite{sompolinsky2016spectre} develop the idea of inclusive blockchains in the form of Directed Acyclic Graphs (DAGs) that allow for higher transaction throughput through a secure, yet more forgiving block-acceptance policy. By virtue of their blockchain design and more forgiving policy, their protocol decreases the inequality in miners' efficiencies (increases fairness). However, \cite{lewenberg2015inclusive, sompolinsky2016spectre} focus on the properties of the new protocol only in the P2P model in particular settings.

\parai{Blockchain models.}
While nearly all \pow{}-blockchain-related papers implicitly assume a model along the lines of what was informally described by \cite{Nakamoto-WhitePaper2008}, few explicitly formalize the model. Of those, \cite{sompolinsky2015secure} looks at the overall system efficiency in the P2P setting, provides efficiency bounds based on the network diameter delay and partial results hinting at the inequality in the two-miners case. Their results, however, do not generalize to more than 2 miners. The authors of \cite{lewenberg2015bitcoin}, on the other hand, compute individual efficiencies for 2 nodes in the P2P setting, and compute the overall system efficiency for $n$ nodes in the coordinated setting provided all the nodes are equidistant. This result is a special case of our more general result, which allows computing both the overall and individual efficiencies in any coordinated setting. To the best of our knowledge, this paper is the first to model the coordinated setting.  \cite{StochasticWANAnalysis} studies the overall system efficiency as a function of network delay, in the P2P setting under the assumption that delays follow an exponential distribution. It provides  expressions for the 2- and 3-miner cases. In contrast, we assume constant delays but focus on more general scenarios. Finally, \cite{blockDeliveryTime,blockDeliveryTime-journal} develop a queuing network model in the P2P setting, but focus on metrics different than ours.


\parai{Simulators.}
There are several simulators to study blockchains (e.g.,~\cite{Aoki-INFOCOM2019,VIBES,Gervais-Github2016}): \cite{Aoki-INFOCOM2019} designed a  simulator to test protocol modifications, and \cite{VIBES} focused on visualizing \pow{} simulations. \cite{gervais2016security}---the work most relevant to ours---analyzed the security and performance implications of various network and consensus parameters of \pow{} blockchains. None of them analyzed the inequality in efficiency induced by network delays.
It was non-trivial to configure these simulators to ignore implementation details irrelevant for studying inequality; hence, they did not suit our purposes.

%% file: sections/model.tex
\section{Models and Efficiency Measures}\label{s:model}

In this section, we present our models of the \pow blockchain, both in the traditional P2P and the coordinated setting.
We also introduce two measures of efficiency (overall and individual).

\input{sections/model-comp}

\input{sections/two-models}
\input{sections/model-justif}


\input{sections/measures}

%% file: sections/model-comp.tex
\subsection{Model Components}\label{ss:model-comp}

Our model of a \pow{} blockchain comprises three components:
(i) a set of miners with their compute capacities;
(ii) the \newterm{hardness} of mining a block; and
(iii) the latency matrix, which implicitly encodes the relative locations of miners in the underlying network and abstracts away all sources of delay.
%

We denote the set of miners who collectively maintain the blockchain as $\Mset =
\{ m_1, \cdots, m_n\}$ with $n\ge 2$.
Each miner $m_i\in \Mset$ is endowed with computational capacity $h_i \in \R_+$,
which is used for solving cryptographic puzzles (essentially, mining blocks).
We assume that $h_i$ is a relative computational capacity (compared to the whole
set of miners), i.e., $\sum_{i=1}^n h_i = 1$.

We represent the hardness of solving the cryptographic puzzle by parameter
$\tau\in\R_+$.
It represents the average time to generate a block with a normalized
computational capacity of $1$ (i.e., for the whole system).
The time taken by miner $m_i\in \Mset$ to generate a block is assumed to be an
exponential random variable\footnote{The exponential assumption is ubiquitous in blockchain models and validated by empirical observations \cite{gervais2016security,decker2013information}.} of parameter $h_i/\tau$ (i.e., of mean $\tau/h_i$),
independent of all other generation times for other blocks and miners.
We denote by $\hti_i = h_i / \tau$ the effective mining rate of miner $m_i$,
i.e., the rate at which the miner discovers new blocks.
Recall that the minimum of $n$ exponential random variables is an exponential
random variable with parameter equal to the sum of the parameters of the $n$ variables.
Then, the time for the whole system to generate a block is an exponential
random variable of parameter $\sum_{i=1}^n \hti_i = \sum_{i=1}^n h_i / \tau =
1/\tau$, which indeed has mean $\tau$.

Miners are typically located at diverse geographical
locations~\cite{Delgado-FC19,Bitnodes-URL2019,Park-IEEE19}, and, hence,
experience delays when exchanging information (e.g., a newly generated block)
with one another.
We assume, hence, a latency of $l_{ij}$ between any two miners $m_i\neq m_j$,
and we denote by $L = [l_{ij}]_{i, j \in \{ 1, \cdots, n\}}$ the matrix of all
latencies.
The matrix $L$ is symmetric with all diagonal elements being zero, but we do not
require that the latency is nonzero between any two miners (i.e., two miners may
be at the same location).
To capture the importance of the time spent in exchanging information between
miners compared to the time spent in generating a block, we define the parameter
$\lambda$ as the ratio between the puzzle hardness $\tau$ and the mean latency $\lmean$: \\[-3mm]
\begin{equation*}
\lambda = \tau / \lmean, \quad \textrm{ where } \quad \lmean = \frac{2}{n(n-1)} \cdot \sum_{i=1}^n\sum_{j>i} l_{ij}.
\end{equation*}
We assume that the graph of all miners is \term{connected}, i.e., all latencies
are finite, and that all latencies $l_{ij}$ correspond to the shortest path
between $m_i$ and $m_j$, i.e., there is no path between $m_i$ and $m_j$ such
that the sum of latencies on that path is smaller than $l_{ij}$.

%% file: sections/two-models.tex
\subsection{Mining and Communication models}\label{ss:two-models}
The miners who maintain the blockchain typically exchange information with one
another using a protocol that can be modeled in two ways:
a decentralized, peer-to-peer model where all miners act as peers communicating
directly with one another, or a quasi-centralized model wherein all
communications transit through a coordinator (i.e., miners never directly interact with
one another).
In this paper, we only consider scenarios with a single coordinator to which all
miners in $\Mset$ are connected.
Whether there exist multiple coordinators that are, in turn, connected in a
peer-to-peer fashion is irrelevant from the perspective of the miners.
In either model, we assume that miners truthfully announce the blocks they
discover, to all other miners or to the coordinator.

\paraib{Peer-to-peer (P2P) model.}
A miner $m_i \in \Mset$ upon discovering a block truthfully appends it to the current longest chain and immediately announces the updated chain to all other miners $m_j$, who receive it after a delay $l_{ij}$.
The miner then continues to mine the next block on this chain.
On receiving a longest-chain update from another miner, miner $m_i \in \Mset$
compares it to their currently held longest chain and acts as follows:
If the received chain is longer than $m_i$'s current chain, $m_i$ discards its
current work, replaces its chain with that received, and starts mining the next
block of that chain.
If the chain received is, on the contrary, shorter than or of equal length as
$m_i$'s current chain, $m_i$ simply discards the chain received, and continues
mining the next block on their current chain.
The system under the P2P model is completely defined by the vector of capacities
$\hb = [h_i]_{i=1, \cdots, n}$, the puzzle hardness $\tau$ and the latency matrix $L$;
or alternatively by the vector of effective capacities $\htib = [\hti_i]_{i=1,
\cdots, n}$ and the latency matrix $L$.

\paraib{Coordinated model.}
The defining characteristic of this model is the presence of a central
coordinator $C$.
We denote by $l_i$ the latency between miner $m_i \in \Mset$ and $C$.
Miners only communicate with the central coordinator, so that the vector $\lb =
[l_i]_{i \in \{1, \cdots, n\}}$ is sufficient to define all latencies in the system.
For consistency in the comparison between the P2P setting with latency matrix
$L$ and the coordinated setting with latency vector $\lb$, we assume that the
coordinator does not introduce a shorter path between any two miners, i.e., for
all $i$ and $j$, $l_{ij} \le l_i + l_j$.\footnote{The assumption does not imply that network paths obey the triangle inequality, but rather that the path connecting the miners is already the best one.}
Unlike the P2P model, the coordinator truthfully and unilaterally maintains the
longest chain.
Upon discovering a block, a miner $m_i \in \Mset$ truthfully and immediately
announces the block to the coordinator and \stress{pauses} until receiving an
updated chain from the coordinator (including a template needed to mine the next block).\footnote{The same pause model is used, e.g., in \cite{lewenberg2015bitcoin}.}
The update chain may be either the chain that $m_i$ extended or another chain,
had a different miner sent a different block to the coordinator before $m_i$.
Miners always mine the next block on the chain returned by the coordinator.
The system under the coordinated model is completely defined by the vector of capacities $\hb = [h_i]_{i=1, \cdots, n}$, the puzzle hardness $\tau$ and
the latency vector $\lb$; or by the vector of effective capacities
$\htib = [\hti_i]_{i=1, \cdots, n}$ and the latency vector $\lb$.



%% file: sections/model-justif.tex
\subsection{The Rationale behind the Models}\label{ss:model-justif}


The miners in a blockchain (per the initial design in~\cite{Nakamoto-WhitePaper2008}) interact with one another through a decentralized P2P network.
Even today, miners in permissionless blockchains such as Bitcoin~\cite{Nakamoto-WhitePaper2008} and Ethereum~\cite{wood2014ethereum} interact over such a P2P network.
The P2P model (in~\S\ref{ss:two-models}) represents this traditional connectivity model and attempts to highlight the factors that affect the efficiency of miners.

%
%

Since the ability of an individual miner to successfully mine a block has
rapidly dwindled, miners typically join a mining pool where a coordinator
harnesses the aggregate compute capacity of a set of miners for mining blocks.
The (mining) pool significantly improves the odds of mining a block, and miners can share the rewards in such a way as to reduce the variance in their revenue.
%
%
The coordinated model (in~\S\ref{ss:two-models}) is meant to succinctly capture the fundamental properties of this mining pool setting, where the only interactions are between the miners and the coordinator.
The miners defer to the coordinator for ascertaining the longest chain in the system, and the coordinator relays the work to the miners and refreshes work periodically (hence the pause).

Today, both the P2P and coordinated communication paradigms are employed in blockchains. In Bitcoin, for instance, four mining pools  account for $56.9\%$ of the blocks mined in aggregate~\cite{btc-shares}. Miners within each mining pool follow a protocol resembling the coordinated model, albeit documentation on the protocol specifics is
sparse~\cite{BitcoinWiki-MPO-URL2020}.
%
%
Mining pool coordinators in turn form a peer-to-peer network to exchange information between one another. We require, therefore, the two models to understand the dynamics both within the mining pools and between different mining pools (or miners).

While in practice there may be multiple sources of communication delays (i.e., propagation times, processing times, etc.), our latency matrix abstracts them away.
Seen differently, our model assumes unlimited bandwidth to abstract away the message sizes.
Our comparison between the P2P and coordinated setting assumes a consistency condition between the two models: $l_{ij} \le l_i + l_j, \forall i,j$.
This condition implies that the coordinator does not introduce a shorter path between any two miners---a natural condition from a theoretical perspective to make the models comparable.
In practice, delays may be lower in the coordinated setting due to smaller messages.
There is, however, a fundamental limit on latency, governed by the speed of light in a medium, which in many cases dominates the communication cost.
Hence our model and consistency condition abstract away this issue (as well as other technical considerations) to formulate a clear mathematical result on the comparison between the overall efficiency in the P2P and coordinated settings. Note also that this result, together with our closed-form analysis of the coordinated setting, provide a valuable lower bound for the P2P setting (see \S\ref{sec.comparison}). 




%% file: sections/measures.tex
\subsection{Measures of Efficiency}\label{ss:measures-of-eff}

We focus on characterizing the implications of network latencies for efficiency.
To this end, we compare the output of the mining model with latency to a
hypothetical case where there is no latency.
Formally, consider a time window of length $T$, and denote by $\hat{B}_i(T)$ (or
simply $\hat{B}_i$) the number of blocks generated by miner $m_i \in \Mset$ that
are included in the longest chain at time $T$.
The longest chain at $T$ corresponds to the chain that would result if we were
to let the system evolve from time $T$ for long enough, after miners stop
mining, i.e., until the system resolves all ambiguities and converges on the
longest chain).
Ties, if any, are broken uniformly at random.

We define the efficiency of miner $m_i \in \Mset$ as\\[-2mm]
\begin{equation}\label{eq.efficiency.ind} \eff_i = \lim_{T\to\infty}\frac{\E \left[
\hat{B}_i(T)\right]}{h_iT/\tau}.
\end{equation}
We use the expectation of the random variable $\hat{B}_i$ to define the
efficiency.
In simulations, however, we use an empirical average over many runs (of sufficient
duration) to estimate the efficiency (see below).
While on a particular realization one might have $\hat{B}_i / (h_iT/\tau) >1$, the
expectation in \eqref{eq.efficiency.ind} clearly satisfies $\eff_i\le 1$.
Furthermore, in the coordinated model, the process $\hat{B}_i(T)$ is a renewal
process and not a Markov process, owing to the pause after discovering a block,
and $\frac{\E \left[ \hat{B}_i(T)\right]}{h_iT/\tau}$ will not be independent of
$T$.
We use, hence, a limit $T\to\infty$ to define the efficiency, which will then be
computed using the elementary renewal theorem (thereby justifying the limit's existence).

The denominator $h_iT/\tau$, in \eqref{eq.efficiency.ind}, represents the average
number of blocks that the miner would have mined within the time window of
length $T$ when mining alone.
In the P2P model, due to the memoryless property of the exponential
distribution, the actual number $B_i$ of blocks mined by $m_i$ during the time
window $[0,T]$ is such that $\E [B_i] = h_iT/\tau$.
In the P2P model, our definition of efficiency in \eqref{eq.efficiency.ind} is,
hence, equivalent to $\E \left[ \hat{B}_i/T\right] / \E \left[ B_i/T\right]$,
which in turn is equivalent to $\E \left[ \hat{B}_i/ B_i\right]$ for a large
enough $T$, since both $\hat{B}_i/T$ and $B_i/T$ converge to constant rate.
Our measure of efficiency for the P2P model is, therefore, equivalent to the
average fraction of blocks mined by $m_i$ that are included in the chain (for
large $T$).
In the coordinated model where miners pause after discovering a block, this
equivalence does not hold and our definition characterizes the inefficiency
introduced by latencies rather than the inefficient use of computing resources.
Lastly, if all latencies are zero, we have $\hat{B}_i=B_i$ and $\E \left[
\hat{B}_i\right] = \E [B_i] = h_iT/\tau$, so that $\eff_i=1$ for all miners in both
models.

Let $\hat{B}$ denote the length of the longest chain at time $T$. Clearly,
$\hat{B} = \sum_{i=1}^n \hat{B}_i$. Then we define the overall system efficiency as\\[-3mm]
\begin{equation}
\label{eq.efficiency.global} \eff = \lim_{T\to\infty} \frac{\E \left[ \hat{B}\right]}{T/\tau}
= \sum_{i=1}^n h_i \eff_i.
\vspace{-1mm}%
\end{equation}
(Recall we assume $\sum_{i=1}^n \!h_i\!\! =\!\! 1$; otherwise we normalize by
$\sum_{i=1}^n \!h_i$.)

Throughout the paper, to measure inequality in a quantity, we use the \term{Gini index}, defined as
half of the relative mean absolute difference;
for a vector $(x_1, \cdots, x_n)$ the Gini index is
$\gamma = (\sum_{i=1}^n \sum_{j=1}^n |x_i - x_j|)/(2 n \sum_{i=1}^n x_i).$
%
Note that $\gamma \in [0, 1]$ and $\gamma=0$ corresponds to no inequality ($x_i = x_j$ for all $i,j$),
while $\gamma=1$ corresponds to the maximum inequality.
We denote by $\gamma_e$ and $\gamma_h$ the Gini indices of individual efficiency $(\eff_1, \cdots, \eff_n)$ and
computational capacity $(h_1, \cdots, h_n)$ accross miners, respectively.
We summarize our notation in Tab.~\ref{tab:notations} (see App.~\ref{app.table}).
In addition, we often use the superscripts $C$ and $P2P$ to distinguish between
the two models:
We use $\eff^C$ and $\eff^{P2P}$ to denote, for instance, the overall system efficiency
in the coordinated and P2P models, respectively.

%% file: sections/simulation.tex
\section{Simulating the Models}\label{s:sim}

The interactions between the different stochastic processes (i.e., miners), especially in the P2P setting, and their effect on the outcomes, notwithstanding the simplicity of the models (\S\ref{ss:two-models}), make it hard to derive analytical solutions for some of the research questions we consider.
To remedy this situation, we developed a discrete-event simulator that simulates the operations of and interactions between miners, and between miners and the
coordinator.
The simulator complements the theoretical model by providing insights into complex P2P scenarios that are not tractable. 
%
We eschew support for features (e.g., transactions or bandwidth constraints) that have no impact on our models.
%
%
In this section, we describe our implementation, elucidate the design choices, and explain simulation configurations (or scenarios) used throughout the paper.

Miners have unit compute capacity in the simulations, and we simulate miners of
varying compute power by
co-locating more than one miner at the same location with zero network latency between them.
%
%
We simulate the P2P model by specifying the compute capacities of miners ($\hb$),
and the latency matrix ($L$) representing the (symmetric) network latencies
between the miners.
By fixing the relative positions of miners via $L$, we can simulate any
arbitrary peer-to-peer network topology.
The simulator also supports the notion of a coordinator.
We simulate the coordinated model by specifying the compute capacities ($\hb$), the latency matrix ($L$) representing the latencies between the miners and the
coordinator position from which the latency vector $\lb$ is deduced.
In addition, we specify the puzzle hardness ($\tau$) and the simulation time.
The latter limits the length of the simulation by specifying either directly the duration of the simulation or, indirectly, the length of the longest chain
generated (in number of blocks).

We simulate mining by sampling the time taken to generate a block from an
exponential distribution with a mean ($\tau$).
We vary $\tau$ to explore different families of blockchains:
$\tau$ values of $10$~minutes and $13$ seconds, for instance, approximates the
mining in Bitcoin and Ethereum, respectively.
The mining of a block in the simulation is independent of the block size or contents, and transactions are excluded.
The time to propagate a block from one miner to another depends only on the
latency between the two miners, which abstracts all sources of delays (see above).
%
%
The miners do \stress{not} re-advertise the blocks they
receive.
Miners also receive blocks without explicitly soliciting them, and they
truthfully follow the protocol. 

We implemented the simulator in Python as a single-threaded application, with
one event loop orchestrating the interactions between the different entities
(e.g., miners) at the appropriate times.
The implementation spans approximately \num{1}k~lines of
code (LoC).
Miners and coordinators are each implemented as a Python class that encapsulates
the characteristics and behavior of the concerned entity.
We added approximately \num{3}k~LoC for configuring the experiments, analyzing, and visualizing
the data.
We release the source code of our implementation as an open source artifact~\cite{our-simulator}. The link also contains a notebook that allows one to reproduce all graphs.

\parai{\textbf{Simulation Scenarios}}\label{ss:scenarios}
A~scenario specifies the relative positions of miners in the
network.
In each scenario, we vary the distribution of compute capacities of miners, succinctly
represented using the Gini index $\gamma_h$.
We use three scenarios, two to simplify the explication of observations and one
modeled after the Bitcoin network to highlight the real-world implications of
our observations.

\begin{figure*}[tb]
    \centering
    \begin{subfigure}[b]{\threecolgrid}
	\begin{subfigure}[b]{\threecolgrid}
	    \centering
		\includegraphics[width={0.7\textwidth}]{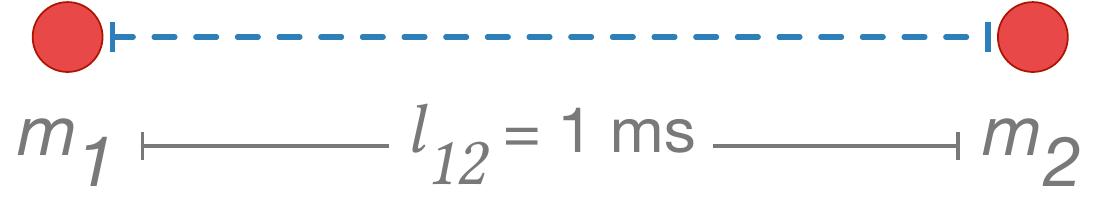}
		\figcap{2-miner P2P scenario}\label{fig:two-node-illust}
	\end{subfigure}
    \begin{subfigure}[b]{\threecolgrid}
        \centering
    	\includegraphics[width={0.7\textwidth}]{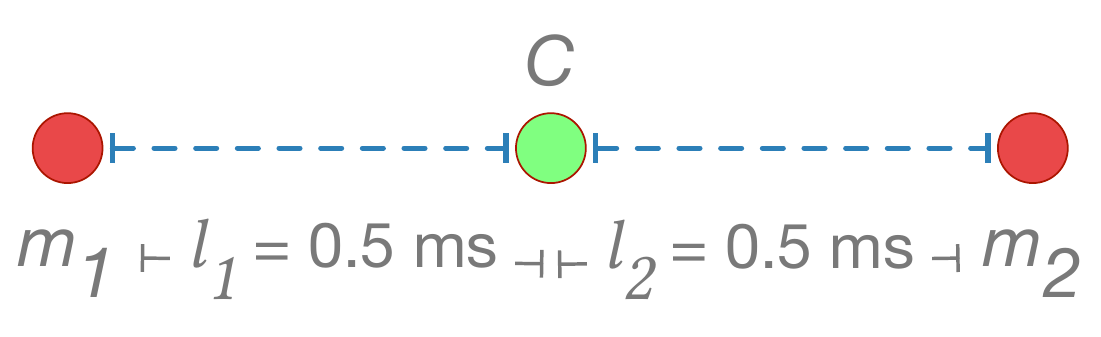}
    	\figcap{2-miner coordinated scenario}\label{fig:two-node-C-illust}
    \end{subfigure}
    \end{subfigure}
	\begin{subfigure}[b]{\threecolgrid}
	    \centering
		\includegraphics[width={0.7\textwidth}]{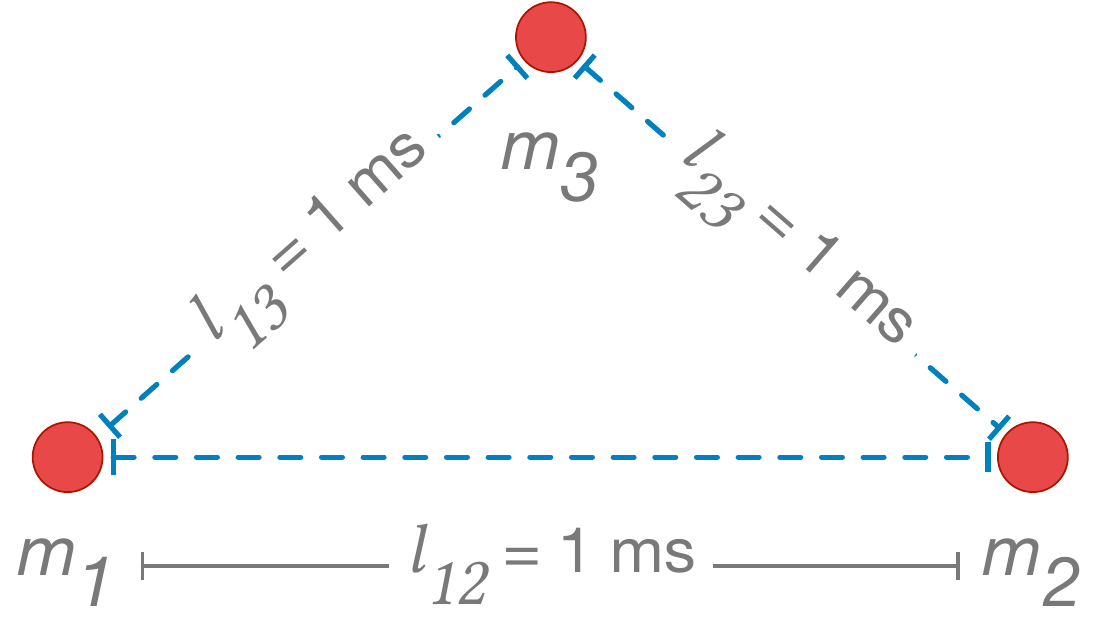}
		\figcap{3-miner P2P scenario}\label{fig:three-node-illust}
	\end{subfigure}
    \begin{subfigure}[b]{\threecolgrid}
        \centering
    	\includegraphics[width={0.7\textwidth}]{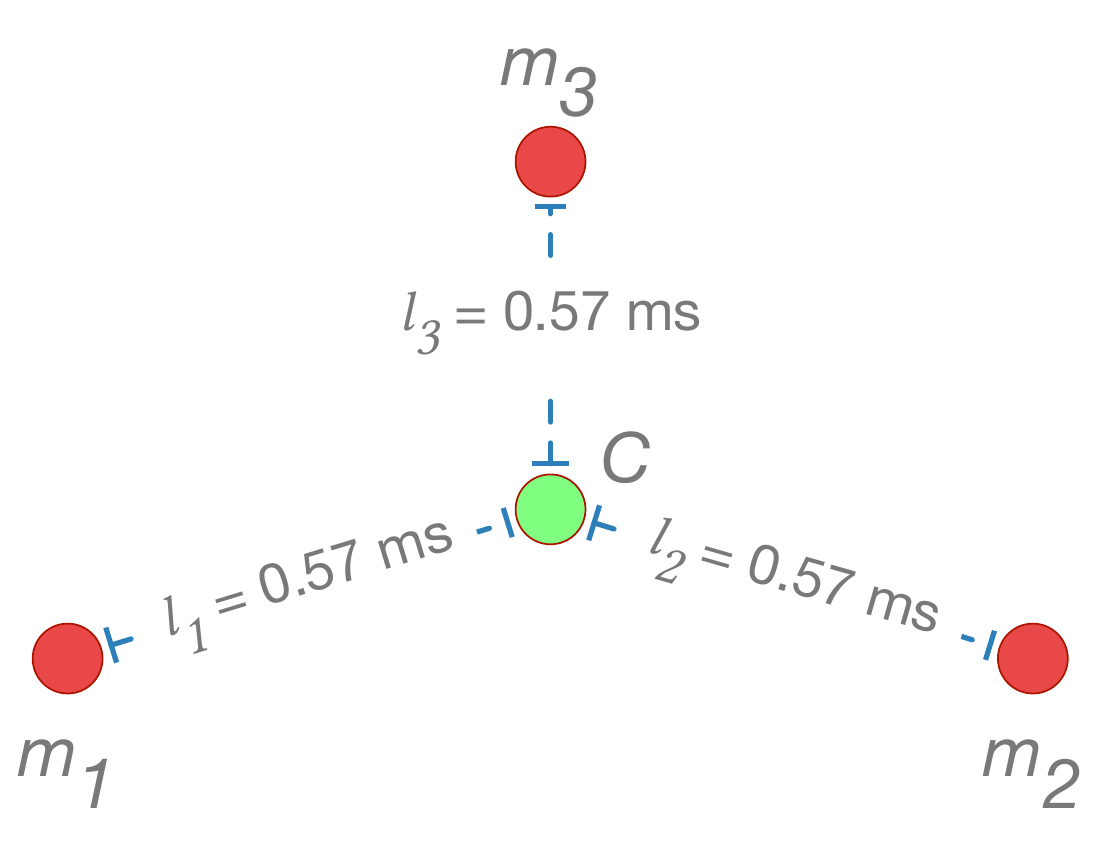}
    	\figcap{3-miner coordinated scenario}\label{fig:three-node-C-illust}
    \end{subfigure}
	\figcap{(P2P \& coord. setting.) 2-miner and 3-miner scenarios with miners in red and the coordinator in green.}\label{fig:sim-scenarios}
\end{figure*}

\parai{Two-miner and three-miner scenarios.}
Fig.~\ref{fig:sim-scenarios} illustrates the 2-miner and 3-miner scenarios
in both the P2P and coordinated settings.
When presenting the simulation results, we annotate the plots with the
appropriate value of $\gamma_h$ to describe the distribution of compute capacities.
In the 2-miner scenario, assigning equal compute capacities to miners will
result in $\gamma_h=0$, whereas assigning them a 40-60\% split of the total compute
capacity will result in $\gamma_h=0.1$.

\parai{Bitcoin-approximation Scenario.}
We mimic Bitcoin's deployment, to a first level of approximation, using a set of
five miners.
We place these miners in cities that are well-known for hosting huge mining
infrastructures~\cite{Aki-WebArticle2018,Magas-WebArticle2018,Tuwiner-WebArticle2019}:
Linthal, Switzerland; Moscow, Russia; Reykjav\'{i}k, Iceland; Sichuan, China;
and Washington D.C., US.
They host the top Bitcoin mining pools (or pool operators), which account for the
majority of mined
blocks~\cite{Blockchain-URL2019,Gencer-FC2018,Khatri-WebArticle2019}.
%
We derive the latency between a city pair using an \stress{optimistic} estimate
of the latency between any two locations (cf. inflation in min. \texttt{ping}
in~\cite{Bozkurt-PAM02017,Chandrasekaran-CoNEXT2015}).
We assume that any city pair is connected by an optical fiber over the shortest
distance between the pair, i.e., geodesic~\cite{Wiki-geodesic-WebArticle2019}.
We divide this distance by the speed of light in glass to obtain the optimistic
latency.
%
%
The resulting mean pairwise latency (or one-way delay) is ${\approx}\ums{52}$, a factor of two lower than empirical observations from~\cite{Gencer-FC2018} (Prior work do not provide, however, pairwise latencies or miners’ locations)---note that it also neglects delays other than the basic ping latency.

\parai{Typical values of $\lambda$ for cryptocurrencies.}
Estimating the value of $\lambda$ (i.e., the ratio of hardness to mean latency) for
different cryptocurrencies is not a straightforward task.
Different \pow{} blockchains use different values for hardness ($\tau$), but one
can empirically measure this value (by measuring, for instance, the average
blocks mined per unit time).
Bitcoin's hardness is \us{600} and that of Ethereum is 10-\us{20} (\us{13} on
average ~\cite{ethstats}).
%
%
Measuring mean latency, in contrast, is more difficult, albeit prior work have
shown how to estimate this
value~\cite{Gencer-FC2018,silva2020impact,decker2013information}.
For Bitcoin, the most recently available estimates, as of April 2017, indicate
that it takes around 1-\us{2} (8-\us{10}) for a block to reach $50\%$ ($90\%$)
of all nodes~\cite{bitcoinstats}.
%
%
%
%
A recent work, for instance, pegs the average block propagation delay for
Ethereum to be at least $\ums{109}$ under ideal conditions~\cite{silva2020impact}.
%
%
Using these latency estimates, we conservatively estimate the $\lambda$ value
for Bitcoin to be between 100-1000 and for Ethereum to be between 10-100.


\parai{Simulation length.}
We run our simulations till the main chain contains $\num{100000}$ blocks, which suffices for obtaining stable estimates of the efficiency metrics (refer~\S\ref{ss:measures-of-eff}).
We also repeat each simulation 10 times to estimate the standard deviation on the efficiency estimates. We observe that the standard deviation is always very small (it is shown as a halo in all plots, but is almost always invisible), hence confirming that the simulation length is sufficient to obtain statistically robust estimates. 



%% file: sections/eff.tex
\section{Overall System Efficiency}\label{s:eff}

In this section, we analyze the overall system efficiency in blockchains with non-zero latency. We begin with the coordinated setting where we have a complete analytical solution for analyzing how latency affects overall utility. We then establish a relationship between the overall system efficiency in the P2P and coordinated settings. We conclude by discussing the optimal position of the coordinator along with other implications of our results. 


\input{sections/eff/coord}

\input{sections/eff/p2p}

\input{sections/eff/p2p-vs-c}

\input{sections/eff/c-pos}

\input{sections/eff/end}

%% file: sections/eff/coord.tex

\subsection{Coordinated Setting}\label{ss:eff-coord}


Miners do not directly interact with one another, but only through the coordinator in this setting(\S\ref{ss:two-models}). Delays are, hence, captured by the vector of latencies from each miner to the coordinator $\lb = [l_i]_{i\in \{ 1, \cdots, n\}}$. We compute the overall system efficiency in this setting for arbitrary parameters ($n$, $\hb$, $\lb$, $\tau$), as formalized in the following theorem.
\begin{theorem}[Overall system efficiency in the coordinated setting]
\label{thm.overall-efficiency-C}
Consider a coordinated model with $n$ miners, computational capacities $\hb$, latency vector $\lb$, and puzzle hardness $\tau$. Recall that $\htib = \hb / \tau$ and let $\lti_i = 2l_i$ for all $i \in \{1, \cdots, n\}$. Without loss of generality, assume that $l_1 \le \cdots \le l_n$. Then the overall system efficiency is 
\begin{equation}
\label{eq.overall-efficiency-C}
    \eff^C(\hb,\lb, \tau) = \eff^C(\htib,\ltib) = \frac{1/\btau}{\hti_1 + \cdots + \hti_n}; \\[-2mm]
\end{equation}
where
\begin{align}
\label{eq.taubar-C}
    \btau =  \sum_{i=1}^n \Bigg\{ &  \frac{e^{\sum_{j=1}^i \hti_j \lti_j}}{\sum_{j=1}^i \hti_j} \cdot 
    \Big[
        \Big(1+ (\textstyle\sum_{j=1}^i \hti_j) \cdot \lti_i\Big) e^{- (\sum_{j=1}^i \hti_j) \cdot \lti_i}\\
        \nonumber & - \Big(1+ (\textstyle\sum_{j=1}^i \hti_j) \cdot \lti_{i+1}\Big) e^{- (\sum_{j=1}^i \hti_j) \cdot \lti_{i+1}}
    \Big]
    \Bigg\}.
\end{align}
Here, we used the convention $l_{n+1} = \infty$ and $l_{n+1} e^{-hl_{n+1}} = 0$ for any $h>0$ to simplify the expression.
\end{theorem}

A complete proof of Thm.~\ref{thm.overall-efficiency-C} is provided in Appendix~\ref{sec.proofs}. The idea is to look at the process $\hat{B}$, which corresponds to the number of blocks mined from the viewpoint of the coordinator. We observe that $\hat{B}$ is a renewal process; then by the elementary renewal theorem we can compute the efficiency as \eqref{eq.overall-efficiency-C}, where $\btau$ is the expected time between two increments of $\hat{B}$, i.e., between two block additions to the chain. To conclude the proof, we show that $\btau$ satisfies \eqref{eq.taubar-C}.
Note that the formula in Thm.~\ref{thm.overall-efficiency-C} assumes that nodes are ordered by (weakly) increasing distance to the coordinator. If that was not the case, we can re-order the nodes without loss of generality.

In Thm.~\ref{thm.overall-efficiency-C}, we write the efficiency as $\frac{1/\btau}{\hti_1 + \cdots + \hti_n}$ to emphasize the compute capacities. Recall, however, that $\hti_1 + \cdots + \hti_n = 1/\tau$ so that the overall efficiency can also be written as $\tau/\btau$. 
While the formula in \eqref{eq.taubar-C} is somewhat involved, Thm.~\ref{thm.overall-efficiency-C} allows the analytical computation of the overall efficiency for arbitrary parameter values. Below, we investigate a number of qualitative observations that can be deduced from this general result. We start with the simple case of $n=2$ miners.
\begin{corollary}[Two-miners case]
\label{cor.overall-efficiency-2-miners}
If $n=2$, then the overall system efficiency reduces to $\frac{1/\btau}{\hti_1 + \hti_2}$ with\\[-2mm]
\begin{equation*}
    \btau = \lti_1 + \frac{1}{\hti_1} \cdot (1-e^{-\hti_1(\lti_2-\lti_1)}) + \frac{1}{\hti_1 + \hti_2} \cdot e^{-\hti_1(\lti_2-\lti_1)}.
\end{equation*}
\end{corollary}

In Cor.~\ref{cor.overall-efficiency-2-miners}, $\lti_1 = 2l_1$ corresponds to the round-trip time from the closest miner $m_1$ to the coordinator (recall that $l_1\le l_2$ by assumption), while $\lti_2-\lti_1 = 2(l_2-l_1)$ corresponds to twice the extra distance from $m_2$ to the coordinator compared to $m_1$. The formula for the two-miners case in Cor.~\ref{cor.overall-efficiency-2-miners} is easy to analyze. For instance, if $l_2 = l_1$, i.e., both miners are equidistant from the coordinator, then we get $\btau = \lti_1 + 1 / (\hti_1 + \hti_2) = \lti_1 + \tau$, which directly states that the overall efficiency is smaller than one and decreases as latency increases.

Equidistant miners yield a number of simplifications, even when $n>2$. Indeed, the relative position of a miner with respect to others in the coordinated setting has no bearing on the overall system efficiency, as long as the latencies between the miners and the coordinator remain unchanged. Miners who are equidistant from the coordinator can, hence, be merged without changing the overall system efficiency, as shown in the next result.
\begin{corollary}[Merging equidistant miners]
\label{cor.merging}
Merging two miners that are equidistant to the coordinator does not change the overall efficiency. Formally, suppose that two miners $m_i$ and $m_{i+1}$ are equidistant from the coordinator, i.e., $l_i = l_i+1$ and let $\hb^{\prime} = [h_1, \cdots, h_{i-1}, h_i+h_{i+1}, h_{i+2}, \cdots, h_n]$ and $\lb^{\prime} = [l_1, \cdots, l_i, l_{i+2}, \cdots, l_n]$. Then $\eff^C(\hb, \lb, \tau) = \eff^C(\hb^{\prime}, \lb^{\prime}, \tau)$. 
\end{corollary}
\begin{proof}
The result follows directly from \eqref{eq.taubar-C} applied to both systems $(\hb, \lb, \tau)$ and
$(\hb^{\prime}, \lb^{\prime}, \tau)$.
\end{proof}

In the case where all $n>2$ miners are equidistant from the coordinator, we deduce a very simple expression for the overall efficiency, as stated in Cor.~\ref{cor.equidistant}. Note that the case of all-equidistant miners is the only case, to our knowledge, for which an expression of the overall system efficiency in the coordinated setting was available in the literature. Unsurprisingly, our expression above coincides with that result (Lemma~2 in \cite{lewenberg2015bitcoin}). 
\begin{corollary}[All-equidistant miners]
\label{cor.equidistant}
If all miners are equidistant from the coordinator, i.e., $l_1 = \cdots = l_n$, then the overall efficiency reduces to\\[-3mm]
\begin{equation*}
    \eff^C(\htib,\ltib) = \tau/\btau; \textrm{ with } \btau = \lti_1 + \tau. 
\end{equation*}
\end{corollary}
\begin{proof}
The result follows directly from Thm.~\ref{thm.overall-efficiency-C} applied in the case where $\lti_1 = \cdots = \lti_n$.
\end{proof}

The expression above also has an intuitive interpretation in light of Cor.~\ref{cor.merging}. If all miners are equidistant from the coordinator, we can merge them all without changing the overall efficiency. We are then left with a single miner of compute capacity $1$ that has a round-trip time of $\lti_1$ to the coordinator. It is therefore natural that the average time between two blocks is $\lti_1 + \tau$ as the single miner has to wait for $\lti_1$ after mining a block.


To conclude, we examine how the overall efficiency varies with the delays. Thm.~\ref{thm.overall-efficiency-C} portrays a somewhat complex dependency on the vector of latencies $\lb$. In the next corollary, however, we show that if the latency from one miner to the coordinator increases (while retaining intact the latency of every other miner to the coordinator), then the overall efficiency decreases. 
\begin{corollary}[Monotonicity]
\label{cor.increase-delay}
If the latency from one miner to the coordinator is increased, everything else being kept equal, the overall system efficiency decreases. Formally, consider a coordinated model with parameters $\hb$, $\lb$, $\tau$ and let, for some $i\in \{1, \cdots, n\}$, $\lb^{\prime}$ be such that $l^{\prime}_j = l_j$ for all $j\neq i$ and $l^{\prime}_i > l_i$. Then $\eff^C(\hb, \lb^{\prime}, \tau) < \eff^C(\hb, \lb, \tau)$.
From this, we also deduce that if all latencies (weakly) increase, i.e., if $\lb^{\prime}$ is such that $l^{\prime}_j \ge l_j$ for all $j$ and $l^{\prime}_i > l_i$ for some $i\in \{1, \cdots, n\}$, then $\eff^C(\hb, \lb^{\prime}, \tau) < \eff^C(\hb, \lb, \tau)$.
\end{corollary}
\begin{proof}
The first point follows directly from \eqref{eq.taubar-C} by noting that $\btau$ increases with $\lb^{\prime}$ instead of $\lb$, hence $\eff^C$ decreases. The second point follows from applying the first sequentially for each miner whose latency is increased. 
\end{proof}

\begin{figure}
  \begin{center}
   \includegraphics[width=1\threecolgrid]{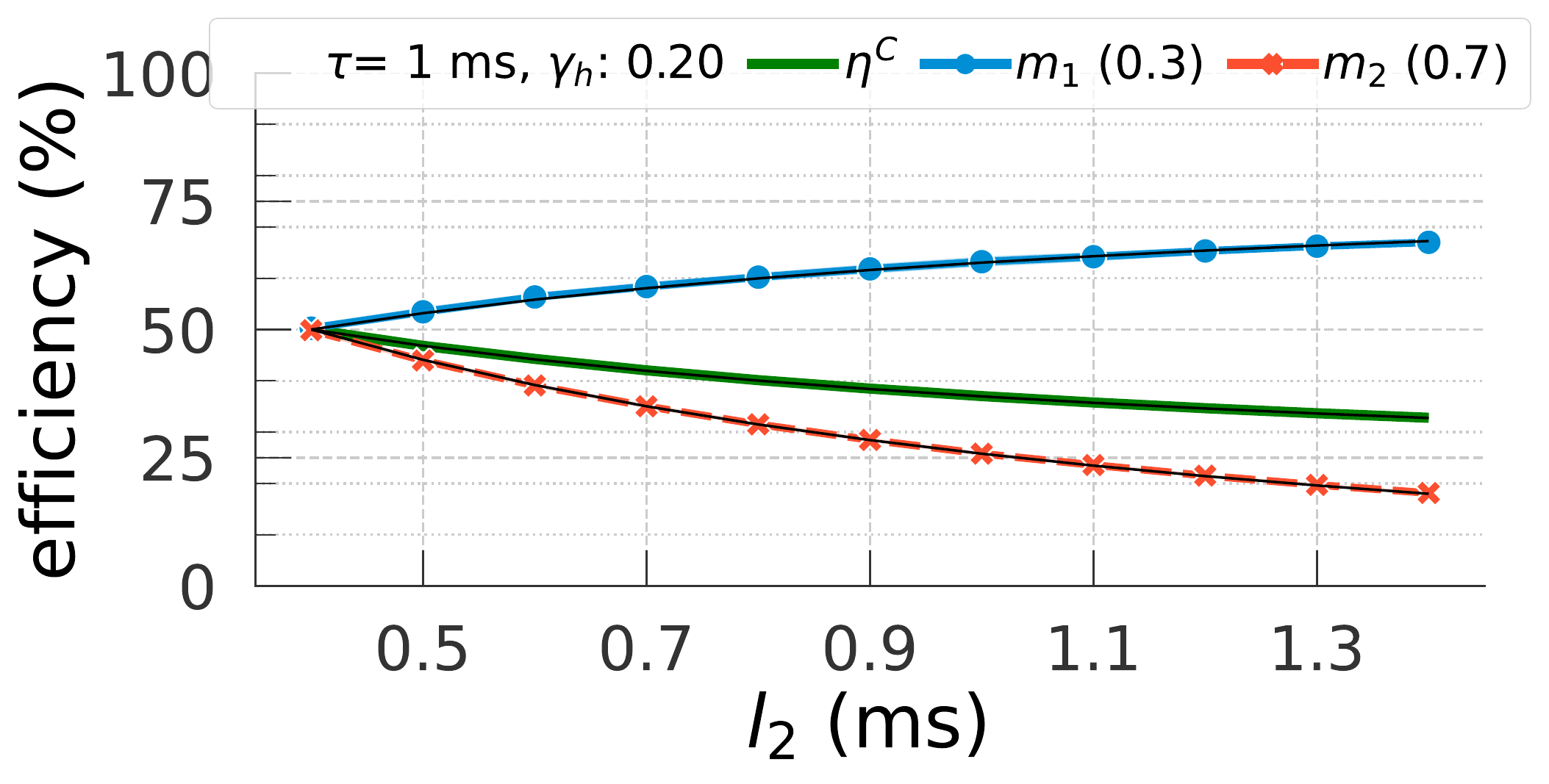}
  \end{center}
  \figcap{(Coordinated setting.) Overall and individual efficiencies as a function of the distance $l_2$ from $m_2$ to the coordinator; in a two-miner scenario with $l_1 = \ums{0.5}$. Here, $h_1 = 0.3$, $h_2 = 0.7$ (hence $\gamma_h = 0.2$), and $\tau = \ums{1}$. The black lines on top of the other curves correspond to the theoretical values from Thm.~\ref{thm.overall-efficiency-C}.
  }\label{fig:theory-result-miner-moves-away}
\end{figure}

%
%
Fig.~\ref{fig:theory-result-miner-moves-away} illustrates the strong dependency of the overall system efficiency on the latency between the miners and the coordinator in a two-miners scenario. We fix the position of $m_1$ and of the coordinator such that $l_1 = \ums{0.5}$, and we move $m_2$ away from $C$ to increase $l_2$ from the equidistant position $l_2 = \ums{0.5}$. Then we observe on Fig.~\ref{fig:theory-result-miner-moves-away} that the overall efficiency decreases as $m_2$ moves away from the coordinator. It also shows that the individual efficiency of $m_1$ increases---we discuss the effect of latency on individual efficiencies in detail in the next section. Finally, Fig.~\ref{fig:theory-result-miner-moves-away} clearly shows the agreement between the theoretical curves and the results from our simulations. We also notice that the standard deviation for the simulation results is so small that it is not visible---hence validating that the length of our simulations is sufficient to obtain statistically robust estimation of the overall system efficiency (cf.~\S\ref{s:sim}). 

%% file: sections/eff/p2p.tex

\subsection{Peer-to-peer Setting}\label{ss:eff-p2p}

\subsubsection{Empirical investigation of the efficiency as a function of latency}


In the P2P setting, miners directly interact with one another through a P2P network without any central coordinator. This configuration makes the efficiency less tractable analytically. The latencies between miners in the P2P network, nevertheless, have implications for the overall system efficiency.
We start by investigating this trade-off through simulations.

\begin{figure*}[tb]
	\centering
	\begin{subfigure}[b]{\threecolgrid}
	    \centering
		\includegraphics[width={\textwidth}]{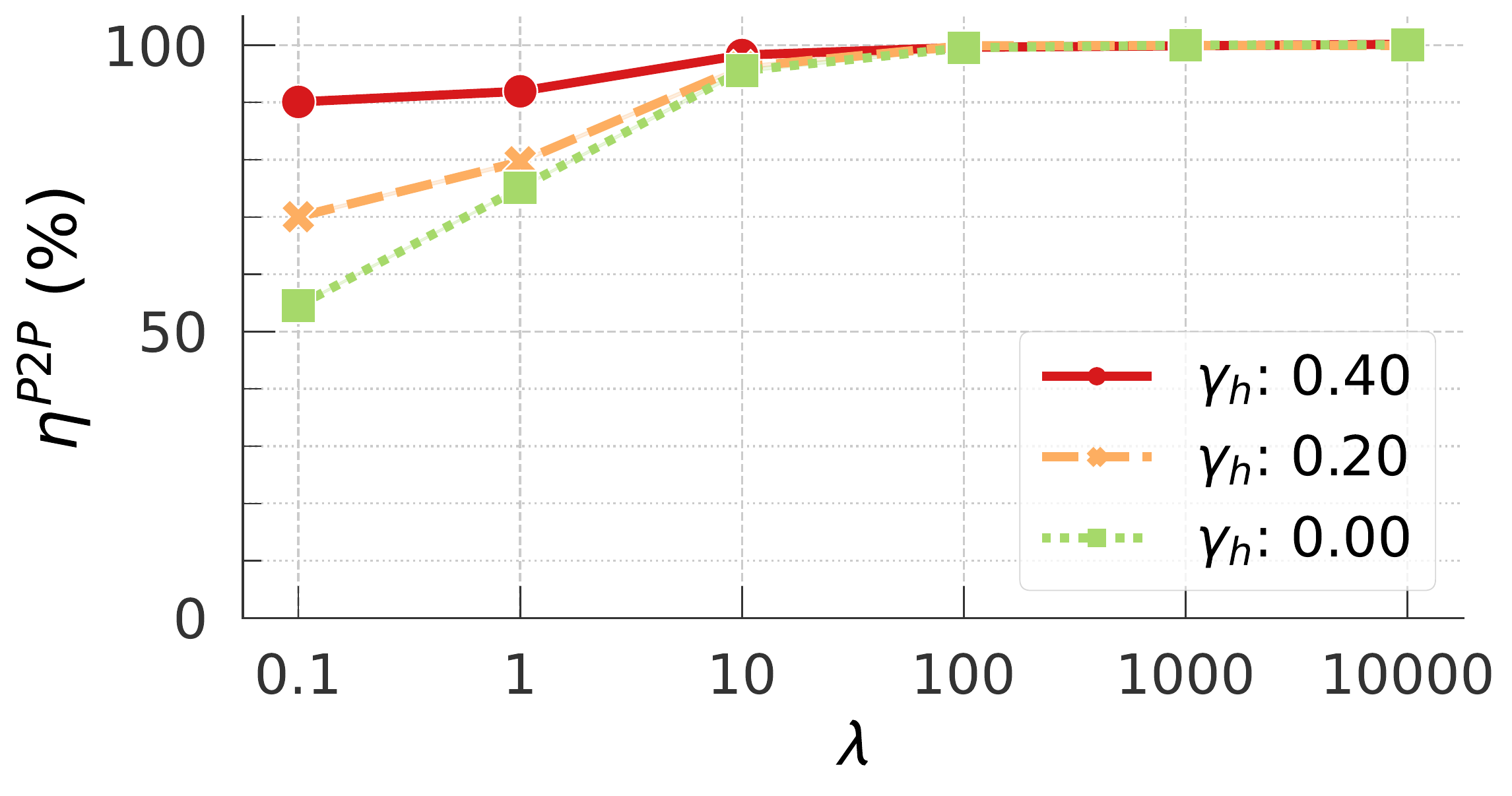}
		\sfigcap{Two miners}\label{fig:two-node-eff}
	\end{subfigure}
	\begin{subfigure}[b]{\threecolgrid}
	    \centering
		\includegraphics[width={\textwidth}]{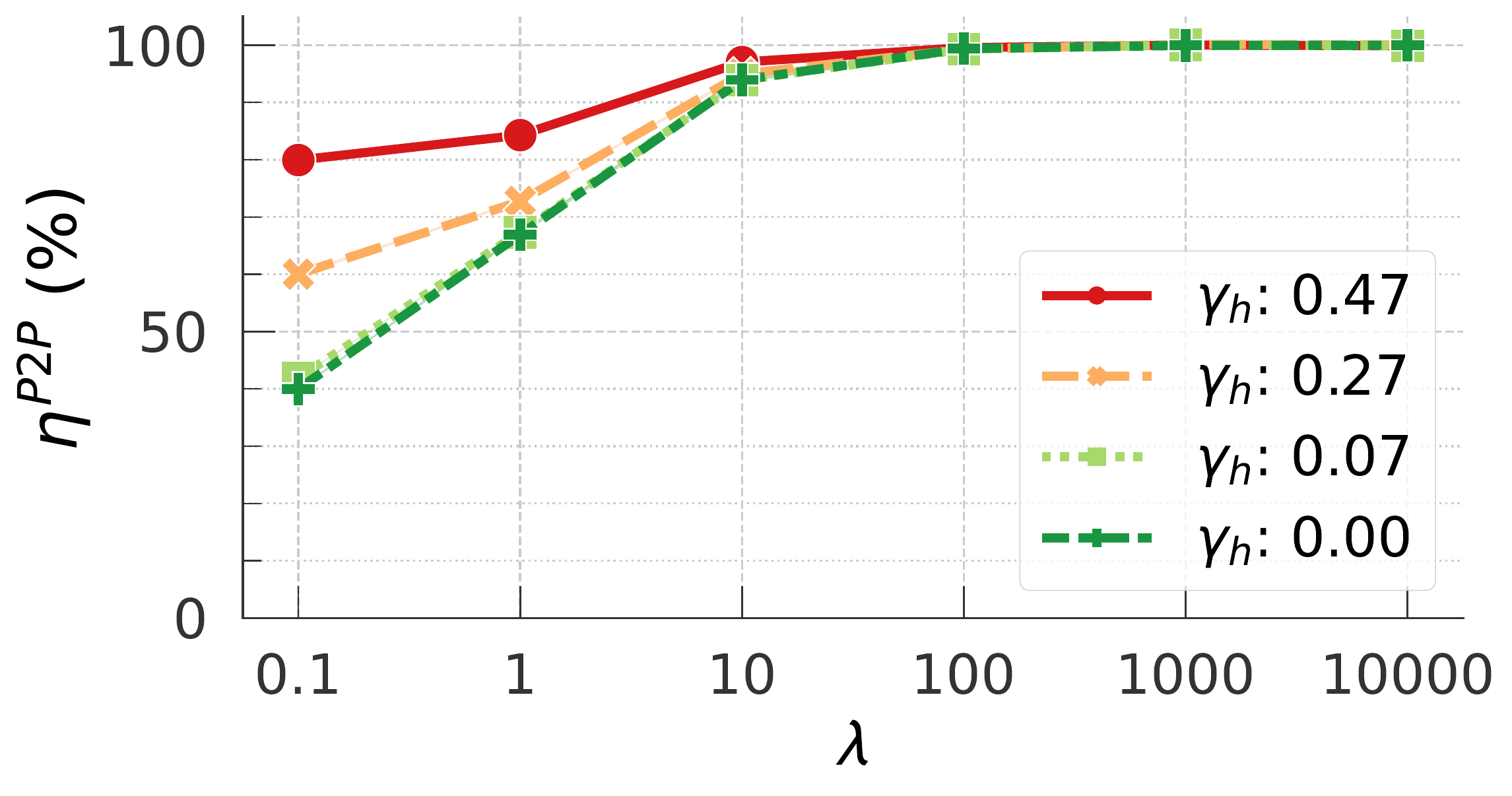}
		\sfigcap{Three miners}\label{fig:three-node-eff}
	\end{subfigure}
	\begin{subfigure}[b]{\threecolgrid}
    	\centering
		\includegraphics[width={\textwidth}]{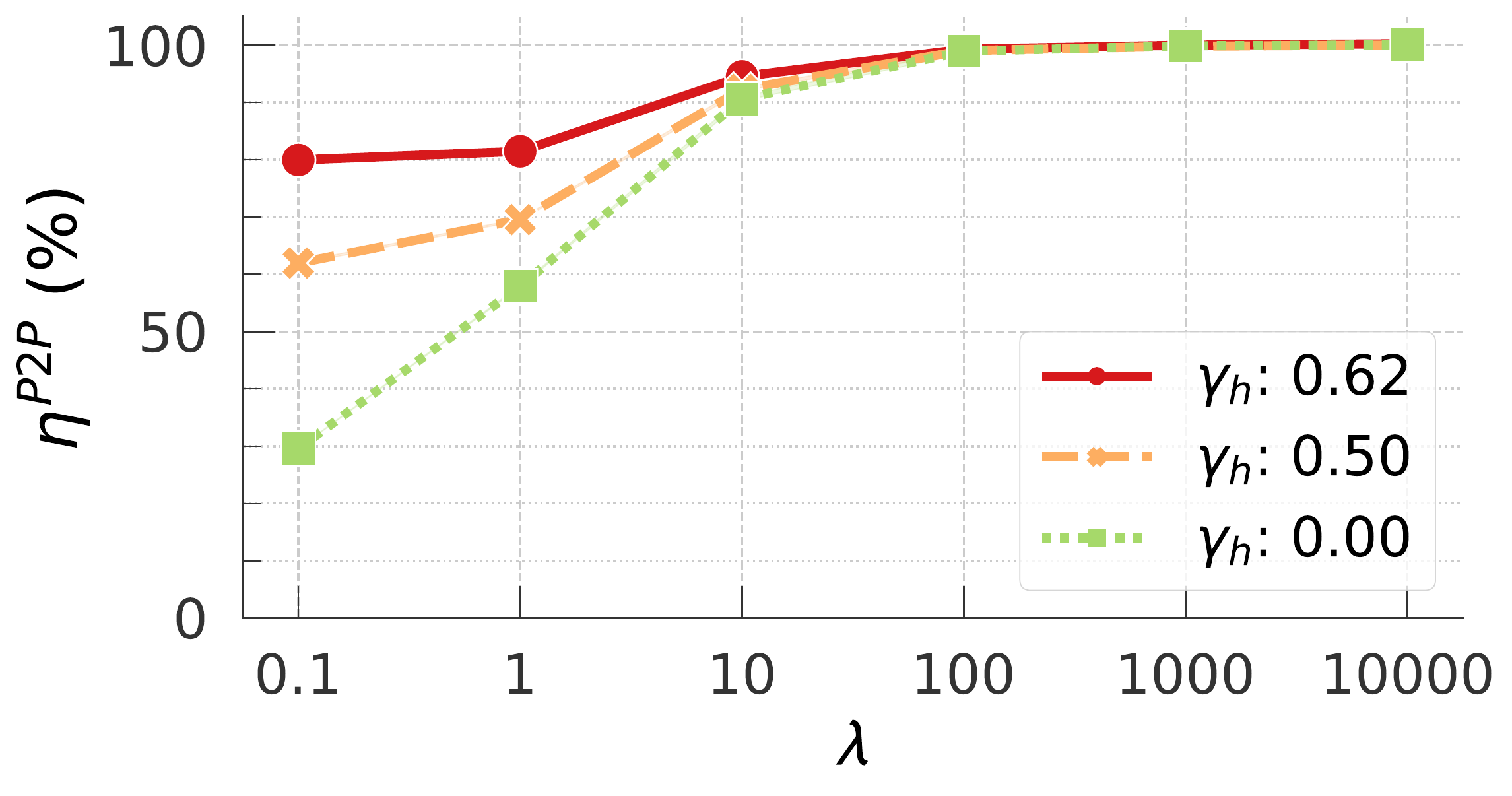}
		\sfigcap{Bitcoin approximation}\label{fig:multi-node-eff}
	\end{subfigure}
	\figcap{(P2P Setting.) Overall system efficiency as a function of $\lambda{}$ (the ratio puzzle hardness to mean latency), for the scenarios described in \S\ref{ss:scenarios} and for various configurations of the compute capacities identified by their $\gamma_h$. 
      }\label{fig:overall-eff}
\end{figure*}

We simulate 3 different scenarios (cf.~\S\ref{ss:scenarios}) and measure the overall system efficiency as a function of $\lambda$ (as shown in Fig.~\ref{fig:overall-eff}).
Recall that $\lambda$ is the ratio of puzzle hardness to mean latency and captures the ratio between the average time to mine a block and the average time to broadcast it.
In both the 2-miner (Fig.~\ref{fig:two-node-eff}) and 3-miner (Fig.~\ref{fig:three-node-eff}) scenarios, the overall system efficiency degrades quickly with decreasing $\lambda$ (i.e., with increasing latency). This observation is consistent under all compute-capacity assignments (succinctly represented by $\gamma_h$), but we restrict the plots to a subset for clarity. When $\lambda{} < 1$, the time spent in mining blocks, on average, is less than that in advertising those blocks. Even in this regime of low $\lambda$ values, a non-uniform, realistic compute-capacity assignment (i.e., $\gamma_h > 0$) ameliorates the overall system efficiency. Perhaps stronger miners (with relatively high compute capacities) outpace the weaker to bring about this improvement. We defer the discussion of such inequalities to \S\ref{s:ineq}.

%
%

Latency across arbitrary endpoints in the Internet is quite high---at least an order of magnitude more than what is theoretically feasible---and reducing latency is a notoriously hard problem~\cite{Singla-HotNets2014}. Such high network latencies make it impractical to reduce hardness ($\tau$), albeit low values of $\tau$ are crucial for increasing the (transaction) throughput. With all other factors held constant, $\tau$ must be drastically increased to improve efficiency.  Bitcoin and Ethereum operate, for instance, with $\tau{}$ values of approximately $10$ minutes and $13$ seconds, respectively, most likely to keep $\lambda{}$ and, consequently, the overall system efficiency high. 
%
%
As we have discussed in \S \ref{ss:scenarios}, Bitcoin and Ethereum operate at $\lambda$ ranges of $100-1000$ and $10-100$ respectively. Per Fig.~\ref{fig:multi-node-eff}, 
$\eff^{P2P}$ at a $\lambda=100-1000$ is close to 100\% and at $\lambda=10-100$ is around 90-99\%, which is in agreement with the estimated $\eff^{P2P}$ of $99\%$ and 90-94\% for Bitcoin and Ethereum respectively~\cite{Gencer-FC2018}.


%% file: sections/eff/p2p-vs-c.tex



\subsubsection{Comparison of the overall efficiency in the P2P and coordinated settings}\label{sec.comparison}

We have seen that, for both the P2P and the coordinated setting, overall efficiency decreases with latency. We now compare the overall system efficiency of the two models. To that end, we fix the number of miners, the vector of computational capacities $\hb$, and the puzzle hardness $\tau$ identical across the models; and we assume that the latency matrix $L$ of the P2P model and the latency vector $\lb$ of the coordinated model are consistent with each other (see~\S\ref{ss:two-models}), i.e., the addition of the coordinator in the coordinated setting does not shorten the path between any two miners in the P2P setting. Our next result shows that, under that consistency condition, the overall system efficiency is always higher in the P2P setting than in the coordinated setting.
\begin{theorem}
\label{thm.overall-efficiency-CvsP2P}
For any $\tau$ and $\hb$ and for any $L$ and $\lb$ such that $l_{ij} \le l_i + l_j$
for all $i$ and $j$, we have\\[-3mm]
\begin{equation*}
\eff^{P2P} (\hb, L, \tau) \ge \eff^C (\hb, \lb, \tau).
\end{equation*}
\end{theorem}
\begin{proof}
This proof relies on constructing, from a base set of independent
Poisson processes on each miner, two coupled sets of block discovery times---one
statistically consistent with the P2P setting and the other with the C setting.
The construction is such that the second set is a strict subset of the first,
and allows us to compare the longest chain obtained from the two sets of
discovery times under the P2P and C settings, respectively.
A complete proof is included in Appendix~\ref{sec.proofs}.
\end{proof}

Thm.~\ref{thm.overall-efficiency-CvsP2P} only requires the consistency condition that $l_{ij} \le l_i + l_j$ for all $i$ and $j$. The requirement implies that, even if the coordinator does not introduce any extra delay between any two miners (which is possible in the two-miners case, for instance, when the coordinator is in between the miners), the result still holds and the overall efficiency is higher for the P2P than the coordinated setting.
An intuitive reason is that miners in the coordinated setting must pause after discovering a block (waiting for information from the coordinator to resume mining). In the P2P setting, this time can lead to a new block discovery, which \emph{can} (although it will not always) end up growing the chain. 
In most cases, however, the path between any two miners via the coordinator will be longer than that in the P2P setting, i.e., the coordinator will add extra delays. These additional delays introduce a second source of efficiency loss in the coordinated setting as described in Cor.~\ref{cor.increase-delay}.


\begin{figure}[tbp]
  \centering
  \begin{subfigure}[b]{1\threecolgrid}
    \centering \includegraphics[width={\textwidth}]{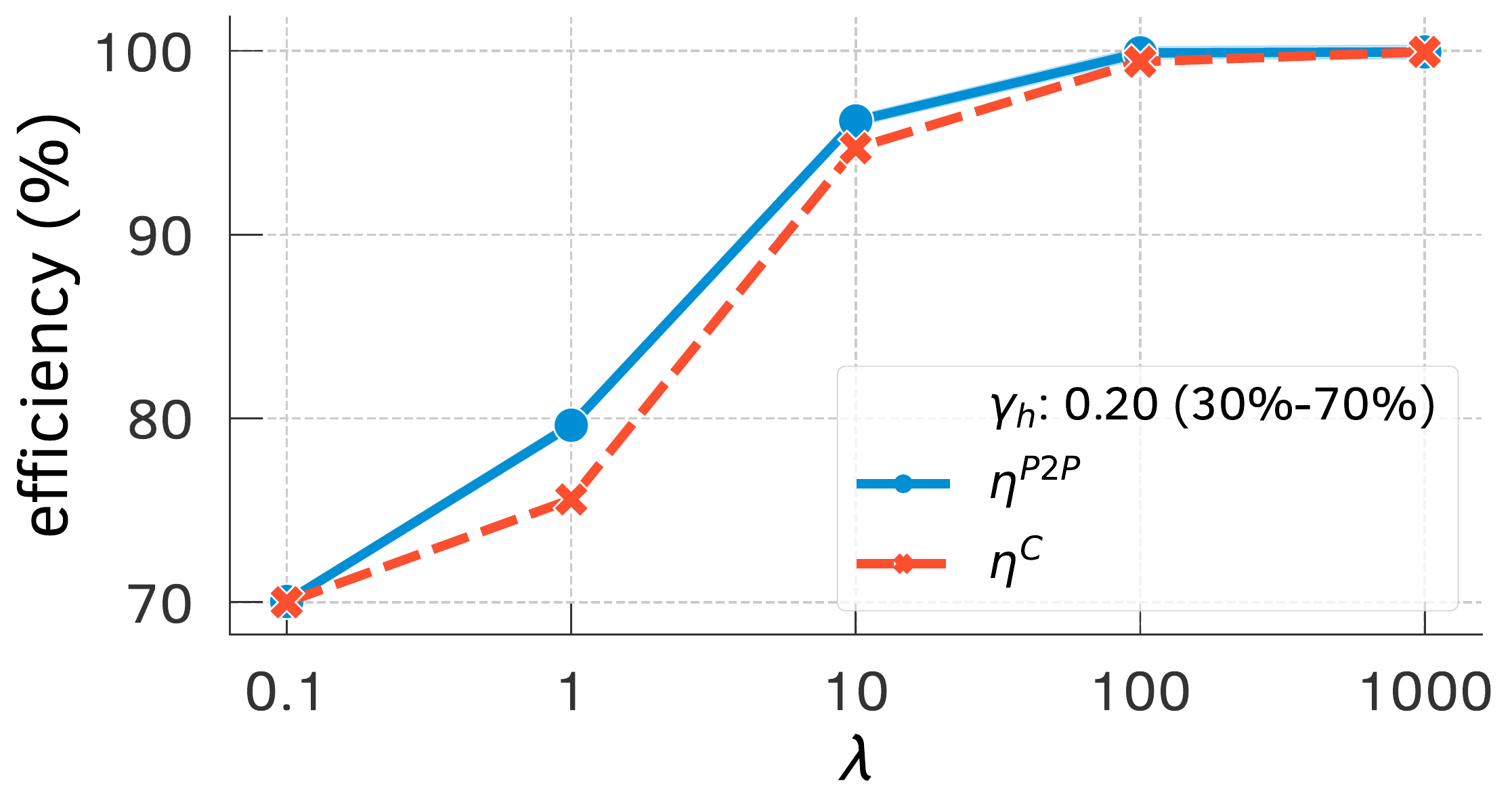}
    \sfigcap{Two miners ($\gamma_h=0.20$)}\label{fig:two-node-p2p-c}
  \end{subfigure}
	\begin{subfigure}[b]{1\threecolgrid}
      \centering \includegraphics[width={\textwidth}]{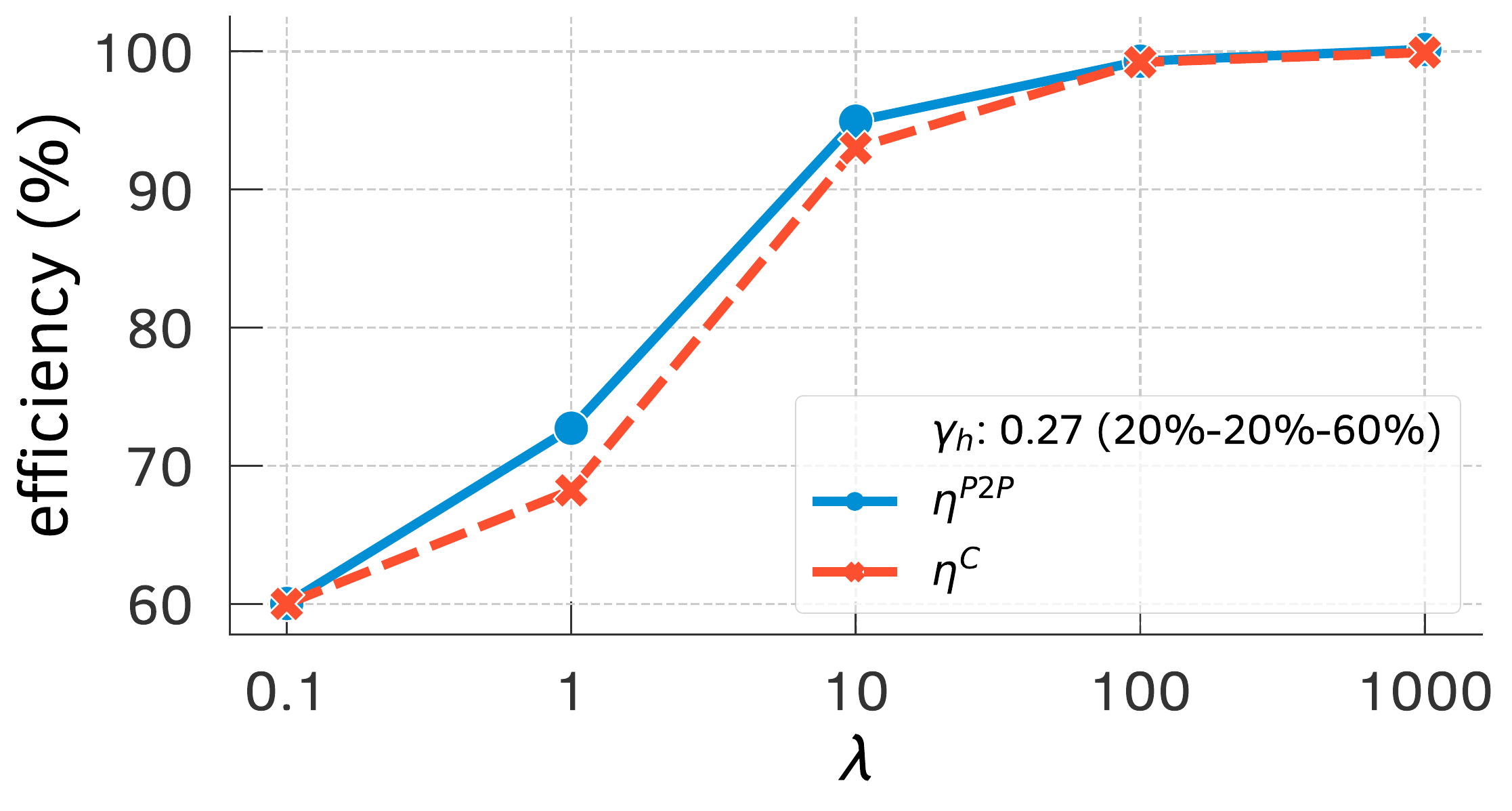}
      \sfigcap{Three miners ($\gamma_h=0.27$)}\label{fig:three-node-p2p-c}
	\end{subfigure}
	\figcap{(P2P \emph{vs.} coordinated setting.) Overall system efficiency as a function of $\lambda$ in the two settings for the scenarios described in \S\ref{ss:scenarios}. For each point in the coordinated setting, the optimal coordinator's position is set. }\label{fig:p2p-vs-c-eff}
\end{figure}

To illustrate the result of Thm.~\ref{thm.overall-efficiency-CvsP2P}, we perform simulations of both the two-miner and three-miner scenarios under different $\lambda$ values and compute-capacity distributions. Fig.~\ref{fig:p2p-vs-c-eff} shows the results for selected configurations. For the coordinated setting, we used the ``best'' position for the coordinator, i.e., the position that results in a maximal overall system efficiency for each point (found via an exhaustive search). We observe that, in most cases, the inequality in Thm.~\ref{thm.overall-efficiency-CvsP2P} is strict; even in the two-miners case where adding the coordinator does not add any extra delay (i.e., where the best position of the coordinator is also in between the two miners). In the next subsection, we delve into what position(s) of the coordinator lead to maximal overall system efficiency and why.

To conclude this subsection, let us observe that the result of Thm.~\ref{thm.overall-efficiency-CvsP2P} can also be interpreted as a lower bound on the overall system efficiency for the P2P setting. Then, to obtain the best possible lower bound for a P2P setting with fixed $\tau$, $\hb$ and $L$, one simply needs to search for the latency vector $\lb$ consistent with $L$ that maximizes the overall efficiency of the coordinated setting, i.e., solve $\max \eff^C (\hb, \lb, \tau)$ over the space of all vectors $\lb$ such that $l_i + l_j \ge l_{ij}$ for all $i$ and $j$. (Note that this problem is different from the problem of the next subsection where we search only in the space of vectors $\lb$ induced by \emph{physically} possible locations of the coordinator.) Such a lower bound is valuable because no closed-form expression exists for the P2P setting; we also observe on Fig.~\ref{fig:p2p-vs-c-eff} that is seems close to the actual efficiency.

%% file: sections/eff/c-pos.tex
\subsection{Optimal Placement of Coordinator}\label{ss:ss:opt-placement}

Where should a coordinator be placed with respect to the miners to maximize the overall system efficiency in the coordinated setting?
This question naturally arises in practice where a mining pool operator (MPO) has to decide where to place the coordinator in their pool for maximizing the overall system efficiency such as those in Bitcoin~\cite{btc-shares}. Unfortunately, as the expression of $\eff^C$ as a function of $\lb$ (or of the coordinator's position) is complex, it is not feasible to find analytically the optimal position of the coordinator. Rather, we use simulations to provide some insights.

\begin{figure*}[tbhp]
  \centering
  \begin{subfigure}[b]{\threecolgrid}
    \includegraphics[width={\textwidth}]{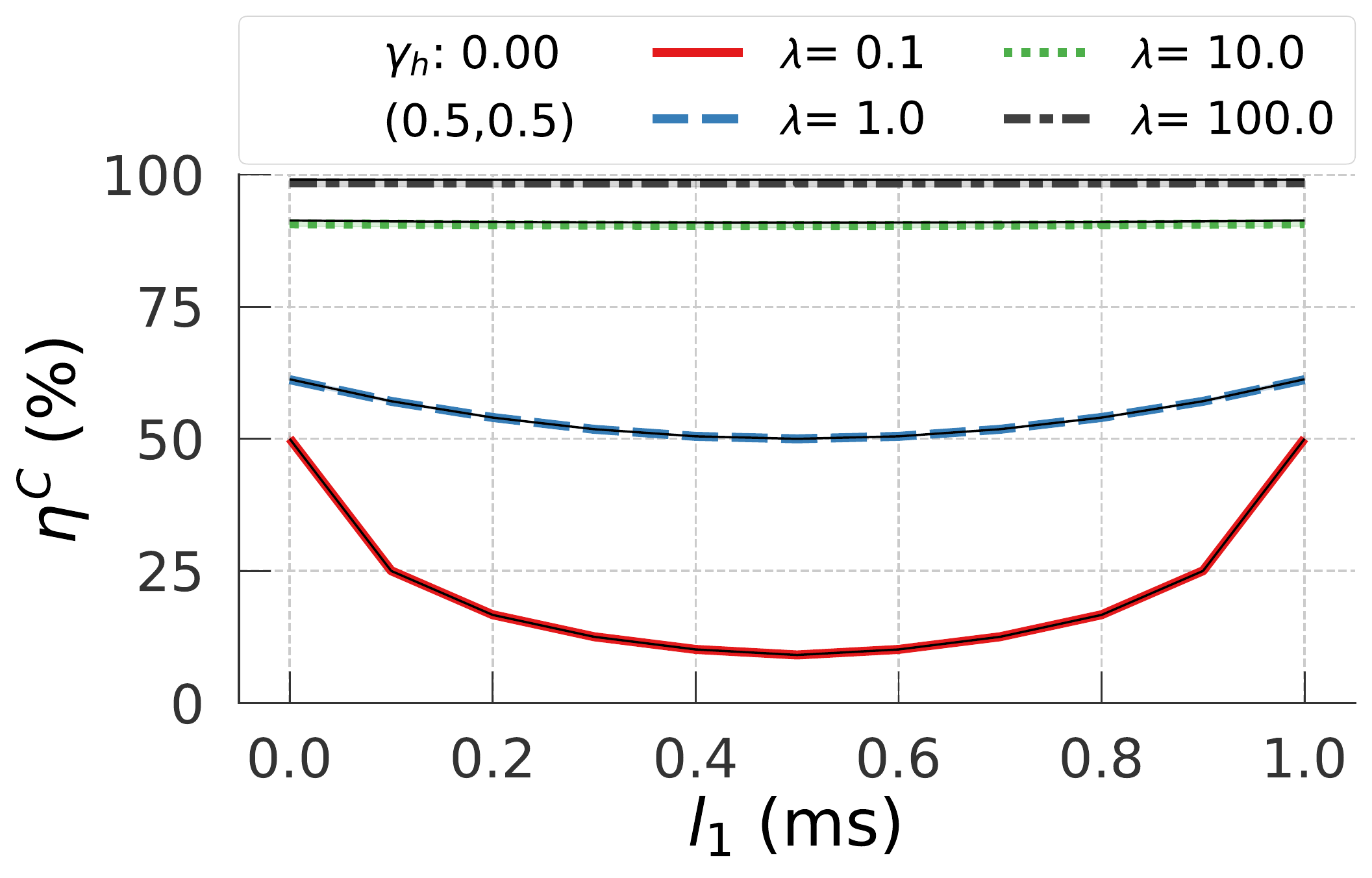}
    \sfigcap{}\label{fig:2m-ec-g0.0}
  \end{subfigure}
  \begin{subfigure}[b]{\threecolgrid}
    \includegraphics[width={\textwidth}]{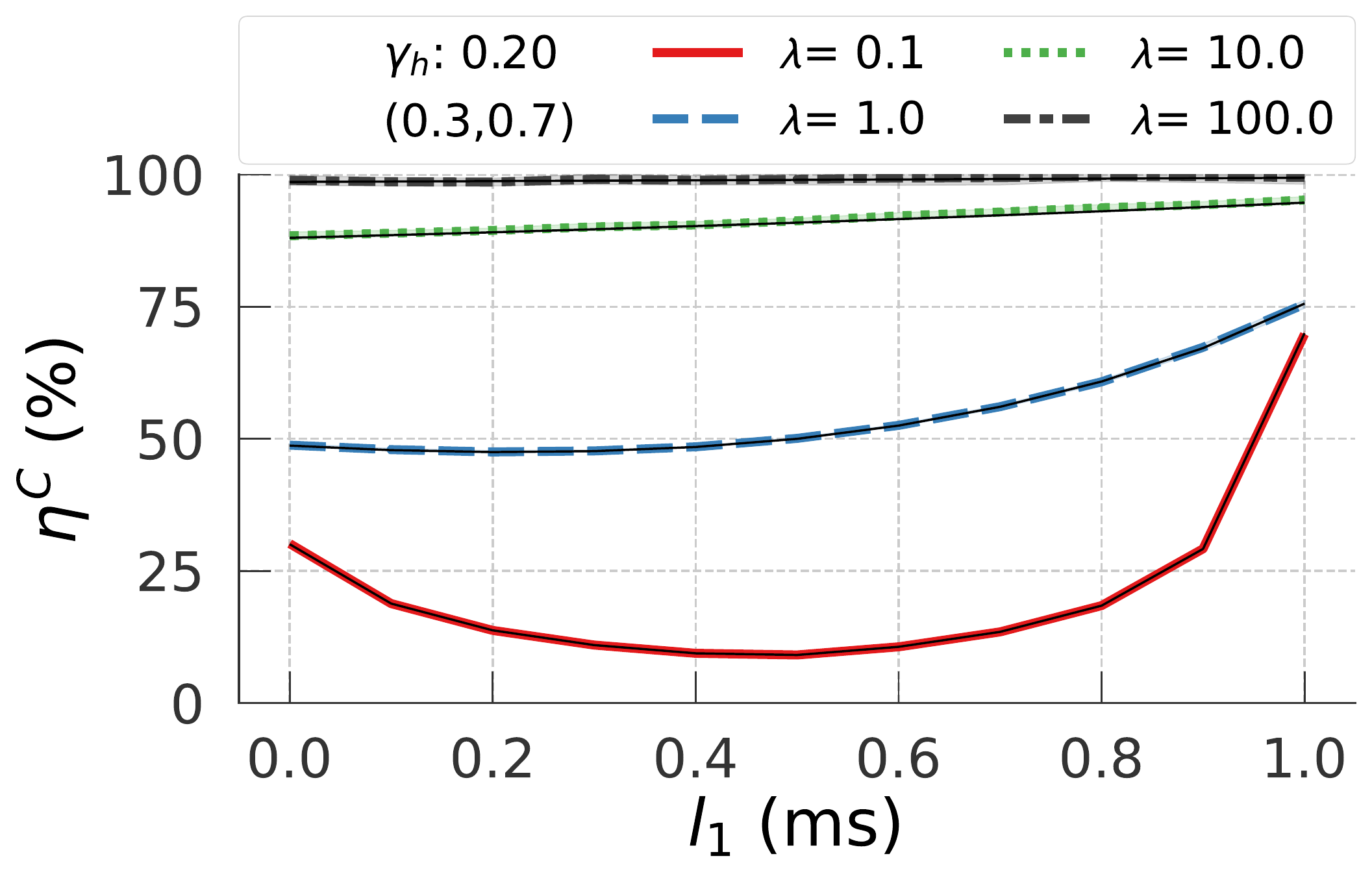}
    \sfigcap{}\label{fig:2m-ec-g0.2}
  \end{subfigure}
  \begin{subfigure}[b]{\threecolgrid}
    \includegraphics[width={\textwidth}]{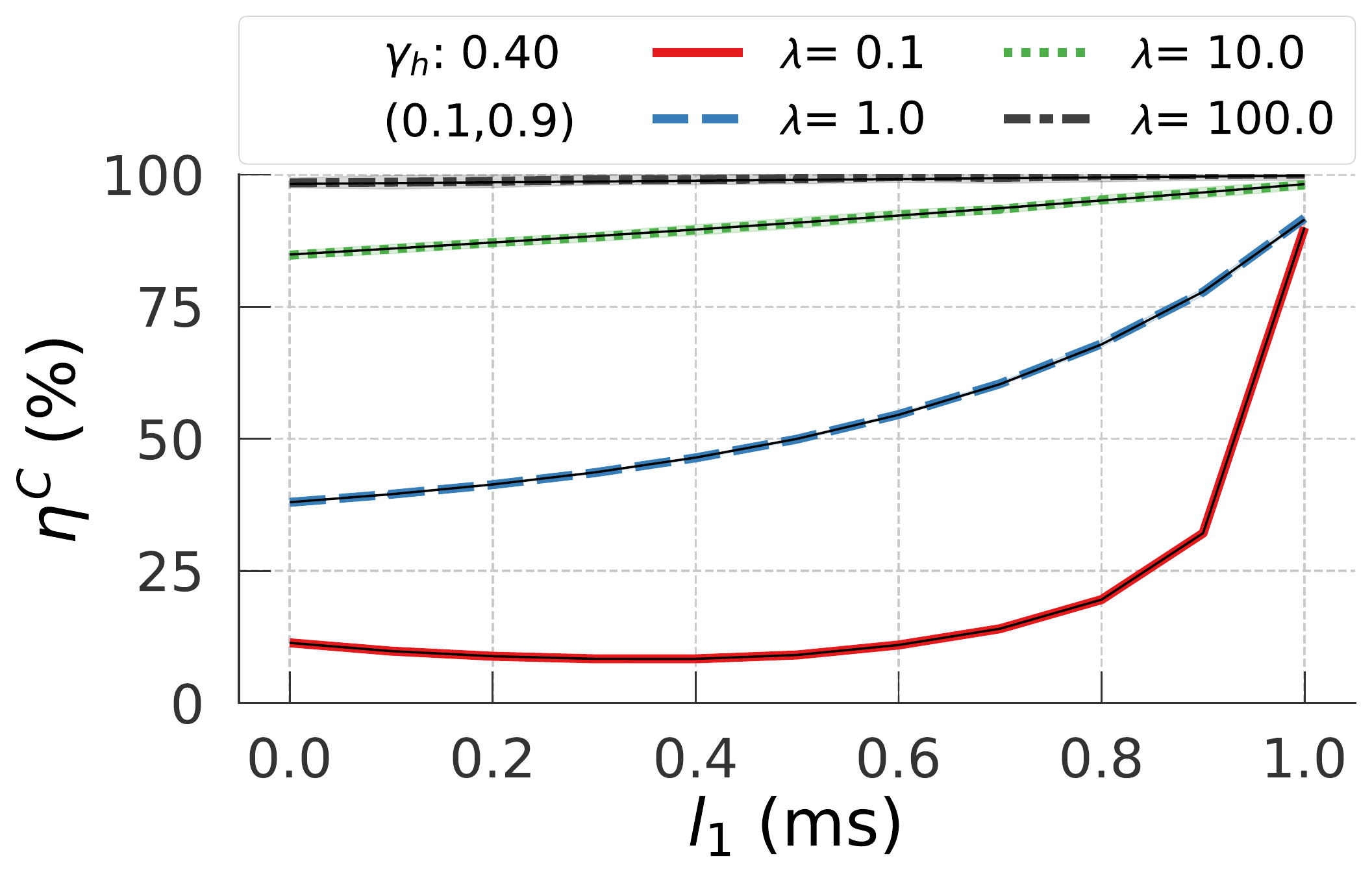}
    \sfigcap{}\label{fig:2m-ec-g0.4}
  \end{subfigure}
  \figcap{(Coordinated setting.) Overall efficiency as a function of the distance $l_1$ from $m_1$ to the coordinator (with $l_1+l_2=\ums{1}$), in the two-miners scenario where miners (a) have equal capacity ($\gamma_h=0$), (b) have a 30\%-70\% split ($\gamma_h=0.2$) and (c) have a 10\%-90\% split ($\gamma_h=0.4$) of the total compute capacity. Note that $m_1$ is the weak miner. The black lines on top of the curves correspond to the theoretical values from Theorem~\ref{thm.overall-efficiency-C}. }\label{fig:2m-ec}
\end{figure*}

Fig.~\ref{fig:2m-ec} shows the results for simulations of the two-miner scenario. We observe that for small values of $\lambda$ (i.e., $\lambda \le 0.1$) with two equally powerful miners (i.e., $\gamma_h = 0$), the overall system efficiency increases when the coordinator moves close to either miner (in Fig.~\ref{fig:2m-ec-g0.0}). When the total compute capacity is unequally distributed (as in the case of Fig.~\ref{fig:2m-ec-g0.2} \& Fig.~\ref{fig:2m-ec-g0.4}), the overall system efficiency peaks as the coordinator moves close to the stronger miner. For high values of $\lambda$, however, this question of optimal placement becomes  moot as the efficiency is close to one for all locations.

\vspace*{0.5em}

\begin{figure*}[tb]
  \centering
  \begin{subfigure}[b]{\fourcolgrid}
    \includegraphics[width={\textwidth}]{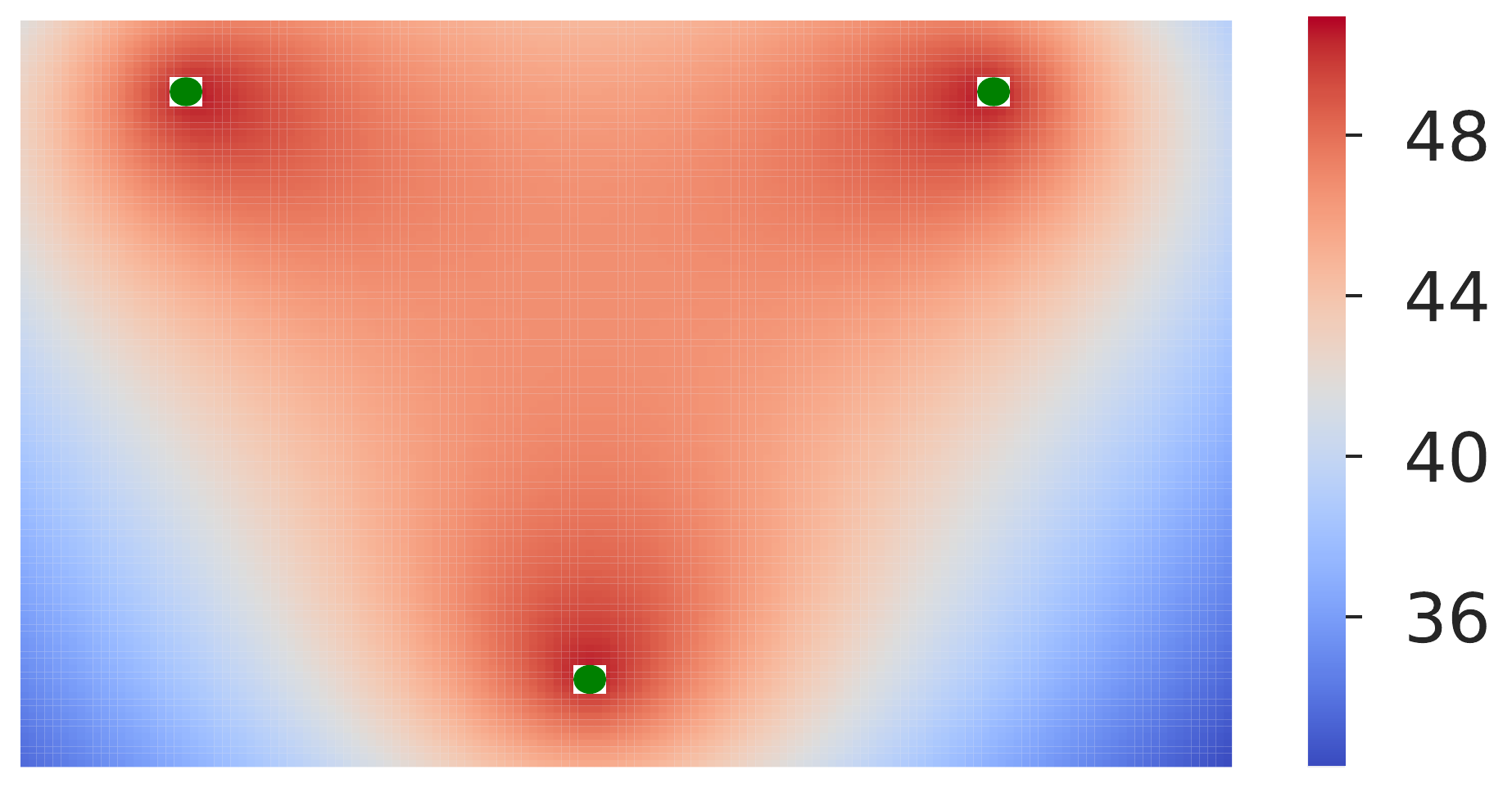}
    \vspace*{2pt}
    \sfigcap{$\gamma_h=0$; $\lambda=1$}\label{fig:3m-ec-g0.0-hm}
  \end{subfigure}
  \begin{subfigure}[b]{\fourcolgrid}
    \includegraphics[width={\textwidth}]{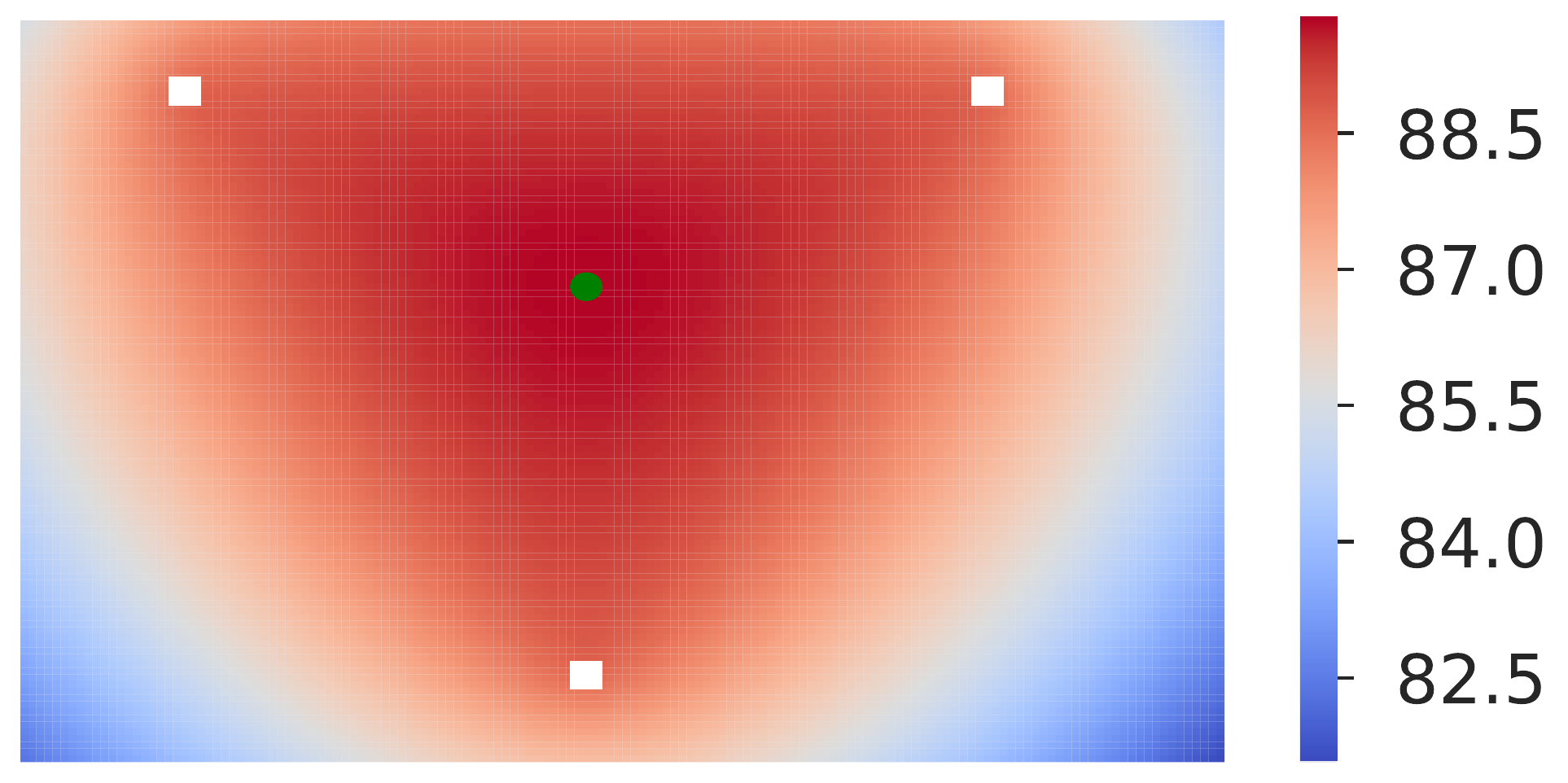}
    \vspace*{2pt}
    \sfigcap{$\gamma_h=0$; $\lambda=10$}\label{fig:3m-ec-g0.0-l10-hm}
  \end{subfigure}
  \begin{subfigure}[b]{\fourcolgrid}
    \includegraphics[width={\textwidth}]{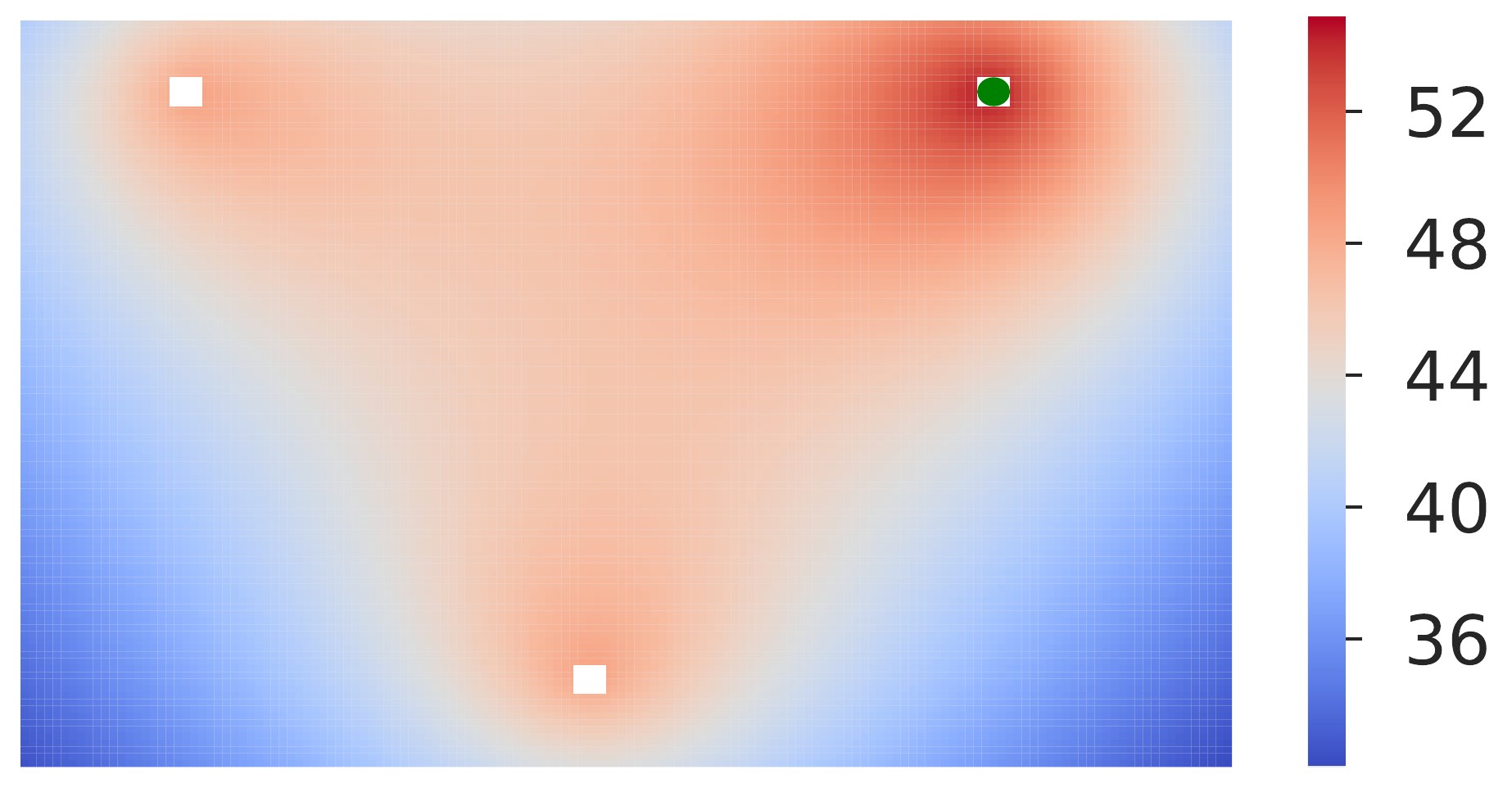}
    \vspace*{2pt}
    \sfigcap{$\gamma_h=0.07$; $\lambda=1$}\label{fig:3m-ec-g0.07-hm}
  \end{subfigure}
  \begin{subfigure}[b]{\fourcolgrid}
    \includegraphics[width={\textwidth}]{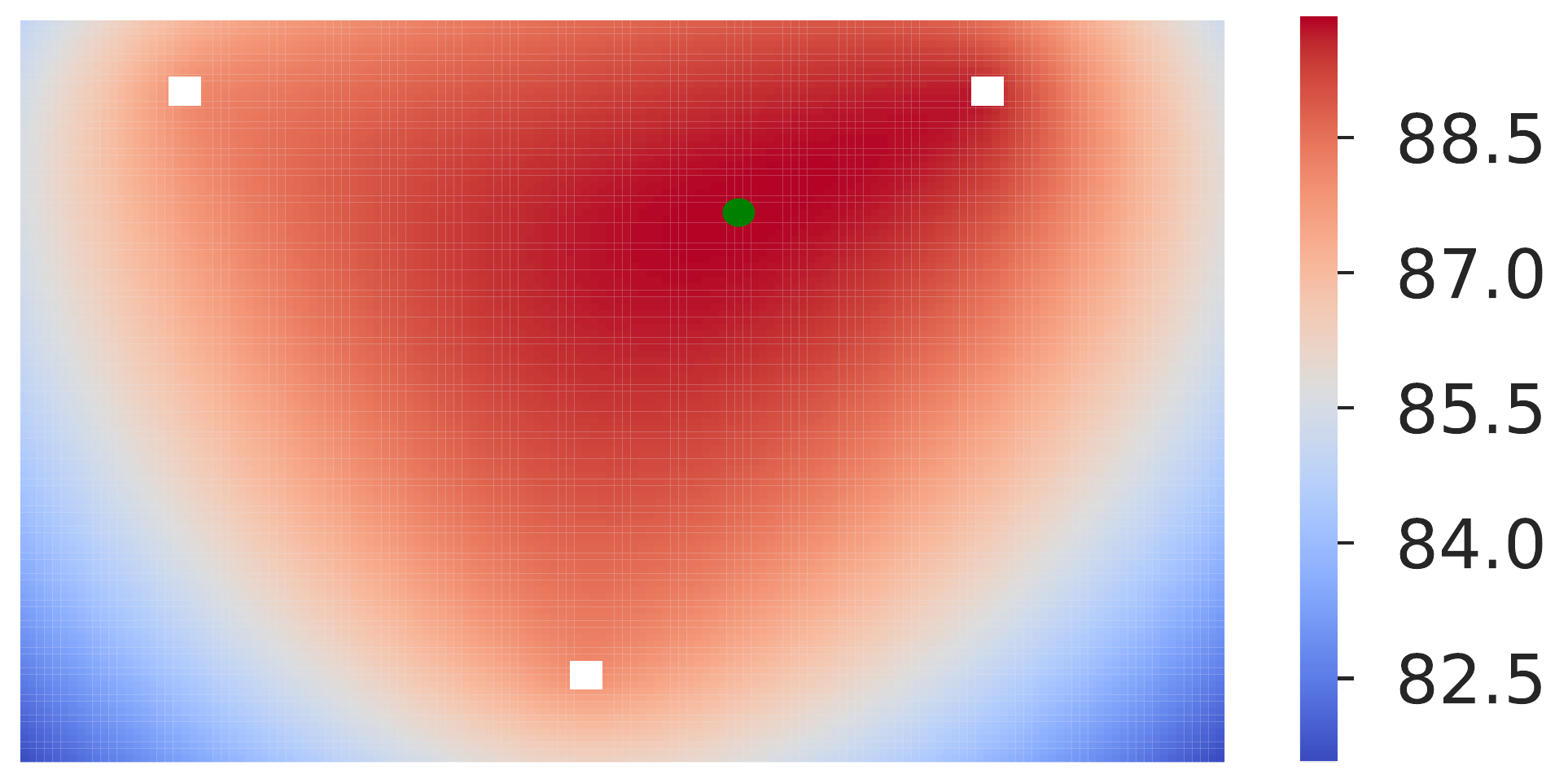}
    \vspace*{2pt}
    \sfigcap{$\gamma_h=0.07$; $\lambda=10$}\label{fig:3m-ec-g0.07-l10-hm}
  \end{subfigure}
  \figcap{(Coordinated setting.) Optimal position of the coordinator in the 3-miner scenario. The color map represents $\eff{}^C$ as a function of the coordinator position in the 2D plane, white squares show miners, and the green circle shows the coordinator position maximizing $\eff{}^C$. The distribution of compute capacities in (a) and (b) is equal, while (c) and (d) have a distribution of 30\%-40\%-30\%.}
  \label{fig:3m-ec-hm}
\end{figure*}

The difficulty in finding the optimal coordinator position becomes more pronounced in the three-miner scenario (Fig.~\ref{fig:3m-ec-hm}). The optimal position(s) might correspond to that of one miner or those of several miners (Fig.~\ref{fig:3m-ec-g0.0-hm} and \ref{fig:3m-ec-g0.07-hm}), but might be completely different in other scenarios. In case of an unequal distribution of compute capacities (Fig.~\ref{fig:3m-ec-g0.07-l10-hm}) the optimal location of the coordinator is closer to the stronger miner, whereas when compute capacities are shared equally across miners (Fig.~\ref{fig:3m-ec-g0.0-l10-hm}) the optimal location is in the middle (i.e., equidistant from all three miners). These simulation results attest to the difficulty in determining the optimal coordinator location with respect to known positions (and capacities) of miners.
%
%

%% file: sections/eff/end.tex

\subsection{Takeaways and Implications}

The overall system efficiency is affected by latencies in both the P2P and coordinated models; but it is always higher in the P2P model. In the coordinated model, finding the optimal location of the coordinator is a non-trivial challenge for general $n$-miner scenario. It may be possible to use a variant of gradient descent based on our closed-form expression, but the non-convexity of the overall system efficiency makes its analysis challenging. 

%% file: sections/ineq.tex
\section{Inequality in Efficiency}\label{s:ineq}
In this section, we turn to the analysis of individual efficiency of miners (and the inequality thereof) in a blockchain with non-zero latency, in both the coordinated and the P2P setting. 


\input{sections/ineq/coord}

\input{sections/ineq/p2p}

\input{sections/ineq/end}


%% file: sections/ineq/coord.tex
\subsection{Coordinated Setting}

We start by investigating how latency affects the individual efficiency of miners in the coordinated setting, in two main steps. As for the overall efficiency, we provide a closed-form expression (Thm.~\ref{thm.individual-efficiency-C}) for computing the efficiency of any individual miner under arbitrary parameters.
\begin{theorem}[Individual efficiency in the coordinated setting]
\label{thm.individual-efficiency-C}
Consider a coordinated model with $n$ miners, with computational capacities $\hb$, latency vector $\lb$, and puzzle hardness $\tau$. Recall that $\htib = \hb / \tau$ and let $\lti_i = 2l_i$ for all $i \in \{1, \cdots, n\}$. Without loss of generality, assume that $l_1 \le \cdots \le l_n$. Then the individual efficiency of miner $m_i\in\Mset$ is\\[-5mm]
\begin{equation}
\label{eq.individual-efficiency-C}
    \eff^C_i(\hb,\lb, \tau) = \eff^C_i(\htib,\ltib) = \frac{p_i/\btau}{\hti_i};\\[-2mm]
\end{equation}
where $\btau$ is given by \eqref{eq.taubar-C} and\\[-3mm]
\begin{equation}
\label{eq.pi-C}
    p_i = \hti_i \cdot\! \sum_{k=i}^n \frac{1}{\hti_1\!+\!\cdots\!+\!\hti_k} \cdot 
    \left( e^{\sum_{j=1}^k -\hti_j (\lti_k-\lti_j)} \!-  e^{\sum_{j=1}^k -\hti_j (\lti_{k+1}-\lti_j)} \right).
\end{equation}
Here, we used the convention $l_{n+1} = \infty$ to simplify the expression.
\end{theorem}

A proof of Thm.~\ref{thm.individual-efficiency-C} is provided in Appendix~\ref{sec.proofs}; it follows similar ideas as in the proof of Thm.~\ref{thm.overall-efficiency-C}.
The term $p_i$ in Thm.~\ref{thm.individual-efficiency-C} corresponds to the probability that a block included in the chain was mined by miner $m_i$. From~\eqref{eq.pi-C}, one can check that $\sum_{i=1}^n p_i = 1$. Then, with $\eff^C_i(\hb,\lb, \tau)$ given by~\eqref{eq.individual-efficiency-C} we recover $\sum_{i=1}^n h_i \eff^C_i(\hb,\lb, \tau) = \eff^C(\hb,\lb, \tau)$, where the overall system efficiency in the coordinated setting, $\eff^C(\hb,\lb, \tau)$, is given by \eqref{eq.overall-efficiency-C}---recall that $\sum_{i=1}^n \hti_i = 1/\tau$.


Using the general result of Thm.~\ref{thm.individual-efficiency-C}, we analyze a number of consequences and special cases. We start with the case of two miners.
\begin{corollary}[Two-miners case]\label{cor.two-miners-ineq}
If $n=2$, then the individual efficiency of $m_i$ is $\frac{p_i/\btau}{\hti_i}$, where \\[-1mm]
\begin{equation*}
    p_1 = 1 - \frac{h_2}{h_1+h_2} e^{-\hti_1 (\lti_2-\lti_1)}, \quad \textrm{ and } \quad p_2 = 1 - p_1.
\end{equation*}
\end{corollary}
\begin{proof}
This directly follows by applying Thm.~\ref{thm.individual-efficiency-C} with $n=2$. 
\end{proof}

We observe in Cor.~\ref{cor.two-miners-ineq} that $p_1 > h_1$ and $p_2 < h_2$ as soon as $l_2>l_1$. This observation implies that, unless the two miners are exactly equidistant from the coordinator, the miner which is farther from the coordinator will have a lower individual efficiency than the other---the former will receive less than its ``fair share'' of blocks in the longest chain.
This result holds \emph{irrespective of the compute capacity of the two miners}.
Extending that idea, we show next that if $n\ge 2$, for any pair of miners that are equidistant from the coordinator the individual miners' efficiencies are identical, regardless of their and of other miners' compute capacities in the system.
\begin{corollary}[Equidistant miners]
\label{cor.equidistant-individual}
Two miners which are equidistant from the coordinator have the same individual efficiency, irrespective of their computational capacity. Formally, suppose that two miners $m_i$ and $m_{i+1}$ are equidistant from the coordinator, i.e., $l_i = l_{i+1}$. Then $\eff^C_i(\hb, \lb, \tau) = \eff^C_{i+1}(\hb, \lb, \tau)$. 
\end{corollary}
\begin{proof}
Since the factor multiplying the effective capacity is the same for $i$ and
$i+1$ in \eqref{eq.pi-C} when $l_i = l_{i+1}$, the individual efficiencies are
identical, even if $\hti_i$ and $\hti_{i+1}$ are
different.
%
%
\end{proof}

The observation in Cor.~\ref{cor.equidistant-individual} trivially extends to the case with more than two miners equidistant from the coordinator. In particular, if all miners are equidistant from the coordinator, then they all have the same (individual) efficiency. This result can also be directly observed from Thm.~\ref{thm.individual-efficiency-C}: if all latencies are equal, then $p_i = h_i$ for all $i$. This result also highlights an equivalence: All miners have the same (individual) efficiency (i.e., get their ``fair share'') if and only if they are equidistant from the coordinator.



The above result states that all miners \stress{must} be equidistant from the coordinator to guarantee fairness. In practice, it is always possible to achieve such a configuration in the coordinated setting, potentially via addition of artificial delays where needed. Now, to optimize the overall efficiency under a no-inequality (or ``fairness'') constraint, it is enough to choose a configuration that minimizes the delay from each miner to the coordinator.

\begin{figure*}[t]
  \centering
  \begin{subfigure}[b]{\threecolgrid}
    \includegraphics[width={\textwidth}]{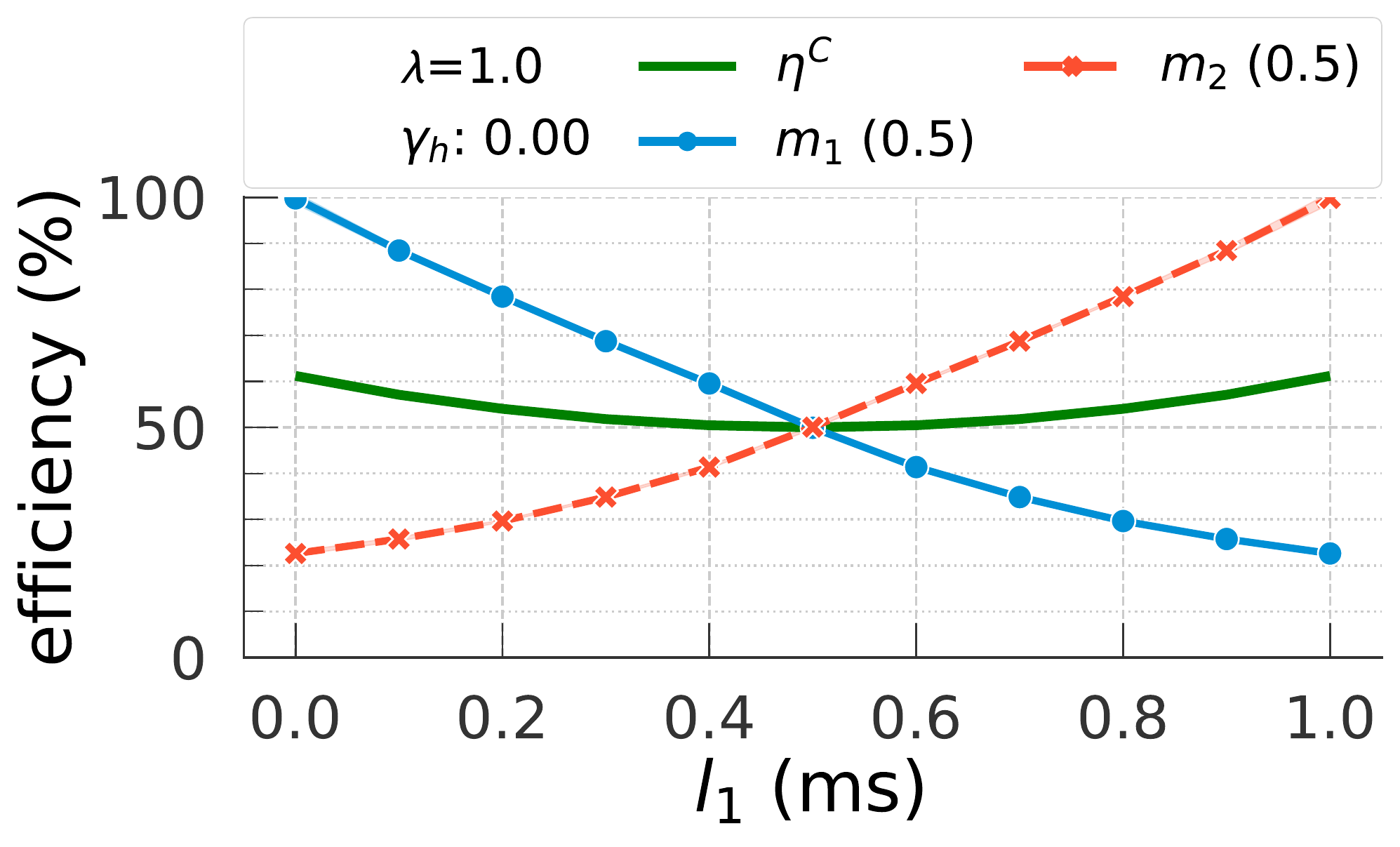}
    \sfigcap{$\gamma_h=0$}\label{fig:coordinated_equalityG0.0}
  \end{subfigure}
  \begin{subfigure}[b]{\threecolgrid}
    \includegraphics[width={\textwidth}]{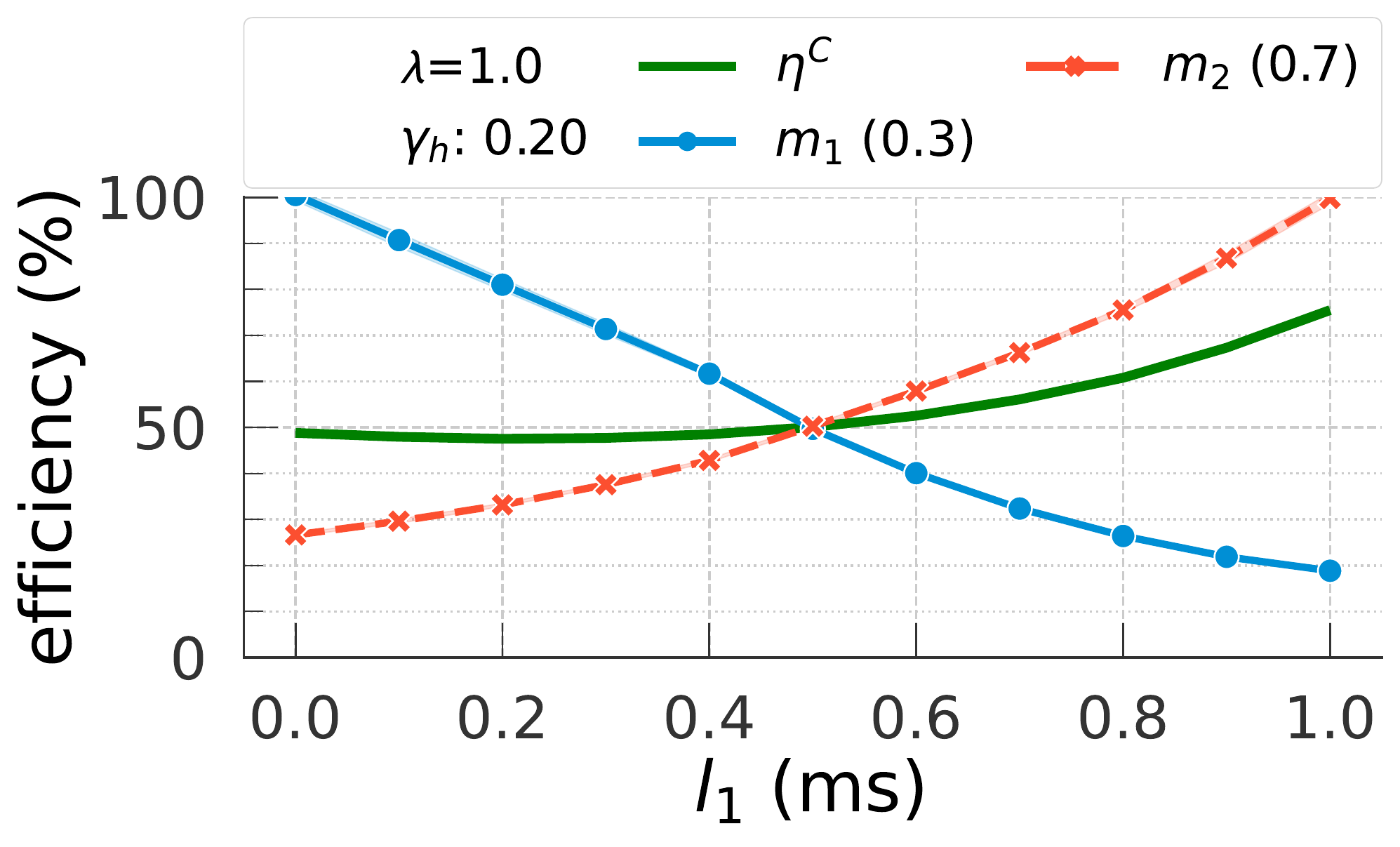}
    \sfigcap{$\gamma_h=0.2$}\label{fig:coordinated_equalityG0.2}
  \end{subfigure}
  \begin{subfigure}[b]{\threecolgrid}
    \includegraphics[width={\textwidth}]{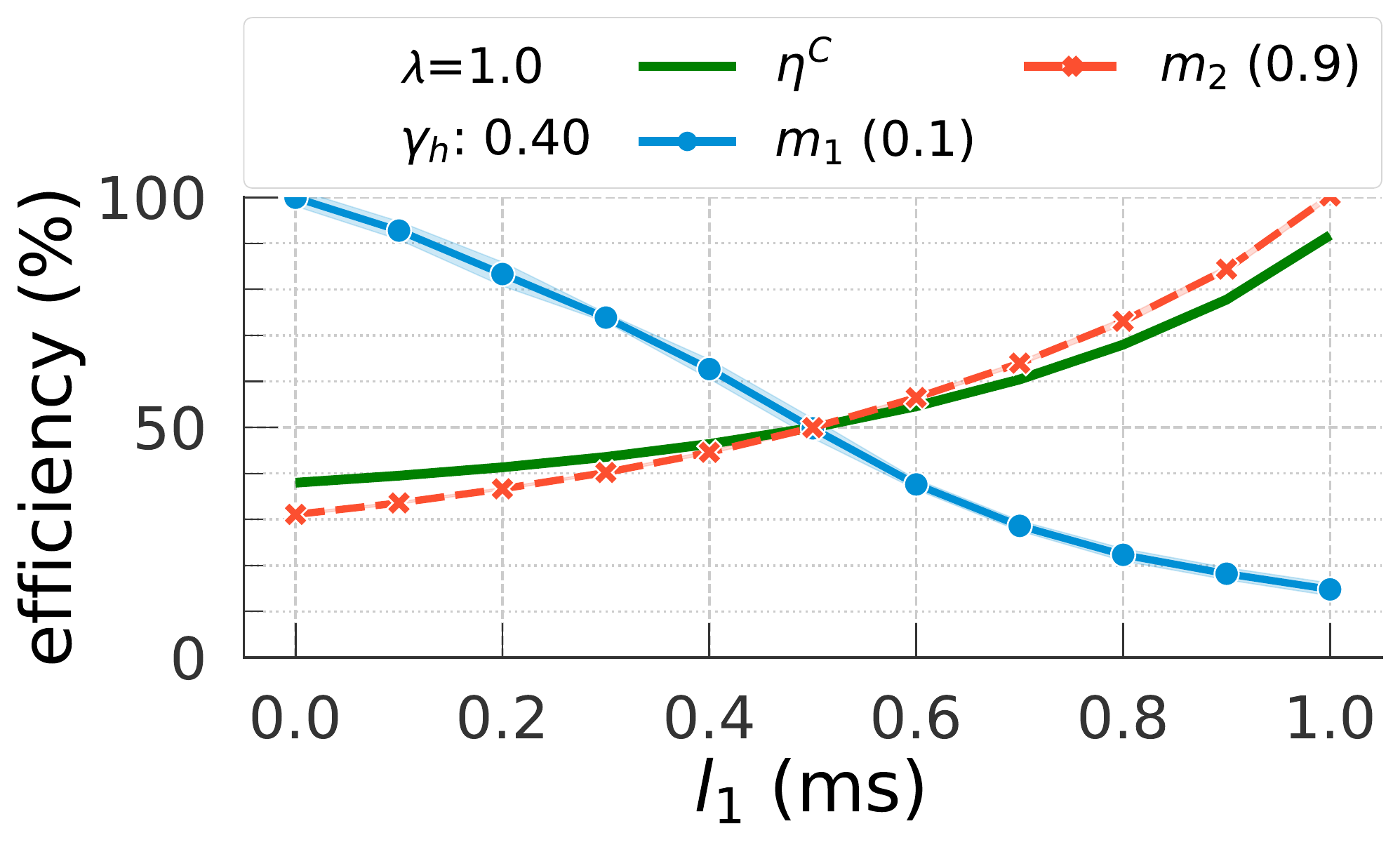}
    \sfigcap{$\gamma_h=0.4$}\label{fig:coordinated_equalityG0.4}
  \end{subfigure}
  \figcap{(Coordinated setting.) Effect of the coordinator's position on the individual efficiencies of miners in the two-miners scenario of \S\ref{ss:scenarios}. The distribution of compute capacity is (a) equal, (b) a 30-70\%, and (c) a 10-90\%. }\label{fig:coord-eq-2m}
\end{figure*}

\begin{figure*}[t]
  \centering
  \vspace{-3mm}
  \begin{subfigure}[b]{\threecolgrid} \includegraphics[width={\textwidth}]{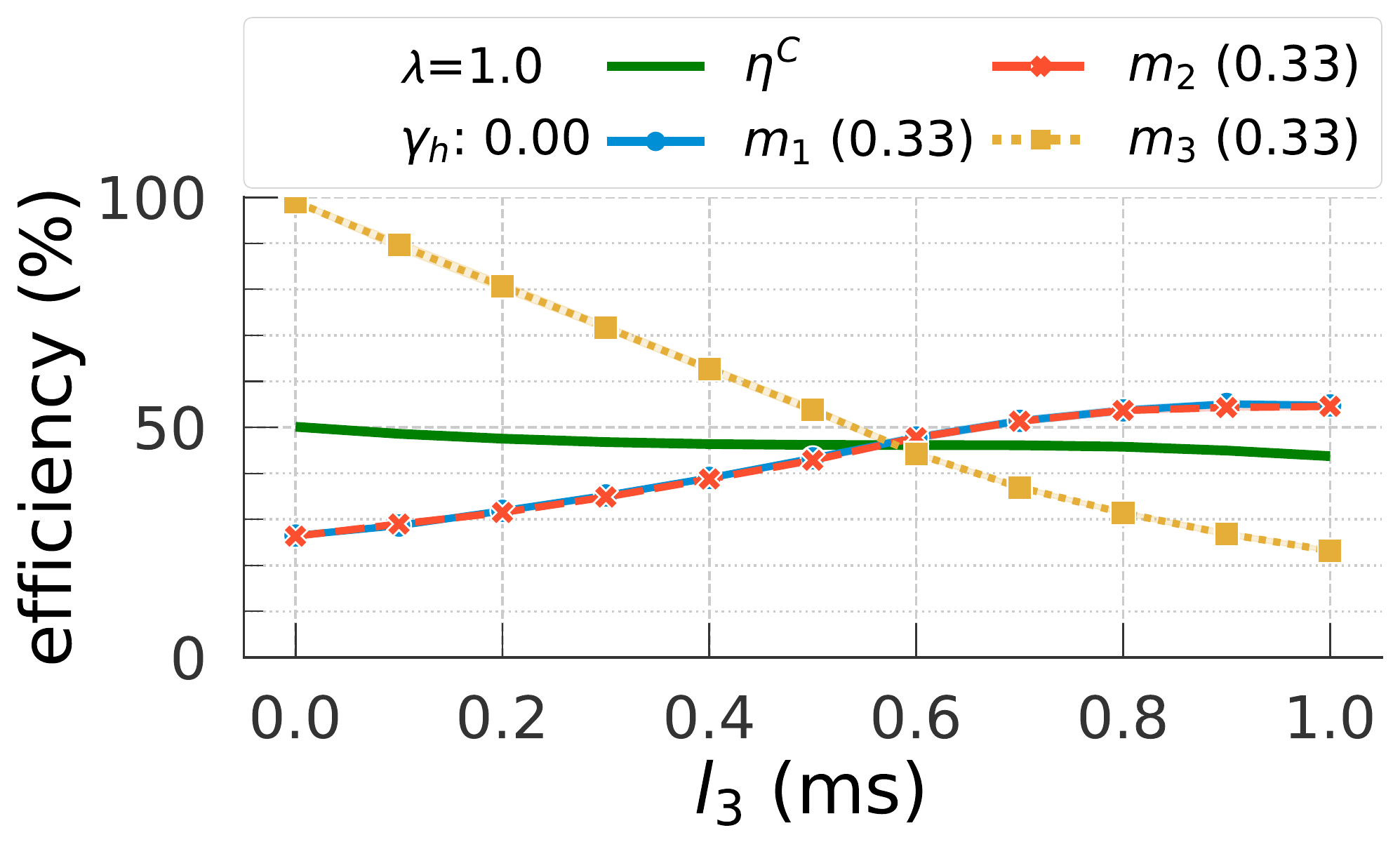}
    \sfigcap{$\gamma_h=0$}\label{fig:3coordinated_equalityG0.0}
  \end{subfigure}
  \begin{subfigure}[b]{\threecolgrid} \includegraphics[width={\textwidth}]{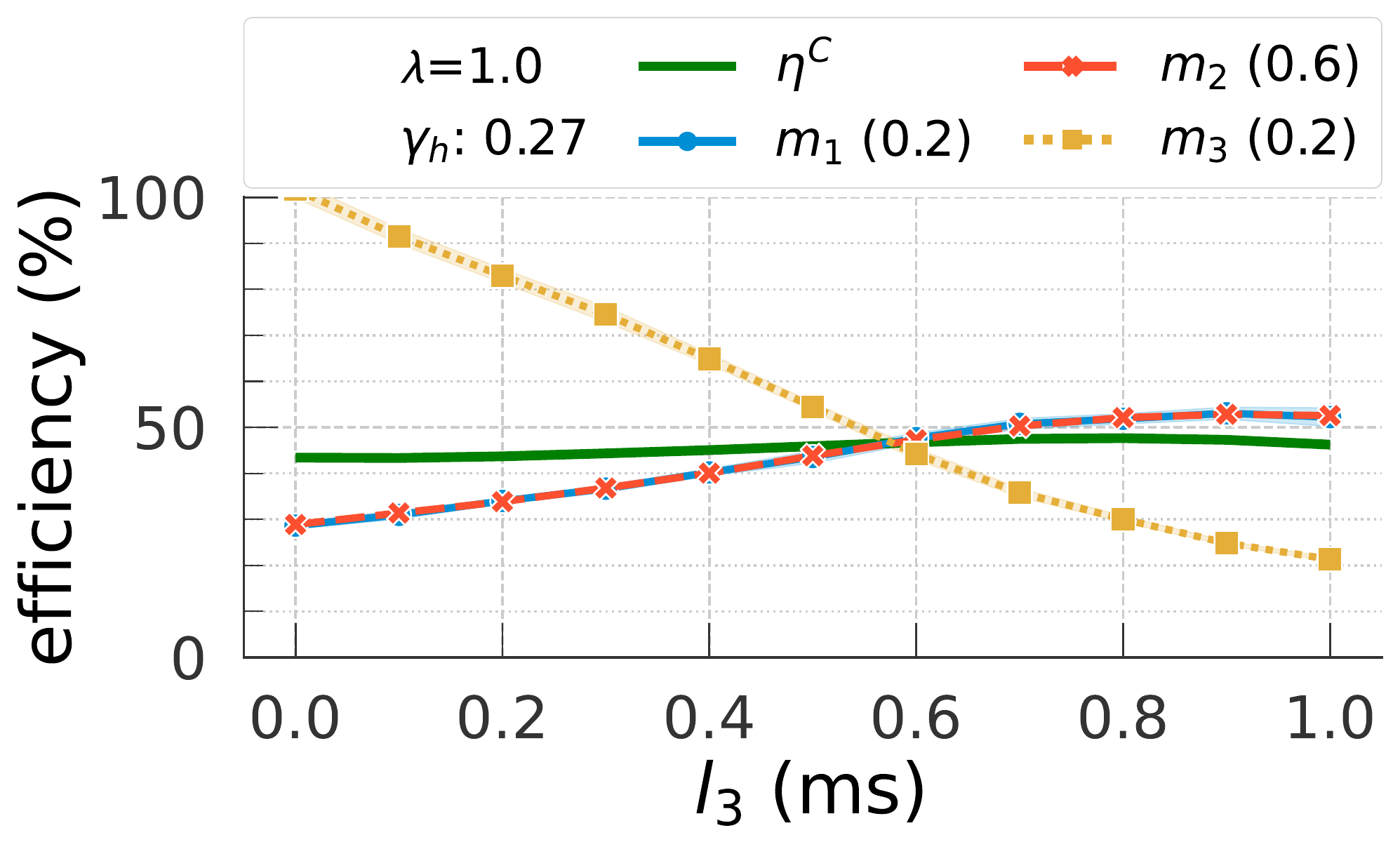}
    \sfigcap{$\gamma_h=0.27$}\label{fig:3coordinated_equalityG0.27}
  \end{subfigure}
  \begin{subfigure}[b]{\threecolgrid} \includegraphics[width={\textwidth}]{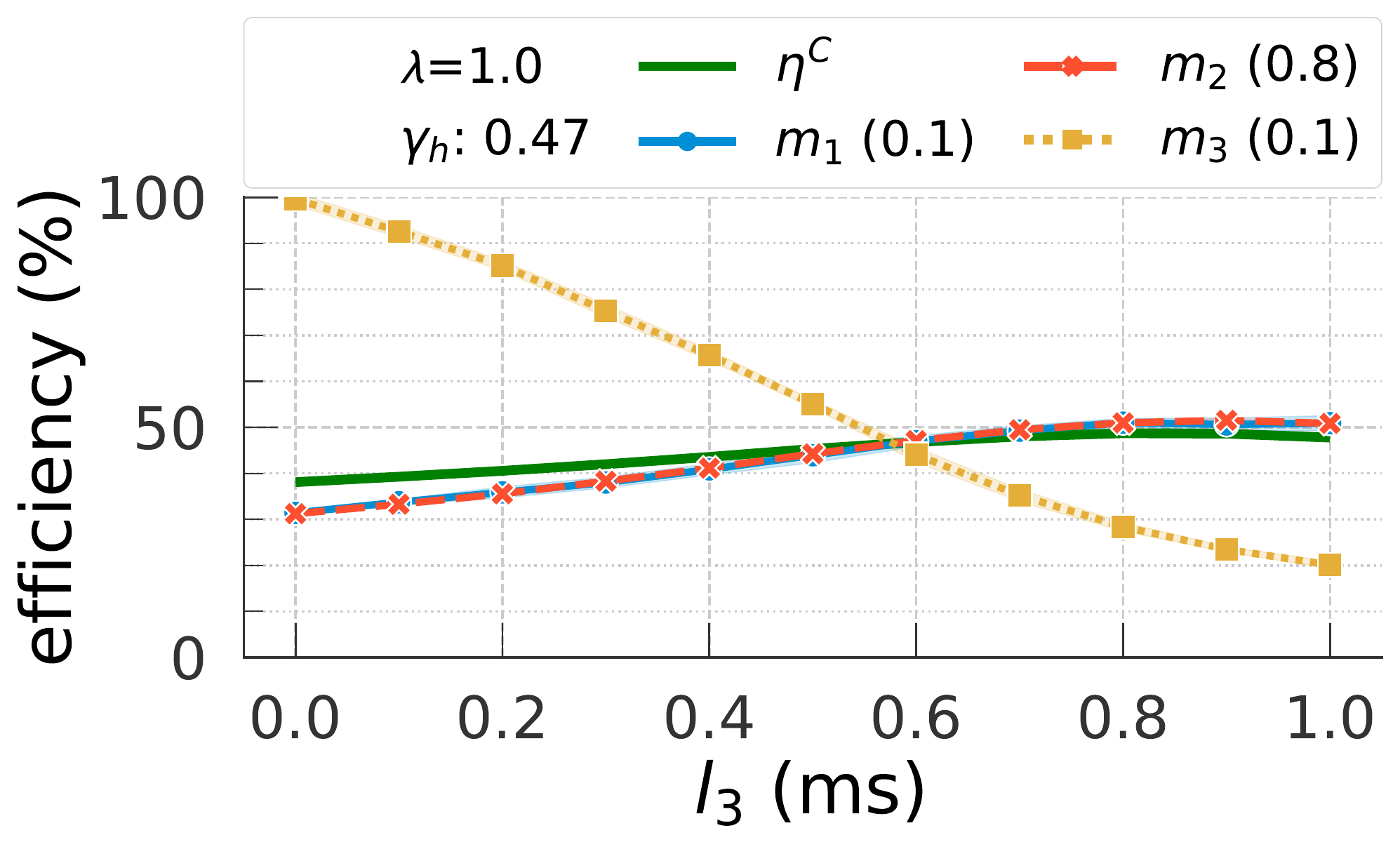}
    \sfigcap{$\gamma_h=0.47$}\label{fig:3coordinated_equalityG0.47}
  \end{subfigure}
  \figcap{(Coordinated setting.) Effect of the coordinator's position on the individual efficiencies in the three-miners scenario of \S\ref{ss:scenarios}. The distribution of compute capacity is (a) equal, (b) a 20-60-20\%, and (c) a 10-80-10\%.}
  \label{fig:coord-eq-3m}
\end{figure*}

To illustrate these theoretical results, we performed simulations of the two-miner and three-miner scenarios with various distributions of compute capacity. The results are presented in Fig.~\ref{fig:coord-eq-2m} and \ref{fig:coord-eq-3m}. We vary the compute capacities of the miners in both scenarios and succinctly annotate each plot with the corresponding $\gamma_h$ value. We observe that, when the coordinator is equidistant from the miners, the individual efficiencies of the miners are identical.
%


%
%

%% file: sections/ineq/p2p.tex
\subsection{P2P Setting} \label{subsec:p2p-setting}
\begin{figure*}[tbp]
  \centering
  \vspace{-3mm}
  \begin{subfigure}[b]{\threecolgrid}
    \includegraphics[width={\textwidth}]{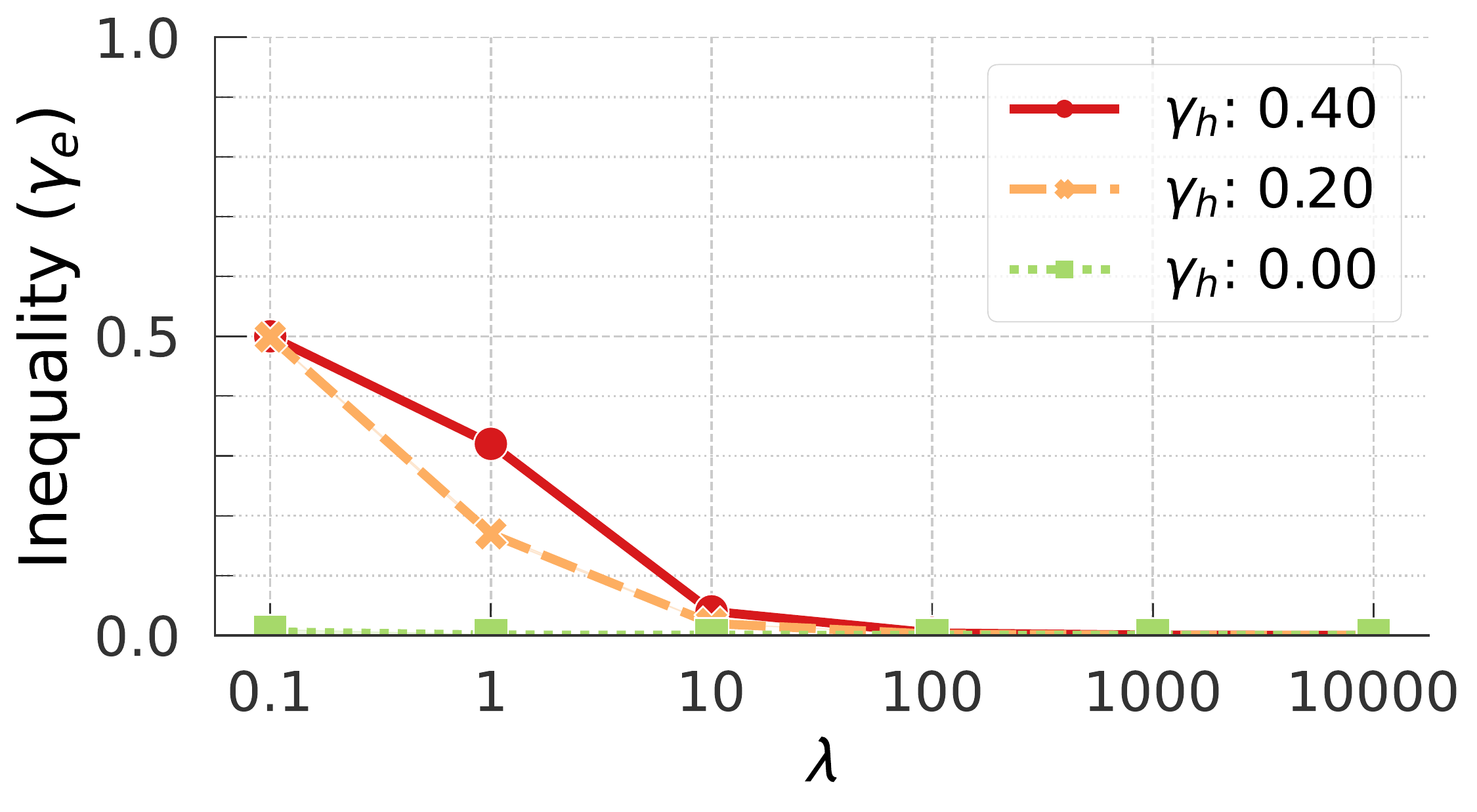}
    \sfigcap{Two miners}\label{fig:two-node-ineq}
  \end{subfigure}
  \begin{subfigure}[b]{\threecolgrid}
    \includegraphics[width={\textwidth}]{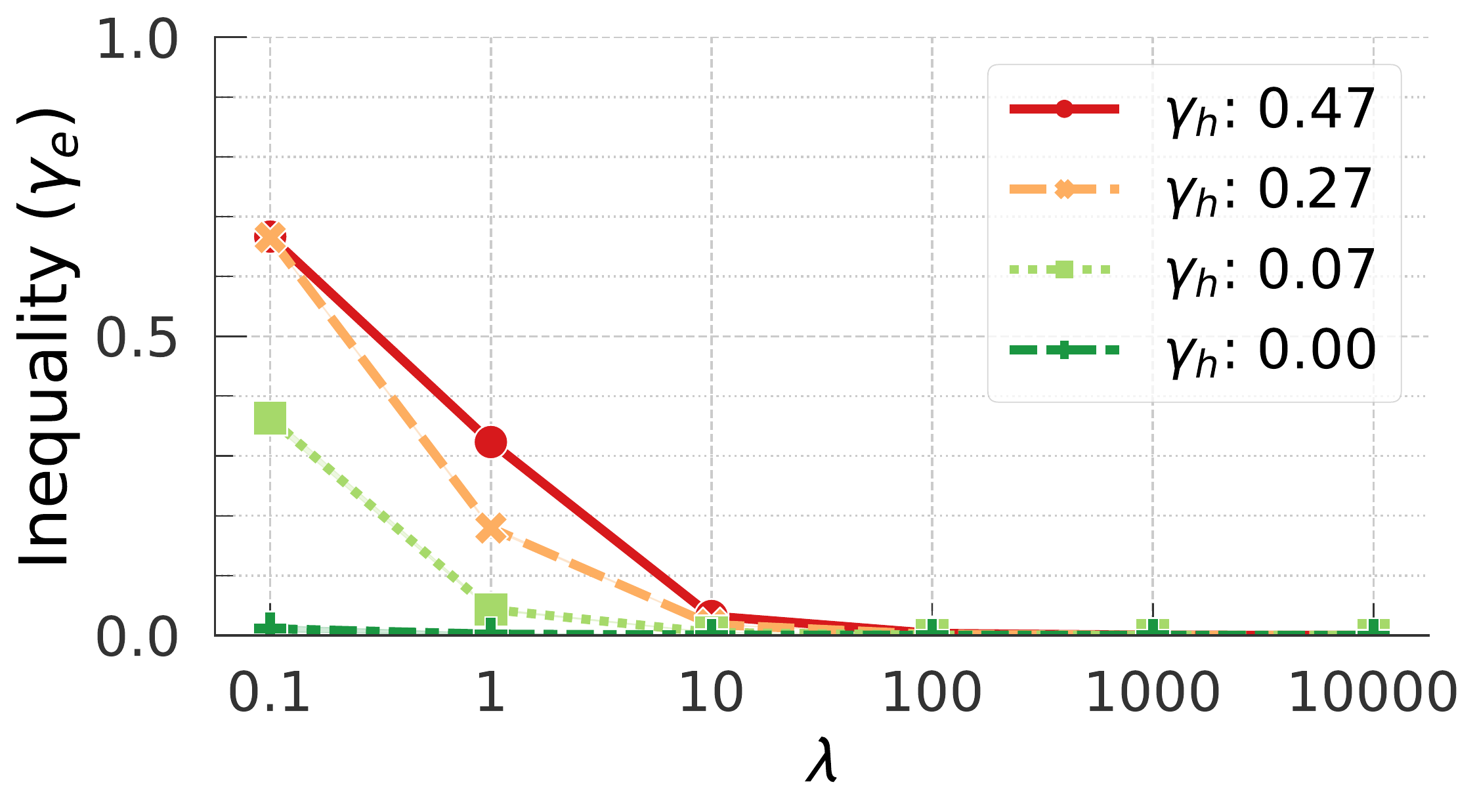}
    \sfigcap{Three miners}\label{fig:three-node-ineq}
  \end{subfigure}
  \begin{subfigure}[b]{\threecolgrid}
    \includegraphics[width={\textwidth}]{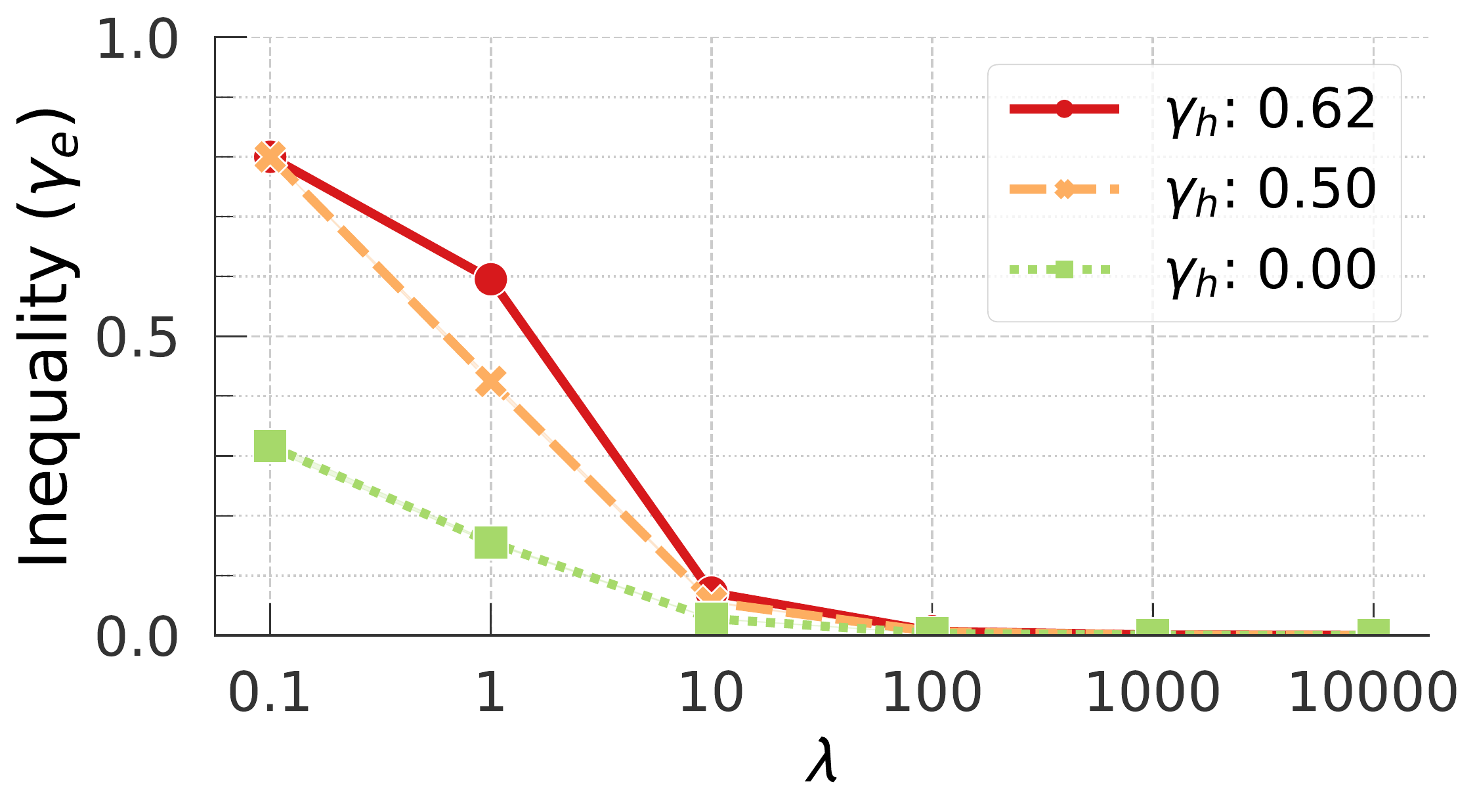}
    \sfigcap{Bitcoin approximation}\label{fig:multi-node-ineq}
  \end{subfigure}
  \figcap{(P2P Setting.) Inequality ($\gamma_e$) in efficiency of miners as a function of $\lambda$ in the scenarios of \S\ref{ss:scenarios} under various configurations of compute capacity.}\label{fig:ineq-in-eff-p2p}
\end{figure*}

\begin{figure*}[tbp]
  \centering
 \vspace{-3mm}
  \begin{subfigure}[b]{\threecolgrid}
    \includegraphics[width={\textwidth}]{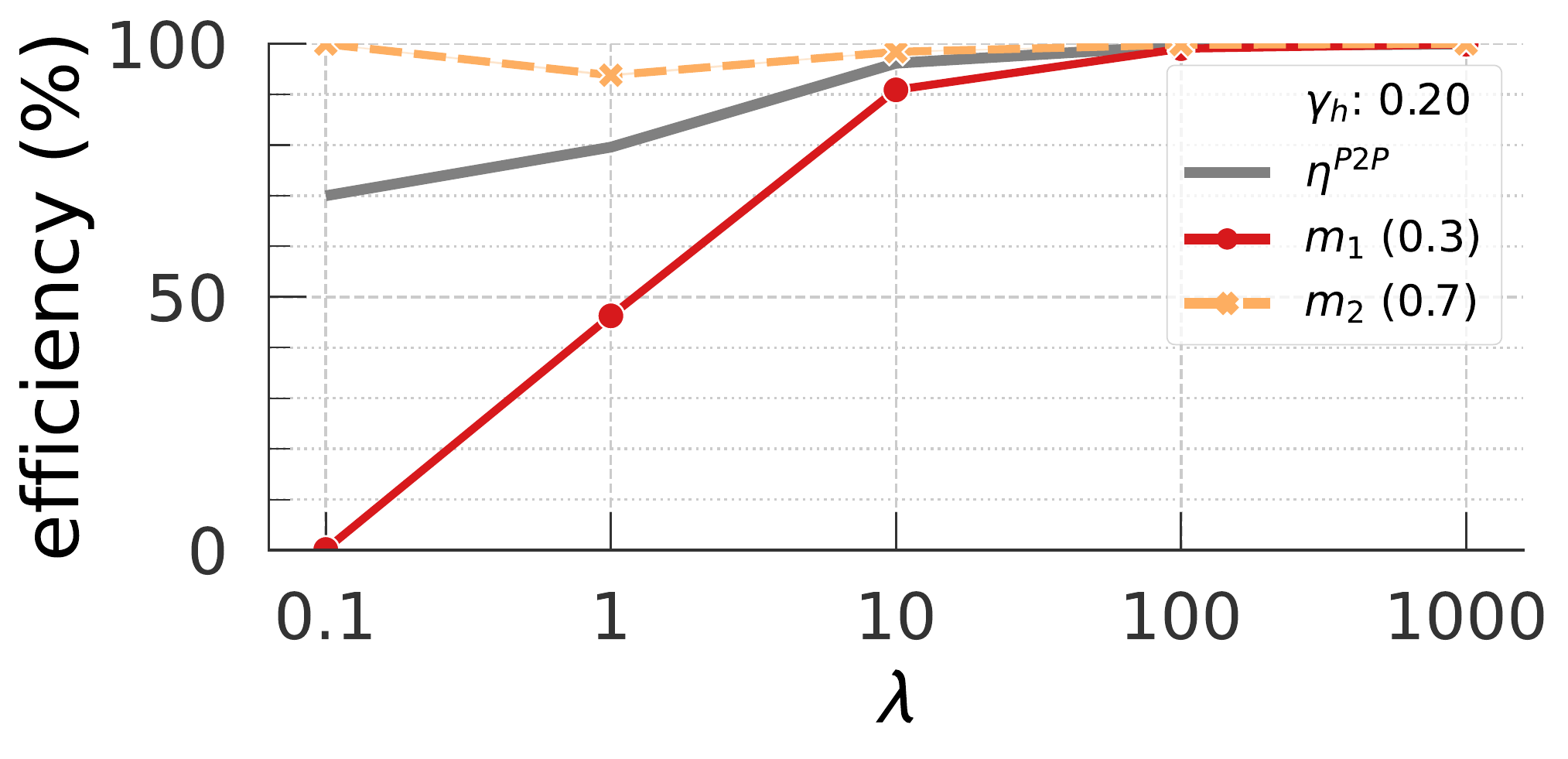}
    \sfigcap{Two miners; $\gamma_h=0.20$}\label{fig:ineq-2m-ind-eff}
  \end{subfigure}
  \begin{subfigure}[b]{\threecolgrid}
    \includegraphics[width={\textwidth}]{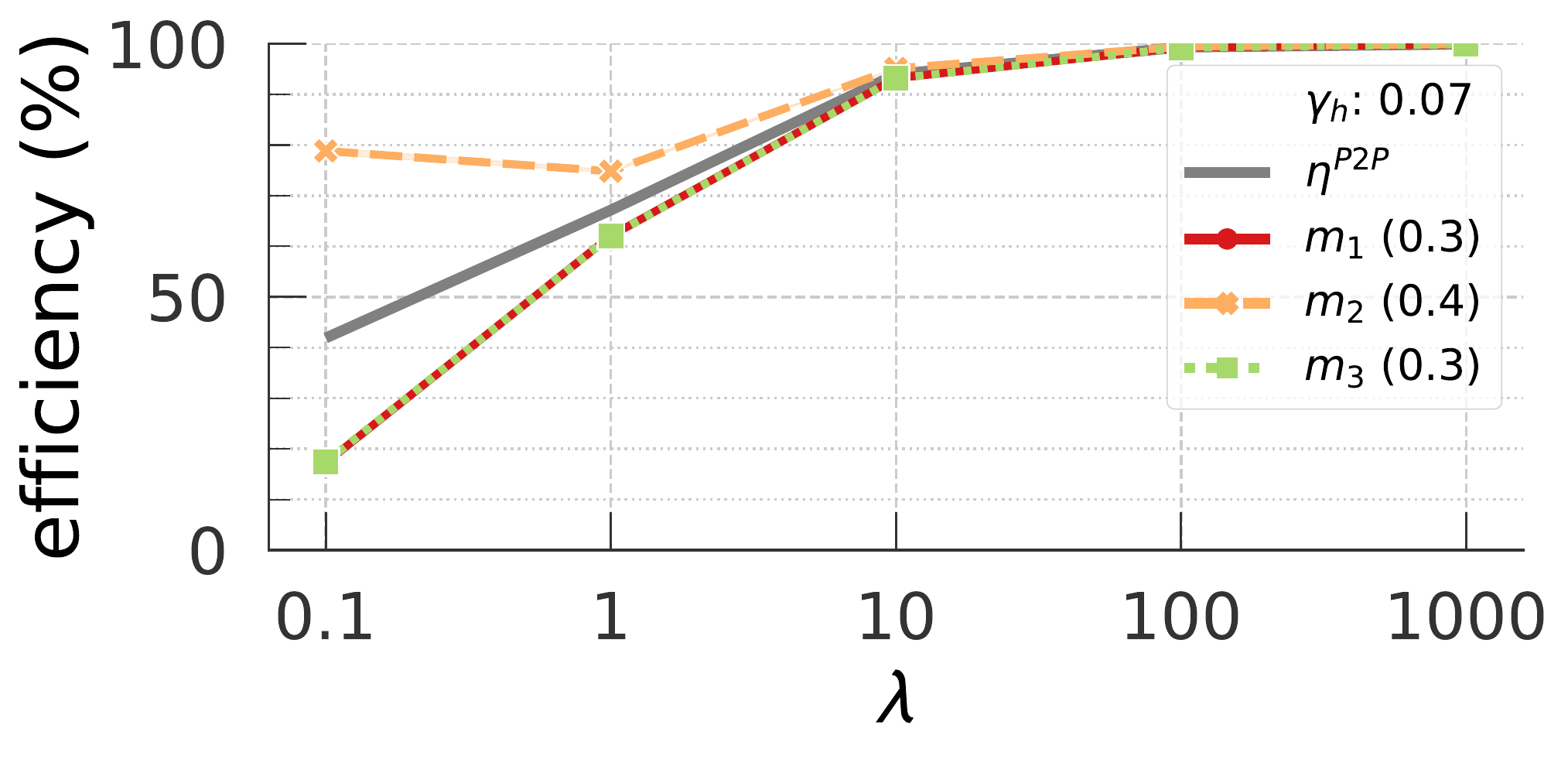}
    \sfigcap{Three miners; $\gamma_h=0.07$}\label{fig:ineq-3m-ind-eff}
  \end{subfigure}
  \begin{subfigure}[b]{\threecolgrid}
 	\includegraphics[width={\textwidth}]{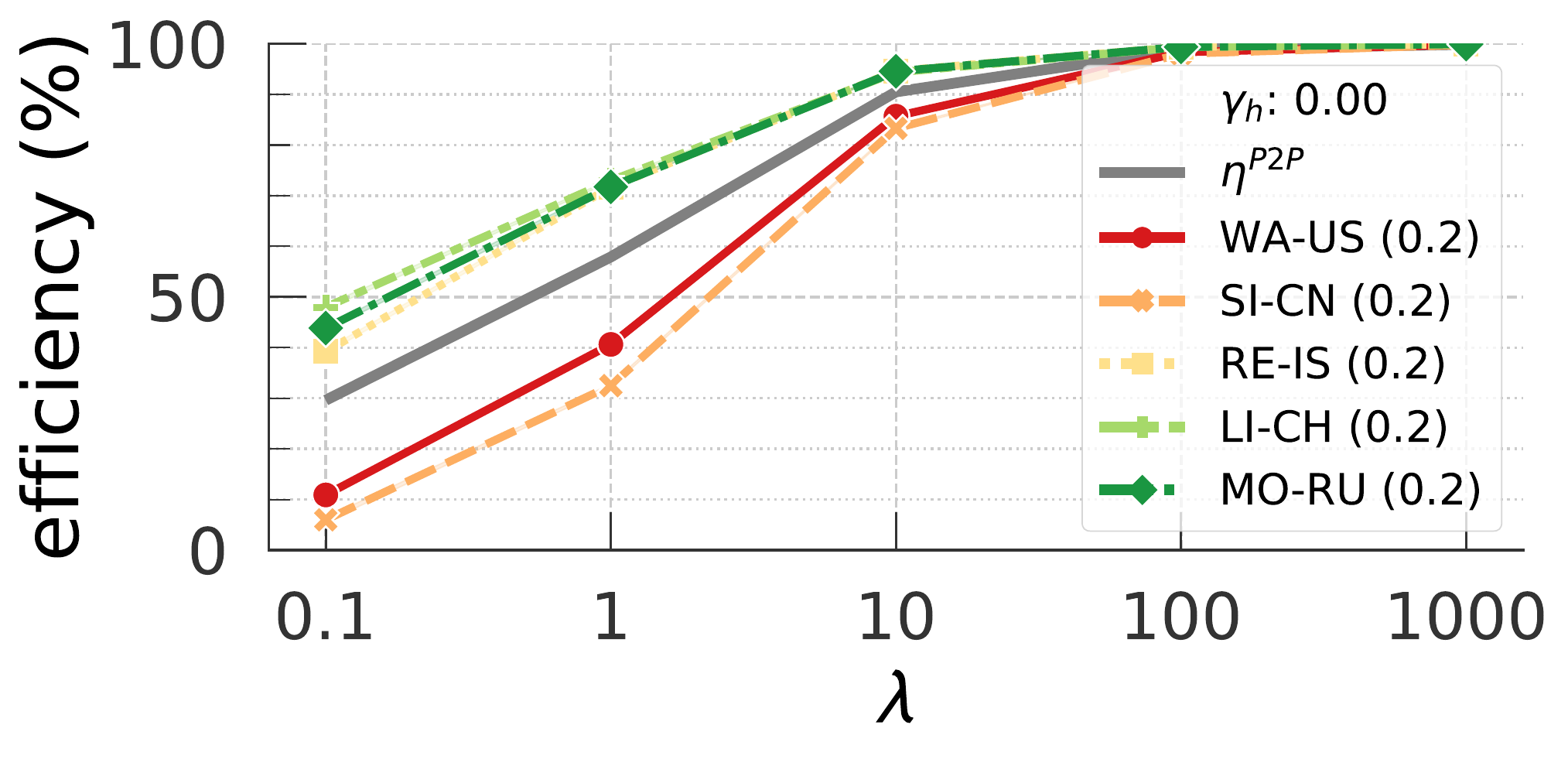}
 	\sfigcap{Bitcoin approx.; $\gamma_h=0.00$}\label{fig:ineq-multi-ind-eff}
  \end{subfigure}
  \figcap{(P2P Setting.) Individual efficiency of miners as a function of $\lambda$ in the scenarios of \S\ref{ss:scenarios} for selected configurations of compute capacity. Each plot corresponds to a single line from the corresponding plot in Fig.~\ref{fig:ineq-in-eff-p2p}.}\label{fig:ineq-p2p-ind-eff}
\end{figure*}

As the computation of individual efficiency in the P2P setting does not lend itself to a theoretical analysis, we rely on simulations to analyze it. To first observe the presence of inequality, we look at the Gini index $\gamma_e$ of individual efficiencies in the scenarios of \S\ref{ss:scenarios}. The results are presented in Fig.~\ref{fig:ineq-in-eff-p2p}. We observe that when the compute capacities of miners are identical ($\gamma_h=0$), the effect of latency on the individual efficiency of miners is also uniform. When $\gamma_h=0$, there is no inequality in efficiency ($\gamma_e$) across miners in the two-miner (Fig.~\ref{fig:two-node-ineq}) and three-miner (Fig.~\ref{fig:three-node-ineq}) scenarios.
The figures, however, also make it readily apparent that any variability in the compute capacity of miners is significantly exaggerated by latency: as $\gamma_h$ increases, $\gamma_e$ deviates from zero (i.e., miners do not experience a fair share of the overall system efficiency).
Fig.~\ref{fig:ineq-2m-ind-eff} and Fig.~\ref{fig:ineq-3m-ind-eff} plot the individual efficiency of miners, instead of the inequality, to emphasize the unfairness in efficiency affected by the network latencies between miners; for clarity, we restrict these plots to only one compute-capacity distribution.
When $\lambda<10$ network latencies have a measurable impact on the individual
efficiency of miners, albeit weak miners suffer a higher drop in efficiency
compared to strong miners---in the extreme case, as $\lambda$ tends towards zero, it appears that only the most powerful miner obtains an efficiency of one; the efficiency of all others tends to reduce to zero.
In the two-miner and three-miner scenarios above, the miners were equidistant from one another. Yet we observed inequality in the efficiency. This observation brings an interesting contrast between the P2P and coordinated setting: Whereas the inequality in efficiency of miners in the coordinated setting depends only on the variability of delay from the coordinator to the different miners, in the P2P setting it is also dictated by the compute capacities of miners.

To explore cases with variability in the latencies in the P2P setting, Fig.~\ref{fig:multi-node-ineq} examines the impact of latency on $\gamma_e$ in the Bitcoin-approximation scenario, where the miners are not equidistant from one another but potentially reflect a realistic deployment (cf.~\S\ref{ss:scenarios}).
Even if the miners were all assigned the same compute-capacity, the differences in pairwise latencies between them results in an unequal, unfair distribution of efficiency across the miners (as shown in Fig.~\ref{fig:ineq-multi-ind-eff}).
These observations offer a new rationale for the use of high $\lambda$ values in practice (e.g., in Bitcoin and Ethereum): A high value of $\lambda$ helps ameliorate the inequality in efficiency across miners.
They also reveal that a~change in the operating parameters, e.g., use of a lower value of $\tau$ for improving the transaction throughput, has significant implications for fairness in efficiency.

\begin{figure}[tbp]
  \centering
  \begin{subfigure}[b]{1\threecolgrid}
	\includegraphics[width={\textwidth}]{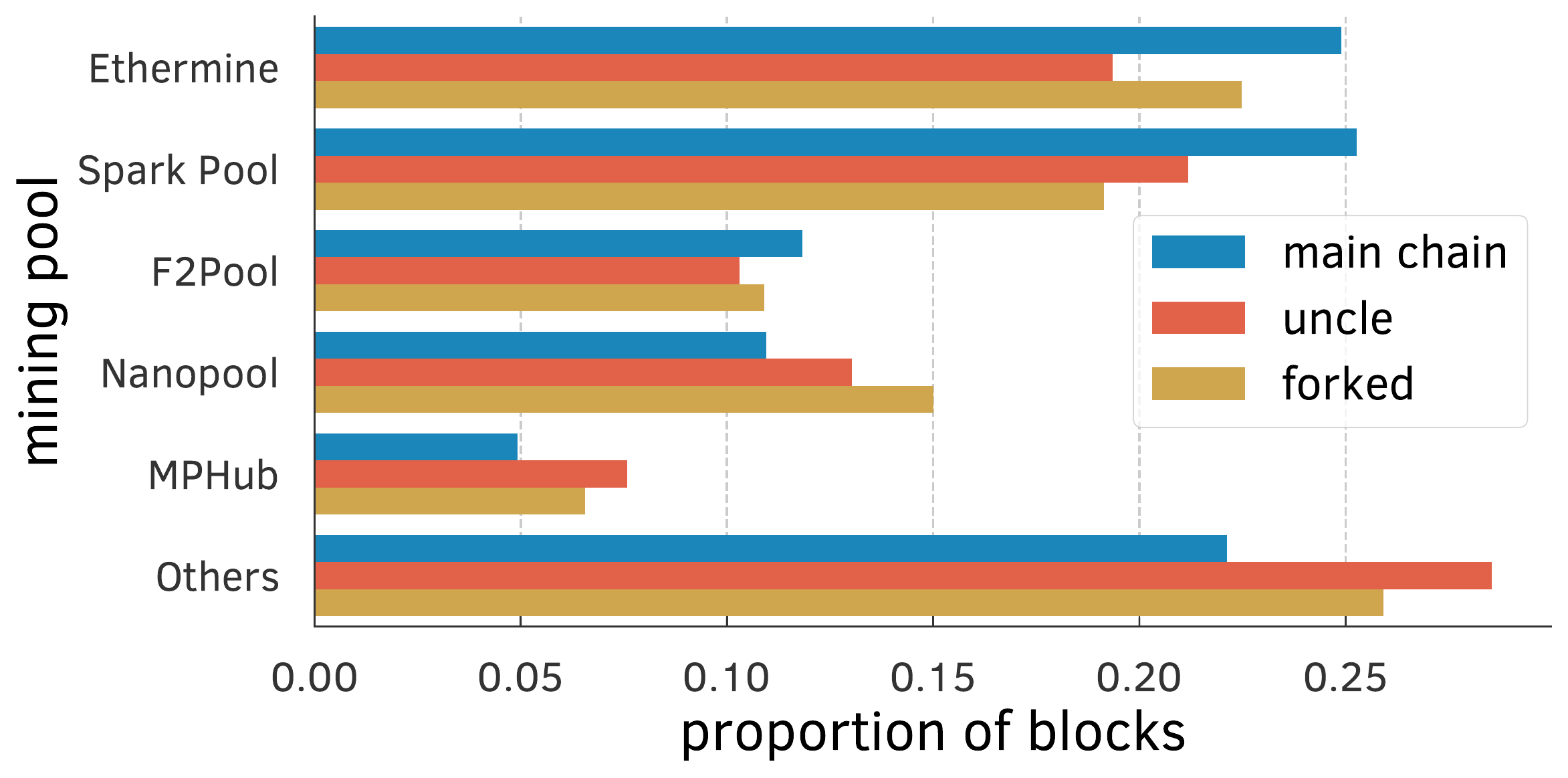}
	\sfigcap{}\label{fig:ethereum-bar}
  \end{subfigure}
  \begin{subfigure}[b]{1\threecolgrid}
    \includegraphics[width={\textwidth}]{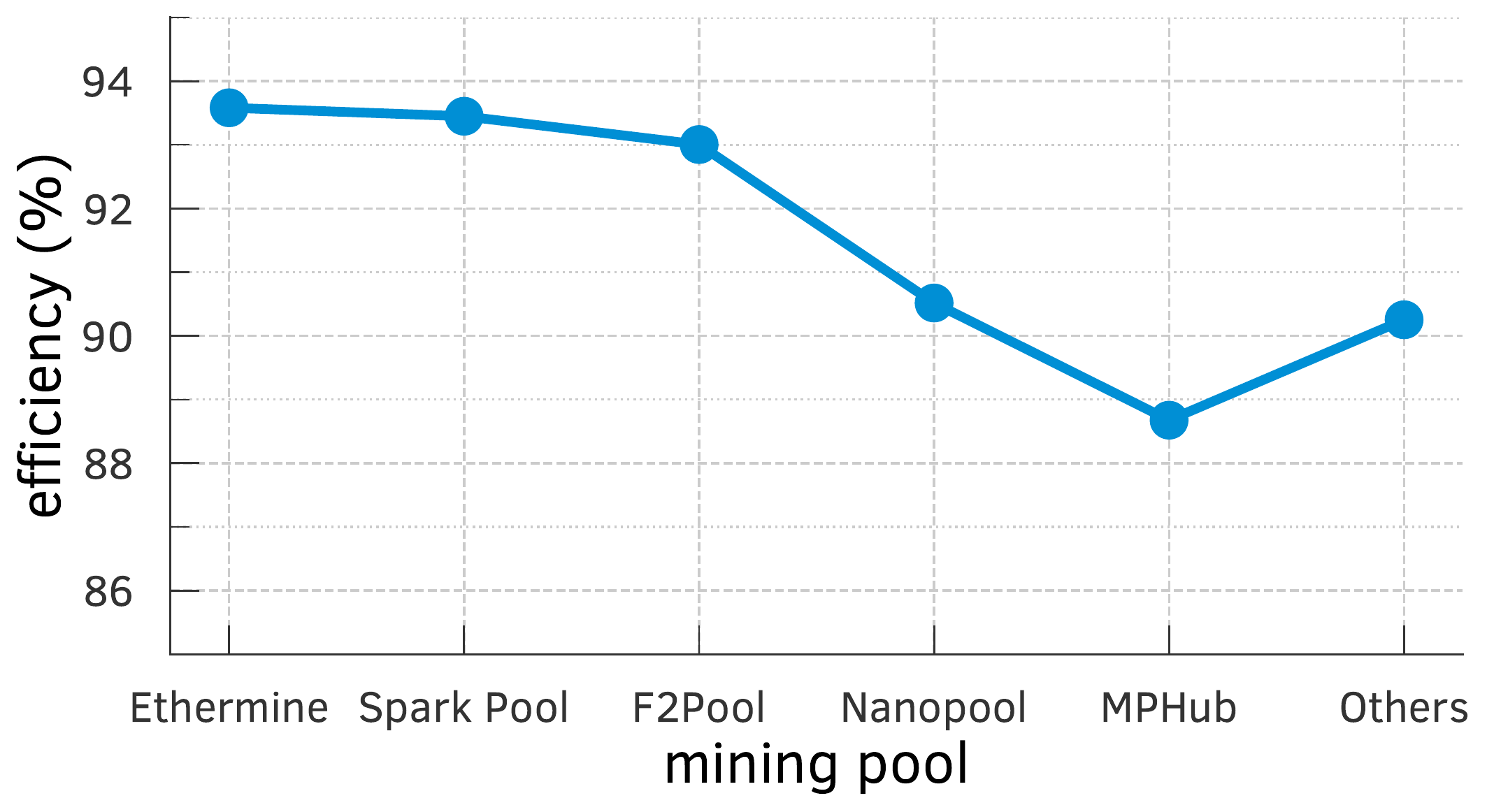}
    \sfigcap{}\label{fig:ethereum-eff}
  \end{subfigure}
  \figcap{Empirical evaluation of efficiency in Ethereum mining pools: (a) Forked and uncled blocks of each pool as a fraction of blocks contributed by it to the longest chain; (b) Estimated efficiency of the different mining pools. 
}
\end{figure}

To supplement the simulation results with empirical observations, we analyzed
the Ethereum blockchain data for the entire year of 2019. 
Specifically, we retrieved all the blocks in the longest (i.e., main) chain as
well as blocks that were ``uncled'' or ``forked'' from Etherscan~\cite{etherscan}.\footnote{In Ethereum, when two blocks are mined roughly at the same time only one will eventually be part of the main chain; but the other still has an opportunity to get included as an uncle---or ommer, see \S4.2 in~\cite{wood2014ethereum}---block (with a reduced reward) that contributes to the security of the main chain. If a block was not included as an uncle, it will become a forked block.} 
%
This data set comprises $\num{2204109}$ blocks in the main chain, $\num{148530}$ uncled blocks, and $\num{39580}$
forked blocks.
We estimate the compute capacities of Ethereum mining pools based on the fraction of blocks contributed by each to the main chain. 
%
Per Fig.~\ref{fig:ethereum-bar} there is a significant variance in compute
capacities of mining pools.
%
%
Given this estimate of compute capacities of miners, we simply ask whether the
loss in contributions of the miners is proportional to their relative share of
compute capacities.
The loss in a miner's contribution of blocks to the main chain is captured by
the count of uncled and forked blocks of that miner.
Fig.~\ref{fig:ethereum-bar} shows that the miners with high compute
capacities experienced proportionally less ``uncling'' and ``forking'' of their blocks compared to the miners with weak compute capacities. Fig.~\ref{fig:ethereum-eff} depicts the efficiency of each mining pool computed as the fraction of the blocks contributed to the main chain relative to the total number of mined blocks (main chain, uncles, and forked blocks) per each mining pool, and it confirms this observation. 

%


To conclude this section, we investigate how the individual efficiency is affected by the position of a miner in the P2P network. We focus here on the case where all compute capacities are the same and use the closeness centrality measure \cite{closenessCentrality,closenessCentrality-impl} to capture how ``well-connected'' a miner is. To explore the relationship between centrality and individual efficiency of a miner, we consider a hypothetical scenario where we place a miner in each of the 240 capital cities in the world. We use an optimistic measure of the latency by assuming all miners are connected with a high speed optical fiber connection. As before, we simulate the blockchain in this P2P setting up to a longest chain of $\num{100000}$ blocks, we repeat the simulation 10 times and take the average individual efficiency for each miner. We then calculate two measures of correlation between the centrality and the individual efficiency vectors: the Pearson and the Spearman correlation coefficients. The Pearson coefficient measures the correlation between the \emph{values} of centrality and individual efficiency (it is the standard correlation coefficient), whereas the Spearman coefficient measures the correlation between the \emph{ranks} of miners according to those two metrics. The results are shown in Tab.~\ref{tab:centrality}. Focusing first on the Pearson correlation, we observe that it is almost zero for large $\lambda$. This result might be due to the fact that at such small delays all miners have efficiency close to one. When $\lambda$ decreases to one, the Pearson correlation coefficient becomes very close to one, meaning that individual efficiency here is essentially proportional to the centrality. When $\lambda$ gets smaller, the Pearson correlation coefficient decreases again. Interestingly though, when looking at the Spearman correlation coefficient, we observe the same trend for $\lambda$ going from large values to one, but we do not observe the decrease for smaller values of $\lambda$ (or at least not as marked as for the Pearson coefficient). This result implies that, for small values of $\lambda$, while the values of centrality and individual efficiency become less correlated, their ranks remain highly correlated. From a deeper investigation, we observed that the result is due to the fact that, as $\lambda$ gets smaller, the curve of individual efficiency as a function of centrality gets highly superlinear: only the most central miners get non-negligible individual efficiency, the others get individual efficiency close to zero. This observation also explains, at least in part, the small decrease of the Spearman coefficient for $\lambda = 0.01$: as many miners have individual efficiency close to zero (and hence close to each other), estimating the rank correlation becomes difficult. To confirm that, we performed simulations with only $15$ cities (instead of $240$ so as to have more reasonable simulation times), varying the length of the simulation. The results are shown in Tab.~\ref{tab:centrality-runs}. We observe indeed that with longer simulations, the decrease in the Spearman correlation coefficient for very small $\lambda$ is much less pronounced. 

\begin{table}
\tabcap{Correlation between the closeness centrality measure and individual miners' efficiencies in a hypothetical setting. In (a), we do simulations of 100K blocks; in (b) we vary the length from 10K to 10M blocks.}
\parbox{.2\textwidth}{
  \vspace{-3pt}
  \subfloat[$240$ cities]{
  \label{tab:centrality}
  \centering
  \resizebox{0.2\textwidth}{!}{
  \begin{tabular}{rcc}
    \toprule
    \multicolumn{1}{c}{\multirow{2}{*}{$\lambda$}} & \textit{Pearson} & \textit{Spearman} \\
    \multicolumn{1}{c}{~} &
    100K & 100K \\
    \midrule
    0.01    & 0.587   & 0.801 \\
    0.1     & 0.862   & 0.975 \\
    1.0     & 0.995   & 0.992 \\
    10.0    & 0.934   & 0.914 \\
    100.0   & 0.308   & 0.296 \\
    1000.0  & 0.062   & 0.077 \\
    10000.0 & 0.060   & 0.055 \\
    \bottomrule
  \end{tabular}
  }
  }
}
\hfill
\parbox{.45\textwidth}{
  \vspace{4pt}
  \subfloat[$15$ cities]{
  \label{tab:centrality-runs}
  \centering
  \resizebox{0.45\textwidth}{!}{
  \begin{tabular}{r|cccc|cccc}
    \toprule
    \multicolumn{1}{c}{\multirow{2}{*}{$\lambda$}} & \multicolumn{4}{c}{\textit{Pearson}} & \multicolumn{4}{c}{\textit{Spearman}} \\
    \multicolumn{1}{c}{~} &
    10K & 100K & 1M & 10M &
    10K & 100K & 1M & 10M \\
    \midrule
0.01 & 0.484 & 0.484 & 0.484 & 0.484 & 0.862 & 0.924 & 0.948 & 0.924 \\
0.1 & 0.791 & 0.791 & 0.792 & 0.792 & 0.967 & 0.96 & 0.961 & 0.964 \\
1.0 & 0.997 & 0.998 & 0.998 & 0.998 & 0.968 & 0.975 & 0.975 & 0.975 \\
10.0 & 0.978 & 0.985 & 0.982 & 0.983 & 0.868 & 0.929 & 0.996 & 0.996 \\
100.0 & 0.644 & 0.92 & 0.954 & 0.975 & 0.643 & 0.839 & 0.857 & 0.921 \\
1000.0 & 0.158 & 0.503 & 0.415 & 0.82 & 0.161 & 0.361 & 0.236 & 0.679 \\
10000.0 & 0.096 & 0.31 & -0.165 & 0.094 & 0.154 & 0.243 & -0.161 & 0.021 \\
    \bottomrule
  \end{tabular}
  }
  }
}
\end{table}

%% file: sections/ineq/end.tex
\subsection{Takeaways and Implications}
We showed that in the P2P setting the distribution of compute capacities and relative locations of miners in the network, captured via the closeness centrality measure, have non-trivial consequences on the inequality in efficiency across miners.
In the coordinated setting, however, inequality in efficiency of miners is simply dictated by the latencies between the miners and the coordinator, and can be alleviated by making all miners equidistant from the coordinator.

%% file: sections/conclusion.tex
\vspace*{-0.3em}
\section{Conclusion} \label{s:conclusion}
\vspace*{-0.3em}
In this paper, we presented closed-form expressions for both the overall
system efficiency and the individual miner efficiency in the coordinated setting,
with an arbitrary number of miners and arbitrary delays.
We derived a lower bound for the overall system efficiency of the P2P setting
using that of the coordinated setting, showing that the former is at least as
high as the latter.

\parai{Hybrid-hierarchical model.}
Although we discussed the P2P and coordinated models as if they are alternatives
for how a blockchain could operate, these models coexist in a real-world system.
The mining pool operator in Bitcoin, for instance, operates both as a
coordinator within their mining pool (coordinated setting) and as a peer,
interacting with other miners and pools, outside of their pool (P2P setting).
We leave a precise modeling of the interactions between the two settings and investigation of the
implications of this hybrid model for efficiency to future work. Yet, we emphasize that our current results provide a number of interesting direct consequences. For instance, our coordinated model with a single coordinator is equivalent to a multi-coordinator setting with zero delay between the coordinators, hence it provides an upper bound on the efficiency of the hybrid model with delay between the coordinators. In the other extreme, clearly the hybrid model can be as efficient as the P2P model if each miner is its own coordinator.

\parai{New notions of efficiency.}
Our analyses focus on one measure of efficiency based on the
computational work performed by miners.
There may exist, however, other notions of efficiency, e.g., focusing on energy consumption.
The results may change under such a notion:
A miner may achieve less than its proportional share of blocks, but still obtain
a better share with respect to the energy consumed.
%
%
We hope our work motivates others to investigate other relevant efficiency
notions.

%% file: sections/acknowledgments.tex
\section*{Acknowledgments}

This research was supported in part by a European Research Council (ERC) Advanced Grant for the project ``Foundations for Fair Social Computing,'' funded under the European Union's Horizon 2020 Framework Programme (grant agreement no. 789373). 

This work has also been partially supported by MIAI @ Grenoble Alpes (ANR-19-P3IA-0003) and by the French National Research Agency through grant ANR-20-CE23-0007.

Finally, we would like to thank Antoine Kaufmann and Isaac Sheff from MPI-SWS for the fruitful discussions and useful feedback.

%% file: sections/app-table.tex
\section{Summary of notation}
\label{app.table}

\begin{table}[h!]
\small
\tabcap{Summary of notations along with their descriptions}\label{tab:notations}
\begin{tabular}{cl}
    \toprule
    \textbf{\textit{Notation}} & \textbf{\textit{Description}} \\
    \midrule
    $m_i$                & $i$-th miner\\
    $\Mset$              & set of miners $\{m_1, \cdots, m_n\}$\\
    $h_i$                & relative computational capacity of miner $m_i$\\
    $\hb$                & vector of relative computational capacities $[h_i]_{i=1, \cdots, n}$\\
    $\hti_i$             & effective computational capacity of miner $m_i$\\
    $\htib$              & vector of effective computational capacities $[\hti_i]_{i=1, \cdots, n}$\\
    $\tau$               & puzzle hardness\\
    $l_{ij}$             & latency between $m_i$ and $m_j$\\
    $L$                  & latency matrix $ [l_{ij}]_{i, j \in \{ 1, \cdots, n\}}$ (for P2P) \\
    $\lmean$             & mean latency $2/(n(n-1)) \cdot \sum_{i=1}^n\sum_{j>i} l_{ij}$\\
    $\lambda$            & hardness to latency ratio $\tau / \lmean$\\
    $l_i$                & latency between miner $m_i$ and $C$\\
    $\lb$                & latency vector $[l_i]_{i\in \{1, \cdots, n\}}$ (for coordinated)\\
    $\lti_i$             & round-trip time $2l_i$ between miner $m_i$ and $C$\\ 
    $\ltib$              & round-trip time vector $[\lti_i]_{i\in \{1, \cdots, n\}}$ (for coordinated)\\
    $T$                  & length of the time window\\
    $\eff_i$                & efficiency of miner $m_i$\\
    $\eff$                  & overall system efficiency\\
    $\gamma_e, \gamma_h$ & Gini index of efficiencies, capacities\\
    \bottomrule
\end{tabular}
\end{table}

%% file: sections/app-proofs.tex
\section{Proofs}
\label{sec.proofs}

\subsection{Proof of Theorem~\ref{thm.overall-efficiency-C}} 

\begin{proof}[Proof of Theorem~\ref{thm.overall-efficiency-C}]
In the coordinated model, miners wait for a message from the coordinator to start mining a block and a block is added to the chain if it is the first on top of the current chain that reaches the coordinator. We consider the process $\hat{B}$ that corresponds to the number of blocks mined from the viewpoint of the coordinator. When $\hat{B}$ increases by one, the coordinator sends a message to all miners, who then start mining the next block upon receiving the message, and $\hat{B}$ increases by one again when the first block reaches the coordinator; and so on. Hence, $\hat{B}$ is a renewal process and by the elementary renewal theorem (and by definition of $\eff$) we have \eqref{eq.overall-efficiency-C}
where $\btau$ is the expected time between two increments of $\hat{B}$. To conclude the proof, we show that $\btau$ satisfies \eqref{eq.taubar-C}. 

$\btau$ is the expected duration between the time when the coordinator sends a new message and the time when it receives the corresponding block back. Let $\tau_1, \cdots, \tau_n$ be the times taken by each miner to discover that block; $\tau_i$ is a random variable of distribution exponential with parameter $\hti_i$. If the block arriving first to the coordinator is from miner $m_1$, then the duration for the coordinator to receive the block after sending the message is $\tau_1 + 2l_1 = \tau_1 + \lti_1$, where $2l_1$ is the communication delay to receive the initial message from the coordinator and to send the block back after mining it. Similarly, the duration is $\tau_i + \lti_i$ if the block is first received from miner $m_i$. Hence we have
\begin{equation*}
    \btau = \btau_1 + \cdots + \btau_n, 
\end{equation*}
where
\begin{equation*}
    \btau_i = \int_{\tau_1 \ge 0} \hti_1 e^{-\hti_1 \tau_1}\ud \tau_1 \;\; \cdots \;\; \int_{\tau_n \ge 0} \hti_n e^{-\hti_n \tau_n}\ud \tau_n \;\; (\tau_i + \lti_i) \cdot \ind_{\tau_i + \lti_i \le \tau_j + \lti_j \textrm{ for all } j\neq i},
\end{equation*} 
where $\ind_{E}$ is $1$ if $E$ holds and $0$ otherwise.

We compute separately each $\btau_i$. We first do a change of variable, for all $i$, $\tau_i + \lti_i \to \tau_i$; this gives
\begin{equation*}
    \btau_i = e^{\sum_{j=1}^n \hti_j \lti_j} \int_{\tau_1 \ge \lti_1} \hti_1 e^{-\hti_1 \tau_1}\ud \tau_1 \;\; \cdots \;\; \int_{\tau_n \ge \lti_n} \hti_n e^{-\hti_n \tau_n}\ud \tau_n \;\; \tau_i \cdot \ind_{\tau_i \le \tau_j \textrm{ for all } j\neq i}.
\end{equation*} 
Then, we isolate the integral on $\tau_i$: 
\begin{align*}
    \btau_i = e^{\sum_{j=1}^n \hti_j \lti_j} 
    \int_{\tau_i \ge \lti_i} \tau_i \hti_i e^{-\hti_i \tau_i}\ud \tau_i 
    & \int_{\begin{subarray}{l}\tau_1 \ge \lti_1 \\ \tau_1 \ge \tau_i\end{subarray}} \hti_1 e^{-\hti_1 \tau_1}\ud \tau_1 \;\; \cdots \;\; \int_{\begin{subarray}{l}\tau_{i-1}\ge \lti_{i-1} \\ \tau_{i-1} \ge \tau_i\end{subarray}} \hti_{i-1} e^{-\hti_{i-1} \tau_{i-1}}\ud \tau_{i-1}\\ 
    & \int_{\begin{subarray}{l}\tau_{i+1}\ge \lti_{i+1} \\ \tau_{i+1} \ge \tau_i\end{subarray}} \hti_{i+1} e^{-\hti_{i+1} \tau_{i+1}}\ud \tau_{i+1} \;\; \cdots \;\; \int_{\begin{subarray}{l}\tau_n\ge \lti_n \\ \tau_n \ge \tau_i\end{subarray}} \hti_n e^{-\hti_n \tau_n}\ud \tau_n.
\end{align*} 
Observe that for $j=1, \cdots, i-1$, condition $\tau_j \ge \lti_j$ is automatically satisfied as soon as $\tau_j \ge \tau_i$ since we integrate $\tau_i$ on the domain $\tau_i \ge \lti_i$ and by assumption $\lti_i \ge \lti_j$. On the other hand, this is not the case for $j=i+1, \cdots, n$. To be able to compute the corresponding integrals we need to split the domain  $\tau_i\ge\lti_i$ into segments $[\lti_k, \lti_{k+1}]$ for $k=i, \cdots, n$ (recall that $l_{n+1}=\infty$). In such a segment, the domain of integration for $\tau_j$ is either $\tau_j \ge \tau_i$ or $\tau_j \ge \lti_j$: 
\begin{align*}
    \btau_i = e^{\sum_{j=1}^n \hti_j \lti_j} \cdot \sum_{k=i}^n 
    \int_{\tau_i \in [\lti_k, \lti_{k+1}]} \tau_i \hti_i e^{-\hti_i \tau_i}\ud \tau_i 
    & \int_{\tau_1 \ge \tau_i} \hti_1 e^{-\hti_1 \tau_1}\ud \tau_1 \;\; \cdots \;\; \int_{\tau_{i-1} \ge \tau_i} \hti_{i-1} e^{-\hti_{i-1} \tau_{i-1}}\ud \tau_{i-1}\\ 
    & \int_{\tau_{i+1} \ge \tau_i} \hti_{i+1} e^{-\hti_{i+1} \tau_{i+1}}\ud \tau_{i+1} \;\; \cdots \;\; \int_{\tau_k \ge \tau_i} \hti_k e^{-\hti_k \tau_k}\ud \tau_k\\
    & \int_{\tau_{k+1}\ge \lti_{k+1}} \hti_{k+1} e^{-\hti_{k+1} \tau_{k+1}}\ud \tau_{k+1} \;\; \cdots \;\; \int_{\tau_n\ge \lti_n} \hti_n e^{-\hti_n \tau_n}\ud \tau_n.
\end{align*} 

Next, using the standard equality $\int_{\tau \ge y} h e^{-h\tau} \ud \tau = e^{-hy}$, we get
\begin{align*}
    \btau_i = e^{\sum_{j=1}^n \hti_j \lti_j} \cdot \sum_{k=i}^n 
    e^{- (\hti_{k+1}\lti_{k+1} + \cdots + \hti_n\lti_n)} \int_{\tau_i \in [\lti_k, \lti_{k+1}]} \tau_i \hti_i e^{-(\hti_1+\cdots+\hti_k) \tau_i}\ud \tau_i
\end{align*} 
Finally, using that 
\begin{equation*}
    \int_{\tau \in [y_1, y_2]} h\tau e^{-h\tau} \ud \tau = \frac{1}{h} \left( (h y_1+1) e^{-hy_1} - (h y_2+1) e^{-hy_2} \right), 
\end{equation*}
we obtain 
\begin{align*}
    \btau_i &= e^{\sum_{j=1}^n \hti_j \lti_j} \cdot \sum_{k=i}^n 
    \frac{\hti_i \cdot e^{- (\hti_{k+1}\lti_{k+1} + \cdots + \hti_n\lti_n)}}{(\hti_1+\cdots+\hti_k)^2} 
    \bigg(  \Big((\hti_1+\cdots+\hti_k) \lti_k+1\Big) e^{-(\hti_1+\cdots+\hti_k)\lti_k} \\
    & \hspace{5.8cm} - \Big((\hti_1+\cdots+\hti_k) \lti_{k+1}+1\Big) e^{-(\hti_1+\cdots+\hti_k)\lti_{k+1}} \bigg) \\
    &= \hti_i \cdot \sum_{k=i}^n 
    \frac{e^{- (\hti_{1}\lti_{1} + \cdots + \hti_k\lti_k)}}{(\hti_1+\cdots+\hti_k)^2} 
    \bigg(  \Big((\hti_1+\cdots+\hti_k) \lti_k+1\Big) e^{-(\hti_1+\cdots+\hti_k)\lti_k} \\
    & \hspace{4.1cm} - \Big((\hti_1+\cdots+\hti_k) \lti_{k+1}+1\Big) e^{-(\hti_1+\cdots+\hti_k)\lti_{k+1}} \bigg).
\end{align*} 

To conclude, we sum the $\tau_i$'s for all $i$'s, exchange the sums on $i$ and $k$ (i.e., $\sum_{i=1}^n \sum_{k=i}^n \cdots = \sum_{k=1}^n \sum_{i=k}^n \cdots$), and merge the different terms with the same denominator $(\hti_1+\cdots+\hti_k)^2$. This yields \eqref{eq.taubar-C}, where we write the only remaining summation index as $i$ instead of $k$. 
\end{proof}

\begin{remark*}
In the proof above, if all latencies are zero (i.e., $\lti_i = 0$ for all $i$), we get $\btau_i = \hti_i / (\hti_1+\cdots+\hti_n)^2$. (Recall that by convention $\lti_{n+1} = \infty$ and $\lti_{n+1} e^{-h \lti_{n+1}} = 0$ for any $h>0$.) This yields $\btau = 1 / (\hti_1+\cdots+\hti_n) = \tau$, consistent with intuition.
\end{remark*}

\subsection{Proof of Theorem~\ref{thm.overall-efficiency-CvsP2P}}

\begin{proof}[Proof of Theorem~\ref{thm.overall-efficiency-CvsP2P}]
We prove the result under the assumption that $l_{ij} = l_i + l_j$ for all $i$ and $j$; that is, the coordinator introduces no extra delay. The result when $l_{ij} \le l_i + l_j$ for all $i$ and $j$ then immediately follows by applying Corollary~\ref{cor.increase-delay}.

Consider $n$ \emph{independent} Poisson point processes $N^{(1)}, \cdots, N^{(n)}$ starting at time $0$ with rates $\hti_1, \cdots, \hti_n$ respectively; and let $(T^{(i)}_j \ge 0)_{j\ge 1}$ be the sequence of increment times of point process $N^{(i)}, i\in\{1, \cdots, n\}$. Based on these $n$ processes, we will derive sequences of block discovery times statistically consistent with the P2P and the C models and show that the longest chain is always longer in the former than in the latter. 

Let $m_i\in\Mset$ be an arbitrary miner. First note that the sequence $(T^{(i)}_j)_{j\ge 1}$ without any modification is statistically consistent with block discovery times in the P2P model. Indeed, consider a time where miner $m_i$ starts mining a new block. This can be either upon discovering a block, or upon receiving a new chain from another miner $m_{j} \neq m_{i}$. Either way, by Markov property, since this start time is independent from the future, the time to the next increment of $N^{(i)}$ is still exponentially distributed with parameter $\hti_i$. Hence, denoting by $\hat{B}^{P2P}_{\{T^{(i)}\}_{i=1, \cdots, n}}(T)$ the size of the longest chain at time $T$ in the P2P model where blocks are discovered at times $(T^{(i)}_j)_{j\ge 1}$, we have 
\begin{equation*}
    \eff^{P2P} (\hb, L, \tau) = \lim_{T\to\infty} \frac{\E \left[ \hat{B}^{P2P}_{\{T^{(i)}\}_{i=1, \cdots, n}}(T)\right]}{T/\tau}.
\end{equation*}

Second, consider the C model. From the sequence $(T^{(i)}_j)_{j\ge 1}$, we construct the set of block discovery times $\{\tilde{T}^{(i)}_j\}_{j\ge 1}$ of miner $m_i$ in the coordinated model by removing all the points from $(T^{(i)}_j)_{j\ge 1}$ that correspond to blocks that ``would not have been discovered'' in the C model due to the pause after discovering a block. Note that it is not possible to simply remove all points within a window of size $\lti_i$ of a previous point at each miner $m_i$; this would remove too many points and no longer be consistent with the C model. Indeed, when a message arrives at $m_i$ from the coordinator announcing a block discovered by $m_{j} \neq m_{i}$, miner $m_i$ resumes mining before the end of the $\lti_i$-duration pause and we need to keep the following point even if it is within $\lti_i$ from the previous point. Concretely, to generate $\{\tilde{T}^{(i)}_j\}_{j\ge 1}$, we simply run the C mining protocol chronologically assuming that blocks are discovered at times $(T^{(i)}_j)_{j\ge 1}$ at each miner $m_i$ but removing each point that is after a discovered block and before the next message from the coordinator (which can be either an ack of $m_i$'s discovered block arriving a fixed $\lti_i$ later, or a message announcing a block discovered by another miner that arrives earlier). Since these messages are at times that do not depend on the future, the time to mine a block is still exponentially distributed with parameter $\hti_i$. Hence, the new set of block discovery times $\{\tilde{T}^{(i)}_j\}_{j\ge 1}$ is statistically consistent with the C model; that is, denoting by $\hat{B}^{C}_{\{\tilde{T}^{(i)}\}_{i=1, \cdots, n}}(T)$ the size of the chain at time $T$ in the C model where blocks are discovered at times $(\tilde{T}^{(i)}_j)_{j\ge 1}$, we have 
\begin{equation*}
    \eff^{C} (\hb, L, \tau) = \lim_{T\to\infty} \frac{\E \left[ \hat{B}^{C}_{\{\tilde{T}^{(i)}\}_{i=1, \cdots, n}}(T)\right]}{T/\tau}.
\end{equation*}

Next, observe that the length of the chain at any miner with discovery times $\{\tilde{T}^{(i)}_j\}_{j\ge 1}$ is the same under the P2P and C model; therefore 
\begin{equation*}
    \hat{B}^{C}_{\{\tilde{T}^{(i)}\}_{i=1, \cdots, n}}(T) = \hat{B}^{P2P}_{\{\tilde{T}^{(i)}\}_{i=1, \cdots, n}}(T).
\end{equation*} 
To see that, consider the sequence of events in the C model and imagine removing the coordinator. Due to our assumption that $l_{ij}=l_i + l_j$ for all $i$ and $j$, a block (or chain) sent from one miner will reach every other miners at exactly the same time in the C and P2P models. The only difference is that some of the chains that would have been blocked by the coordinator will now be forwarded to the other miner in the P2P model. However, in such cases, when the chain reaches the other miner, it will not make it switch because it will be of the same length or smaller than the chain held by that miner.

To conclude, observe that $\{\tilde{T}^{(i)}_j\}_{j\ge 1}$ is a subset of $\{T^{(i)}_j\}_{j\ge 1}$ for all miners $m_i\in\Mset$. Hence the longest chain obtained under the P2P model with discovery times $\{\tilde{T}^{(i)}_j\}_{j\ge 1}$ is smaller than or equal to that obtained under the P2P model with discovery times $\{T^{(i)}_j\}_{j\ge 1}$; that is
\begin{equation*}
     \hat{B}^{P2P}_{\{\tilde{T}^{(i)}\}_{i=1, \cdots, n}}(T) \le \hat{B}^{P2P}_{\{T^{(i)}\}_{i=1, \cdots, n}}(T). 
\end{equation*}
Combining the four equations above concludes the proof. 
\end{proof}

\subsection{Proof of Theorem~\ref{thm.individual-efficiency-C}}

\begin{proof}[Proof of Theorem~\ref{thm.individual-efficiency-C}]
As in the proof of Theorem~\ref{thm.overall-efficiency-C}, we consider the process $\hat{B_i}$ (the number of blocks from $m_i$ included in the chain) as a function of time. This process is a renewal reward process where the reward is one if a block is ``won'' by miner $m_i$ and zero otherwise. Hence, applying the elementary renewal theorem to $\hat{B}_i$ (and given the definition of $\hat{B}_i$ in \eqref{eq.efficiency.ind}) immediately gives \eqref{eq.individual-efficiency-C} where $p_i$ is the probability that the first block coming back to the coordinator is from miner $m_i$. To conclude the proof we show that $p_i$ satisfies \eqref{eq.pi-C}. 

The probability $p_i$ can be written as 
\begin{equation*}
    p_i = \int_{\tau_1 \ge 0} \hti_1 e^{-\hti_1 \tau_1}\ud \tau_1 \;\; \cdots \;\; \int_{\tau_n \ge 0} \hti_n e^{-\hti_n \tau_n}\ud \tau_n \;\; \ind_{\tau_i + \lti_i \le \tau_j + \lti_j \textrm{ for all } j\neq i}. 
\end{equation*}
We compute it similarly to the computation of $\btau_i$ is the proof of Theorem~\ref{thm.overall-efficiency-C} (it is the same expression with the $\tau_i$ in the integral). We first do a change of variable, for all $i$, $\tau_i + \lti_i \to \tau_i$; this gives
\begin{equation*}
    p_i = e^{\sum_{j=1}^n \hti_j \lti_j} \int_{\tau_1 \ge \lti_1} \hti_1 e^{-\hti_1 \tau_1}\ud \tau_1 \;\; \cdots \;\; \int_{\tau_n \ge \lti_n} \hti_n e^{-\hti_n \tau_n}\ud \tau_n \;\; \ind_{\tau_i \le \tau_j \textrm{ for all } j\neq i}.
\end{equation*} 
Then, we isolate the integral on $\tau_i$: 
\begin{align*}
    p_i = e^{\sum_{j=1}^n \hti_j \lti_j} 
    \int_{\tau_i \ge \lti_i} \hti_i e^{-\hti_i \tau_i}\ud \tau_i 
    & \int_{\begin{subarray}{l}\tau_1 \ge \lti_1 \\ \tau_1 \ge \tau_i\end{subarray}} \hti_1 e^{-\hti_1 \tau_1}\ud \tau_1 \;\; \cdots \;\; \int_{\begin{subarray}{l}\tau_{i-1}\ge \lti_{i-1} \\ \tau_{i-1} \ge \tau_i\end{subarray}} \hti_{i-1} e^{-\hti_{i-1} \tau_{i-1}}\ud \tau_{i-1}\\ 
    & \int_{\begin{subarray}{l}\tau_{i+1}\ge \lti_{i+1} \\ \tau_{i+1} \ge \tau_i\end{subarray}} \hti_{i+1} e^{-\hti_{i+1} \tau_{i+1}}\ud \tau_{i+1} \;\; \cdots \;\; \int_{\begin{subarray}{l}\tau_n\ge \lti_n \\ \tau_n \ge \tau_i\end{subarray}} \hti_n e^{-\hti_n \tau_n}\ud \tau_n.
\end{align*} 
We break up the domain  $\tau_i\ge\lti_i$ into segments $[\lti_k, \lti_{k+1}]$ for $k=i, \cdots, n$ (recall that $l_{n+1}=\infty$):
\begin{align*}
    p_i = e^{\sum_{j=1}^n \hti_j \lti_j} \cdot \sum_{k=i}^n 
    \int_{\tau_i \in [\lti_k, \lti_{k+1}]} \hti_i e^{-\hti_i \tau_i}\ud \tau_i 
    & \int_{\tau_1 \ge \tau_i} \hti_1 e^{-\hti_1 \tau_1}\ud \tau_1 \;\; \cdots \;\; \int_{\tau_{i-1} \ge \tau_i} \hti_{i-1} e^{-\hti_{i-1} \tau_{i-1}}\ud \tau_{i-1}\\ 
    & \int_{\tau_{i+1} \ge \tau_i} \hti_{i+1} e^{-\hti_{i+1} \tau_{i+1}}\ud \tau_{i+1} \;\; \cdots \;\; \int_{\tau_k \ge \tau_i} \hti_k e^{-\hti_k \tau_k}\ud \tau_k\\
    & \int_{\tau_{k+1}\ge \lti_{k+1}} \hti_{k+1} e^{-\hti_{k+1} \tau_{k+1}}\ud \tau_{k+1} \;\; \cdots \;\; \int_{\tau_n\ge \lti_n} \hti_n e^{-\hti_n \tau_n}\ud \tau_n.
\end{align*} 

Finally using the standard integral expression for the exponential distribution we get
\begin{align*}
    p_i & = e^{\sum_{j=1}^n \hti_j \lti_j} \cdot \sum_{k=i}^n 
    e^{- (\hti_{k+1}\lti_{k+1} + \cdots + \hti_n\lti_n)} \int_{\tau_i \in [\lti_k, \lti_{k+1}]} \hti_i e^{-(\hti_1+\cdots+\hti_k) \tau_i}\ud \tau_i \\ 
    & = e^{\sum_{j=1}^n \hti_j \lti_j} \cdot \sum_{k=i}^n 
    e^{- (\hti_{k+1}\lti_{k+1} + \cdots + \hti_n\lti_n)} \cdot \frac{\hti_i}{\hti_1+\cdots+\hti_k} \cdot 
    \left( e^{-(\hti_1+\cdots+\hti_k) \lti_{k}} - e^{-(\hti_1+\cdots+\hti_k) \lti_{k+1}} \right) \\ 
    & = \sum_{k=i}^n 
    e^{\hti_1\lti_1 + \cdots + \hti_k\lti_k} \cdot \frac{\hti_i}{\hti_1+\cdots+\hti_k} \cdot 
    \left( e^{-(\hti_1+\cdots+\hti_k) \lti_{k}} - e^{-(\hti_1+\cdots+\hti_k) \lti_{k+1}} \right) \\
    & = \sum_{k=i}^n \frac{\hti_i}{\hti_1+\cdots+\hti_k} \cdot 
    \left( e^{\sum_{j=1}^k -\hti_j (\lti_k-\lti_j)}  -  e^{\sum_{j=1}^k -\hti_j (\lti_{k+1}-\lti_j)} \right). 
\end{align*} 
This is exactly \eqref{eq.pi-C}, which concludes the proof.
\end{proof}